\documentclass{emulateapj}

\usepackage{lscape}

\def\ifm#1{\relax\ifmmode#1\else$\mathsurround=0pt #1$\fi}

\def\msun{M_{\odot}}

\def\ltsima{$\; \buildrel < \over \sim \;$}
\def\lsim{\lower.5ex\hbox{\ltsima}}
\def\gtsima{$\; \buildrel > \over \sim \;$}
\def\gsim{\lower.5ex\hbox{\gtsima}}
\def\eg{{\it e.g.}\,}
\def\ie{{\it i.e.}\,}
\def\th{^{\rm th}}

\def\sigmamedian{273}	%%note this is sigma e2!
\def\zdmedian{0.494}
\def\zdlower{0.16}
\def\zdupper{0.16}
\def\zdslacsmedian{0.189}
\def\zdslacslower{0.066}
\def\zdslacsupper{0.096}
\def\zdlsdmedian{0.810}

\def\zsmedian{1.199}
\def\zslower{0.18}
\def\zsupper{0.22}
\def\zsslacsmedian{0.606}
\def\zsslacslower{0.157}
\def\zsslacsupper{0.201}
\def\zslsdmedian{3.263}
\def\zslsdlower{0.61}
\def\zslsdupper{0.61}

\def\zd{z_{\rm d}}
\def\zs{z_{\rm s}}
\def\zspdf{z_{\rm s,pdf}}

\def\Ds{D_{\rm s}}
\def\Dds{D_{\rm ds}}
\def\Sigmacrit{\Sigma_{\rm crit}}

\def\paperI{Paper~I{}}

\def\Reff{R_{\rm eff}}
\def\Reff2{R_{\rm eff}/2}
\def\Reff{R_{\rm eff}}
\def\REin{R_{\rm Ein}}
\def\MEin{M_{\rm Ein}}
\def\fdm{f_{\rm DM}}
\def\fdmsre2{f^{\rm Salp}_{{\rm DM,}\Reff2}}
\def\fdmcre2{f^{\rm Chab}_{{\rm DM,}\Reff2}}
\def\fdmbar{f_{{\rm DM,}0}}

\def\hst{{\it HST}{ }}

\def\sigmasie{\sigma_{\mathrm{SIE}}}
\def\sigmae2{\sigma_{\rm e2}}
\def\galfit{{\tt galfit}{ }}

\def\magz{m_{\rm z}}
\def\magi{m_{\rm i}}
\def\magr{m_{\rm r}}
\def\magg{m_{\rm g}}
\def\magu{m_{\rm u}}

\def\pr{{\rm Pr}}
\def\pixels{\mathbf{d}}

\def\CaII{Ca\,{\sc ii}}
\def\NaII{Na\,{\sc ii}}
\def\HeII{He\,{\sc ii}}
\def\Mgb{Mg\,{\sc ii}}
\def\Hb{H$\beta$}
\def\OII{[O\,{\sc ii}]}
\def\OIII{O\,{\sc iii}]}
\def\CIII{C\,{\sc iii}]}
\def\CIV{C\,{\sc iv}}

\def\Ntot{17}
\def\Ndyn{11} % number of sl2s lenses gold + silver
\def\Nnohst{6} % number of sl2s lenses that still lack HST data.
\def\meangammap{2.16}
\def\errplusmeangammap{0.09}
\def\errminusmeangammap{0.09}
\def\scattergammap{0.25}
\def\errplusscattergammap{0.10}
\def\errminusscattergammap{0.07}
\def\meanfdm{0.42}
\def\errplusmeanfdm{0.08}
\def\errminusmeanfdm{0.08}
\def\scatterfdm{0.20}
\def\errplusscatterfdm{0.09}
\def\errminusscatterfdm{0.07}
\def\meanfdmchab{0.68}
\def\errplusmeanfdmchab{0.04}
\def\errminusmeanfdmchab{0.06}
\def\scatterfdmchab{0.11}
\def\errplusscatterfdmchab{0.06}
\def\errminusscatterfdmchab{0.04}
 
\def\meanvgammap{2.22}
\def\errplusmeanvgammap{0.17}
\def\errminusmeanvgammap{0.21}
\def\gradvgammap{-0.16}
\def\errplusgradvgammap{0.48}
\def\errminusgradvgammap{0.51}
\def\scattervgammap{0.23}
\def\errplusscattervgammap{0.09}
\def\errminusscattervgammap{0.06}
\def\meanallvgammap{2.12}
\def\errplusmeanallvgammap{0.03}
\def\errminusmeanallvgammap{0.04}
\def\gradallvgammap{-0.25}
\def\errplusgradallvgammap{0.10}
\def\errminusgradallvgammap{0.12}
\def\scatterallvgammap{0.17}
\def\errplusscatterallvgammap{0.02}
\def\errminusscatterallvgammap{0.02}
\def\meanvfdm{0.38}
\def\errplusmeanvfdm{0.12}
\def\errminusmeanvfdm{0.50}
\def\gradvfdm{0.07}
\def\errplusgradvfdm{0.36}
\def\errminusgradvfdm{0.36}
\def\scattervfdm{0.18}
\def\errplusscattervfdm{0.08}
\def\errminusscattervfdm{0.07}
\def\meanallvfdm{0.27}
\def\errplusmeanallvfdm{0.06}
\def\errminusmeanallvfdm{0.06}
\def\gradallvfdm{0.36}
\def\errplusgradallvfdm{0.18}
\def\errminusgradallvfdm{0.24}
\def\scatterallvfdm{0.13}
\def\errplusscatterallvfdm{0.02}
\def\errminusscatterallvfdm{0.02}
\shorttitle{Lens Mass Structure}
\shortauthors{Ruff et~al.}

\def\umel{2}
\def\iap{3}
\def\ucsb{1}
\def\kipac{4}

\begin{document}

\title{The SL2S Galaxy-scale Lens Sample.  II. Cosmic evolution of
dark and luminous mass in early-type galaxies}

\author{Andrea~J.~Ruff\altaffilmark{\ucsb,\umel}$^{*}$}
\author{Rapha\"el~Gavazzi\altaffilmark{\iap}}
\author{Philip~J.~Marshall\altaffilmark{\ucsb,\kipac}}
\author{Tommaso~Treu\altaffilmark{\ucsb}$^{\dag}$}
\author{Matthew~W.~Auger\altaffilmark{\ucsb}}
\author{Florence~Brault\altaffilmark{\iap}}
%\author{the SL2S collaboration}

% Marshall, Treu, Ruff:
\altaffiltext{\ucsb}{Physics department, University of California, Santa Barbara, CA 93106, USA} 
% Ruff:
\altaffiltext{\umel}{University of Melbourne, Parkville 3010, Victoria, Australia}
% Gavazzi, Brault:
\altaffiltext{\iap}{Institut d'Astrophysique de Paris, UMR7095 CNRS - Universit\'e
 Pierre et Marie Curie, 98bis bd Arago, 75014 Paris, France}
\altaffiltext{\kipac}{KIPAC, P.O. Box 20450, MS29, Stanford, CA 94309, USA}
\altaffiltext{*}{{\tt aruff@unimelb.edu.au}}
\altaffiltext{$\dag$}{{Packard Research Fellow}}

%-------------------------------------------------------------------------------

\begin{abstract}

We present a joint gravitational lensing and stellar-dynamical analysis of 11
early-type galaxies (median deflector redshift $\zd=0.5$) from Strong Lenses
in the Legacy Survey (SL2S).  
Using newly measured redshifts and stellar velocity dispersions from 
Keck spectroscopy with lens models from Paper I, we derive the total mass 
density slope inside the Einstein radius for each of the 11 lenses.  The
average total density slope is found to be 
$\langle\gamma'\rangle$$\,=\,$$\meangammap^{+\errplusmeangammap}_{-\errminusmeangammap}$ 
($\rho_{\rm tot}\propto r^{-\gamma'}$), with an intrinsic scatter of  
$\scattergammap^{+\errplusscattergammap}_{-\errminusscattergammap}$.  We also
determine the dark matter fraction for each lens within half the effective
radius, $\Reff/2$ and find the average projected dark matter mass fraction to
be $\meanfdm^{+\errplusmeanfdm}_{-\errminusmeanfdm}$ with a scatter of 
$\scatterfdm^{+\errplusscatterfdm}_{-\errminusscatterfdm}$ for a
\citeauthor{Salpeter1955} IMF.
By combining the SL2S results with those from the Sloan Lens ACS
Survey (median $\zd=0.2$) and the Lenses Structure and Dynamics survey
(median $\zd=0.8$), we investigate cosmic evolution of $\gamma'$ and
find a mild trend
$\partial\langle\gamma'\rangle/\partial\zd$$\,=\,$$\gradallvgammap^{+\errplusgradallvgammap}_{-\errminusgradallvgammap}$.
This suggests that the total density profile of massive galaxies has
become slightly steeper over cosmic time. If this result is confirmed
by larger samples, it would indicate that dissipative processes played
some role in the growth of massive galaxies since $z\sim1$.
\end{abstract}


%-------------------------------------------------------------------------------

\section{Introduction}\label{sect:intro}

Early-type (\ie elliptical and lenticular) galaxies in the local
universe are considered to be simple objects
\citep[\eg][]{B+S93,Mer99,Cio09a}. In the central few kpc most of
their mass is dominated by stars, while at larger radii there is
convincing evidence -- at least for the most massive systems -- that
dark matter halos are dominant. Their stellar populations are relatively
simple, dominated by old stars with little or negligible star
formation \citep[\eg][]{Ren06}. Remarkably, many global properties,
ranging from the chemical composition of their stars to their size,
luminosity and mass of the central black hole, correlate tightly with
their stellar velocity dispersion, $\sigma$
\citep[\eg][]{Ber++05,GFS09a}.

In spite of this suspected simplicity, their formation and evolution
are still poorly understood, and therefore they are the subject of
many observational and theoretical investigations. A number of
observational facts have proved difficult to explain by theoretical
models. These include: i) the tightness of the empirical correlations
with $\sigma$ \citep[\eg,][]{Ber++05,GFS09a,Nip++09}; ii) the
so-called downsizing trend of their stellar populations, \ie the
correlation between mean stellar age and present day stellar mass
\citep[\eg,][]{Tho++05,Tre++05,vDW++05,Jun++05,diS++05}; iii) the
evolution of the upper end of their mass function since $z\sim1$
\citep[\eg,][]{BEC05,BTE07,CDR06,vdW++09,Hop++10d}; iv) the unusually
compact size of high redshift massive red galaxies
\citep[\eg,][]{Tre++98,Dad++05,vDo++08,SLA09,Cas++09,Man++10,New++10}. From
a theoretical standpoint, it is clear that understanding the interplay
between baryons, black holes, and dark matter is essential to develop
a scenario that can quantitatively reproduce all observations. The
physical processes that need to be accurately modeled in a successful
theory appear to include black hole accretion and the related energy
and momentum feedback \citep[\eg,][]{Cro++06,COP09}, dry and wet
major mergers \citep[\eg,][]{K+S06,CLV07,Rob++06c}, as well as minor
mergers and dry accretion of minor satellites
\citep[\eg,][]{NJO09,Hop++10d}.

Most of the observational studies, including those listed in the
previous paragraph, are concerned with global parameters of early-type
galaxies. An entirely new line of investigation can be opened up if we
are able to dissect early-type galaxies and map their internal
dynamical structure as a function of cosmic time.  By decomposing the
internal mass distribution of early-type galaxies into luminous and
dark components we can start addressing the following questions, the %%%%
answers to which would provide essential clues as to their formation
and evolution. How and when is mass assembled to form early-type
galaxies?  How are baryons converted into stars and accumulated inside
dark matter halos? Is the mass density profile of early-type galaxies
comparable to that observed in numerical simulations? Do isolated
early-type galaxies undergo internal structural and dynamical
evolution? As numerical simulations of early-type galaxies become more
and more realistic, detailed knowledge of their internal structure
(\eg their distribution functions) will provide more and more
stringent tests of the current paradigm of structure formation
\citep[\eg][]{Mez++03,Naa++07,O++07,L+O10}.

Great progress in answering these questions has been achieved in the
past few years with the systematic study of early-type galaxies acting
as strong gravitational lenses. For these systems, strong lensing
provides an absolutely calibrated measurement of mass at a fiducial
radius (the Einstein radius), which is typically comparable in size to
the effective radius. By combining this mass tracer with traditional
diagnostics such as stellar velocity dispersion
\citep{Mir95,N+K96,T+K02a}, and stellar mass maps from multicolor
imaging and/or spectroscopy, one can break many of the degeneracies
inherent to each method alone, including the mass-anisotropy
degeneracy, bulge-halo degeneracy, and the stellar mass/initial mass
function (IMF) degeneracy \citep[\eg][]{K+T03,T+K04,Tre++10}.
Additional information can be gathered with the addition of
weak-lensing \citep{Gav++07,J+K07,Lag++09}, although at the moment
this is not possible for individual galaxies. The SLACS team applied
this methodology to a sample of more than 80 early-type galaxies
\citep{Bol++06,Bol++08a,Aug++09}, which have been shown to be
indistinguishable from equally massive non-lensing early-type galaxies
in terms of their internal properties and environment
\citep{Bol++06,Tre++06,Tre++09}.

Among the most relevant findings of SLACS is that the total mass-density
profile $\rho_{\rm tot}\propto r^{-\gamma'}$ of early-type galaxies is close
to isothermal with $\gamma'=2.085^{+0.025}_{-0.018}$ with intrinsic scatter
less than 0.1 \citep{Koo++09,Bar++09,Aug++10}, even though neither the
stars nor the dark matter obey a simple power law profile \citep[see
also][]{WBB04}. This ``bulge-halo conspiracy'', similar to the disk-halo
conspiracy found for spiral galaxies \citep[\eg][]{v+S86}, has implications
both for lensing studies but also for galaxy formation studies. Since the total
mass density profile is preserved by dry-mergers \cite[][]{Deh05,KZK06,NTB09},
and dark matter-only profiles are not isothermal \citep[\eg][]{Nav++10}, the
isothermal nature has to be established through dissipational processes
\citep[\eg,][]{Koo++06}.  The other main finding of the SLACS survey is that 
(assuming a constant stellar IMF) the fraction of dark matter $\fdm$ within a
fixed fraction of the effective radius increases with galaxy mass or stellar
velocity dispersion \citep[see also][]{J+K07,Gri++08b,Cardone09,Cardone10}.
Possible explanations for this trend are: varying
efficiency in converting baryons into stars as a function of halo mass, varying
inner slope of the dark matter halo with mass, and evolutionary processes. For
example, dry mergers can increase the fraction of dark matter within the
effective radius \citep[\eg,][]{NTB09}.

The SLACS, however, sample is limited to low redshift by the selection
function of the parent sample of SDSS luminous galaxies.  Therefore,
evolutionary studies with the SLACS sample are limited to
a short baseline. Suitable samples of strong lenses are smaller at high redshift.   
In fact, most of the strong lensing
galaxies known to date at $z>0.4$ are too faint and/or dominated by
strongly lensed bright quasars to allow for detailed kinematic studies
of the deflector. Of the handful of exceptions
\citep[\eg,][]{Fau++08,Lag++09}, only a few of them have published
lensing and dynamical analysis \citep{T+K02a,Ohy++02,T+K04,Suy++10}.
These early studies found results similar to SLACS, \ie the internal
slope $\gamma'$ is close to isothermal, and the central dark matter
fraction appears to increase with mass as found by the LSD Survey
\citep{T+K04}. However, the results also hint that things may have been
different at $z\sim1$ when the universe was less than half its present
age: the scatter in $\gamma'$ might have been larger \citep{T+K04} and
perhaps even the average might have been different
\citep{Koo++06}. Unfortunately, current samples beyond $z\sim0.4$ with
lensing and dynamical data are too small to probe the evolution of
the internal structure of early-type galaxies, its scatter and trends
with mass at the same time.

In this paper we present the first detailed study of a sample of early-type
lens galaxies identified by the Strong Lenses in the Legacy survey (SL2S).  We
use new spectroscopic data obtained at the W.M. Keck telescopes, in combination
with multicolor photometry from the CFHT Legacy Survey to infer dynamical and
stellar mass for the deflector galaxies. This imaging data-set is complemented
by \hst imaging that allowed to confirm the lensing nature of the SL2S
candidates and detailed lens modeling. The newly measured source and deflector
redshifts are combined with gravitational lens models from \paperI\ (Gavazzi et
al., in preparation) to infer lensing masses.  We determine the total mass
density profile of the early-type lens galaxies (quantified by $\gamma'$) and
their central dark matter fraction ($\fdm$).  The median redshift of the sample
is $z=\zdmedian$, providing an ideal complement to the earlier SLACS and LSD
samples. By combining the three samples we investigate evolutionary trends in
these quantities.

This paper is organized as follows. In Section~\ref{sect:survey} we describe
the SL2S survey, and how the galaxy-scale lens candidates were selected. We
then present our spectroscopic measurements in Section~\ref{sect:spec} before
combining them with lens model parameters to investigate the mass structure of
massive galaxies since redshift 0.9 in Section~\ref{sect:lsd}. In
Section~\ref{sect:mstar} we incorporate stellar mass estimates in order to
separate the dark and luminous components of the lens galaxies. The cosmic
evolution of the total mass density slope and dark matter fraction are
discussed in Section~\ref{sec:evolution}. After a brief discussion of our
results in Section~\ref{sect:discuss} we conclude in Section~\ref{sect:concl}.
Throughout this paper magnitudes are given in the AB system.  We assume a
concordance cosmology with matter and dark energy density $\Omega_m=0.3$,
$\Omega_{\Lambda}=0.7$, and Hubble constant H$_0$=70 km s$^{-1}$Mpc$^{-1}$.

%-------------------------------------------------------------------------------

\section{Strong Lenses in the Legacy Survey: SL2S}\label{sect:survey}

In this section we describe our feeder survey, the CFHT Legacy Survey, and how
we identified our new sample of galaxy scale lenses at intermediate redshift. A
more detailed description is given in \paperI. 

Strong lensing candidates were selected from the Canada-France-Hawaii Telescope
Legacy Survey (CFHTLS)\footnote{See
\url{http://www.cfht.hawaii.edu/Science/CFHLS/} and links therein for a
comprehensive description}.  In brief, the survey consists of two main
components of sufficient depth and image quality to be interesting for lens
searching\footnote{{\tt http://terapix.iap.fr/rubrique.php?id$\_$rubrique=259}}.
Both are imaged in the $u^*$, $g$, $r$, $i$ and $z$ bands with the 1~deg$^2$
field-of-view Megacam Camera. The multi epoch Deep survey covers 4 pointing of
1~deg$^2$ each. Two different image stacks were produced: D-85 contains the
85\% best seeing images whereas the D-25 only includes the 25\% best seeing
images.  For finding lenses we only considered these latter, better resolution
stacks.  They reach a typical depth of $u^*\simeq 26.18$, $g=25.96\simeq
25.47$, $r\simeq 25.43$, $i\simeq 25.08$ and $z\simeq 24.57$ (80\% completeness
for point sources) with typical FWHM point spread functions of $0\farcs75$,
$0\farcs69$, $0\farcs64$, $0\farcs62$ and $0\farcs61$, respectively.  The Wide
survey is a single epoch imaging survey, covering some 171 deg$^2$ in 4 patches
of the sky. It reaches a typical depth of $u^*\simeq 25.35$, $g\simeq 25.47$,
$r\simeq24.83$, $i\simeq 24.48$ and $z\simeq 23.60$ (AB mag of 80\%
completeness limit for point sources) with typical FWHM point spread functions 
of $0\farcs85$, $0\farcs79$, $0\farcs71$, $0\farcs64$ and $0\farcs68$,
respectively. Because of the greater area, the Wide component is our main
provider of lens candidates.

Images from both the Deep and Wide survey modes were analyzed to find strong
lens candidates using several algorithms, as described by \citet{Cab++07} at
the group and cluster mass scales, and by Gavazzi et al. (in preparation) at
the galaxy scale with the {\tt RingFinder} algorithm. The ring-detecting
algorithm is aimed at detecting compact rings around centers of isolated
galaxies ($<10^{13} h^{-1}M_{\odot}$), and works by focusing on the achromatic
image excesses around early-type lens galaxies that are indicative of the
presence of lensed arcs. For each of a sample of pre-selected bright ($i_{\rm
AB} \le 22.5$) red galaxies, a scaled, PSF-matched version of the $i$-band
cutout image was subtracted from the $g$-band image of the same system. The
rescaling in this operation is performed such that the early-type galaxy light
is efficiently removed, leaving only objects with an SED different from that of
the target galaxy. These (typically) blue residuals are then characterized with
an object detector, and analyzed for their position, ellipticity, and
orientation, and those showing characteristic properties of lensed arcs are kept as
lens candidates. A sample of several hundred good candidates were visually
inspected, and ranked for follow-up with {\em HST}. 

Currently, 65 CFHT galaxy-scale lens candidates have been observed firstly with ACS,
then WFPC2, and finally with WFC3 as snapshot programs over cycles 15, 16 and
the ongoing cycle 17. Details of these observations are given in \paperI.
Approximately 50\% of the lens candidates were confirmed as lenses in this way.
The sources are all faint blue galaxies, with very few showing signs of an
active nucleus.
Those with the most convincing lens models (see \paperI) were selected for
spectroscopic follow up to obtain high precision redshifts for lens and source
galaxies, and lens galaxy velocity dispersions. These observations are
described in the next section.

%-------------------------------------------------------------------------------

\section{Spectroscopic observations}\label{sect:spec}

%%%%%%%%%%%%%%%%%%%%%%%%%%%%%%%%%%
\begin{figure*}
\centering\includegraphics[width=0.9\linewidth]{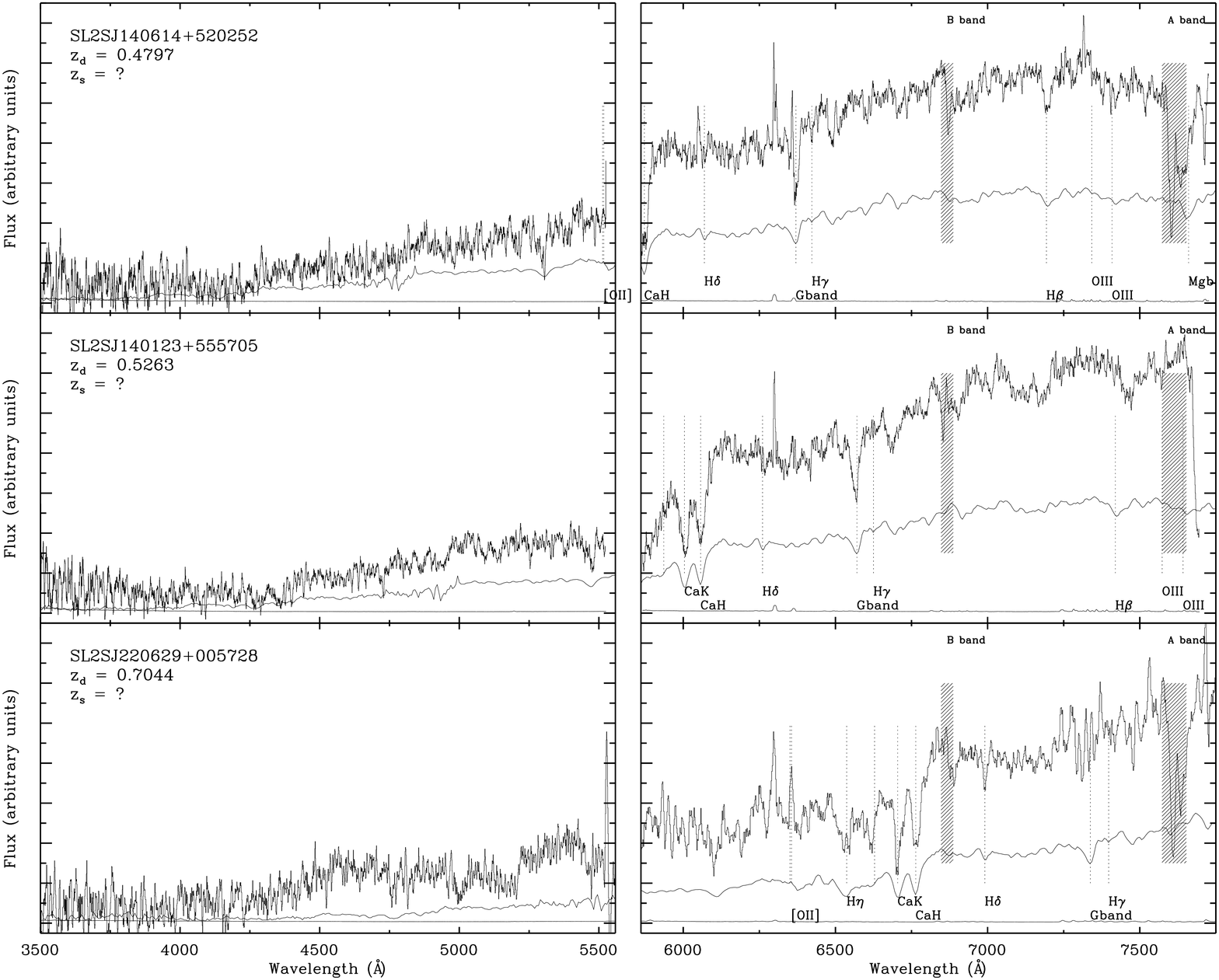}\\
\caption{\label{fig:spec1} Keck/LRIS spectra for 17 SL2S lenses. An
elliptical galaxy template is shown in a thin black line beneath each
spectrum. The dotted black lines mark features of the lens galaxy, while
source emission lines are highlighted in red. 
The shaded regions indicate A and B band telluric absorption features.
}
\end{figure*}
%%%%%%%%%%%%%%%%%%%%%%%%%%%%%%%%%%

%%%%%%%%%%%%%%%%%%%%%%%%%%%%%%%%%%
\begin{figure*}
\figurenum{\ref{fig:spec1}}
\centering\includegraphics[width=0.9\linewidth]{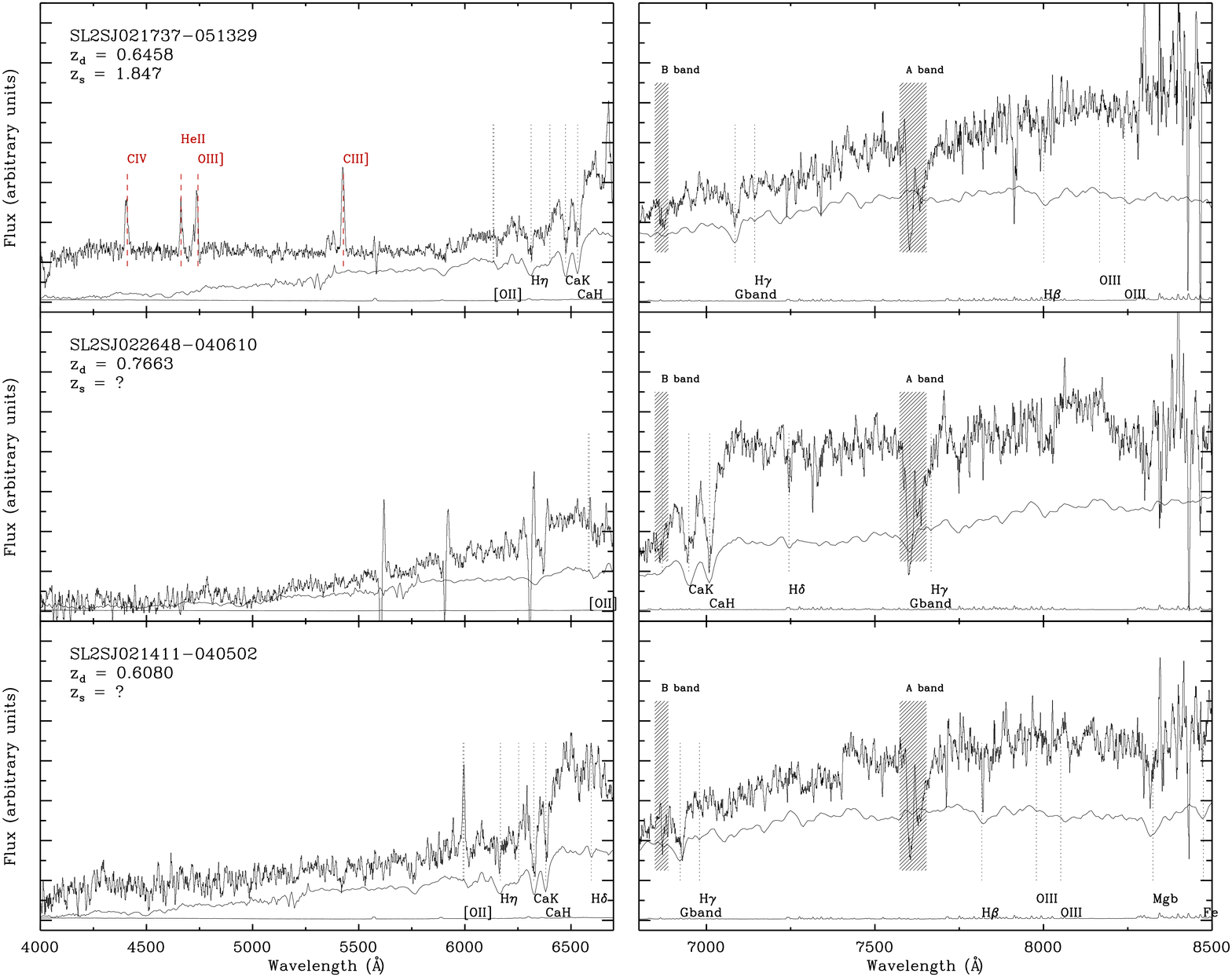}\\
\caption{{\it continued.}}
\end{figure*}
%%%%%%%%%%%%%%%%%%%%%%%%%%%%%%%%%%

%%%%%%%%%%%%%%%%%%%%%%%%%%%%%%%%%%
\begin{figure*}
\figurenum{\ref{fig:spec1}}
\centering\includegraphics[width=0.9\linewidth]{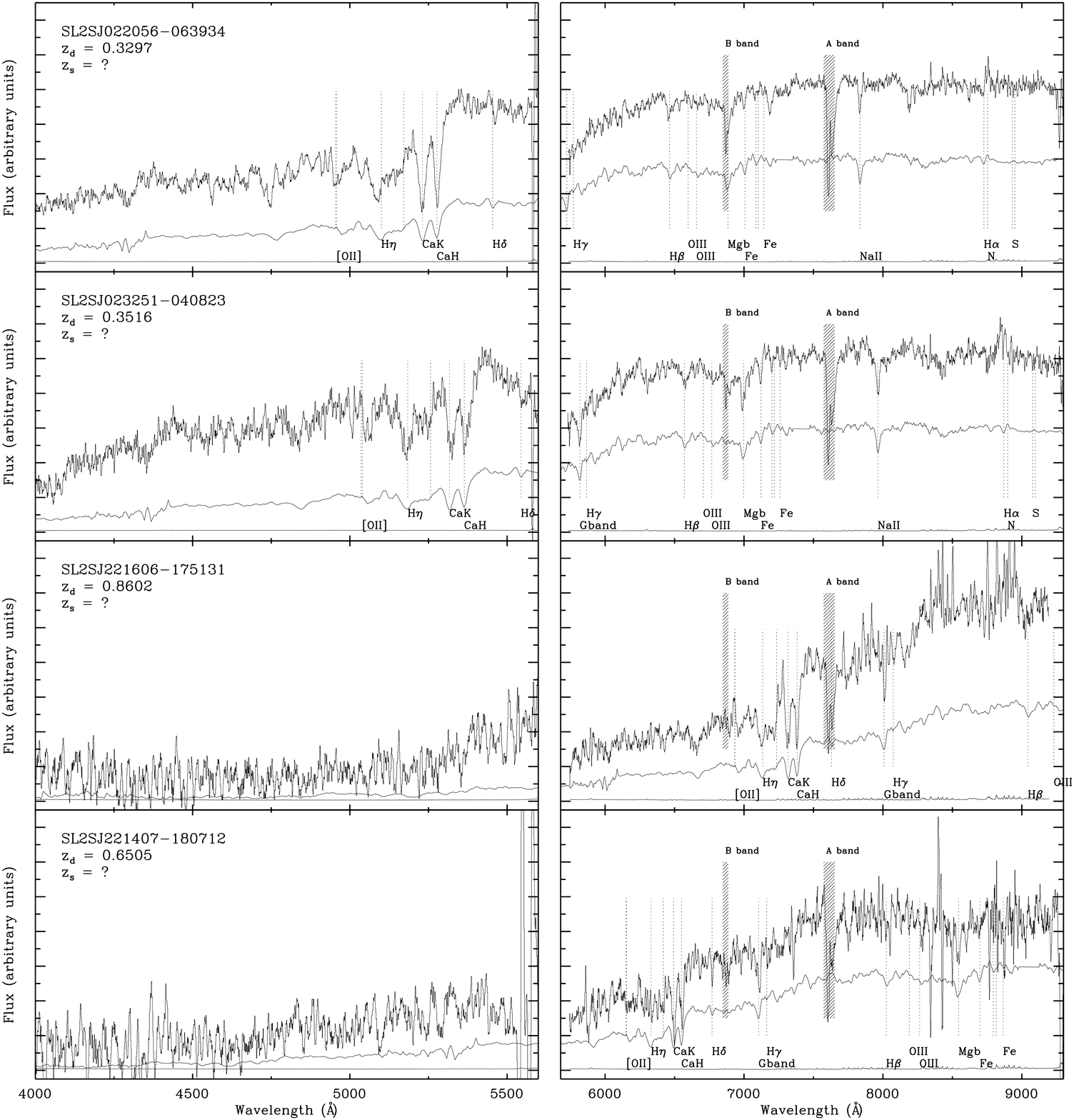}\\
\caption{{\it continued.}}
\end{figure*}
%%%%%%%%%%%%%%%%%%%%%%%%%%%%%%%%%%

%%%%%%%%%%%%%%%%%%%%%%%%%%%%%%%%%%
\begin{figure*}
\figurenum{\ref{fig:spec1}}
\centering\includegraphics[width=0.9\linewidth]{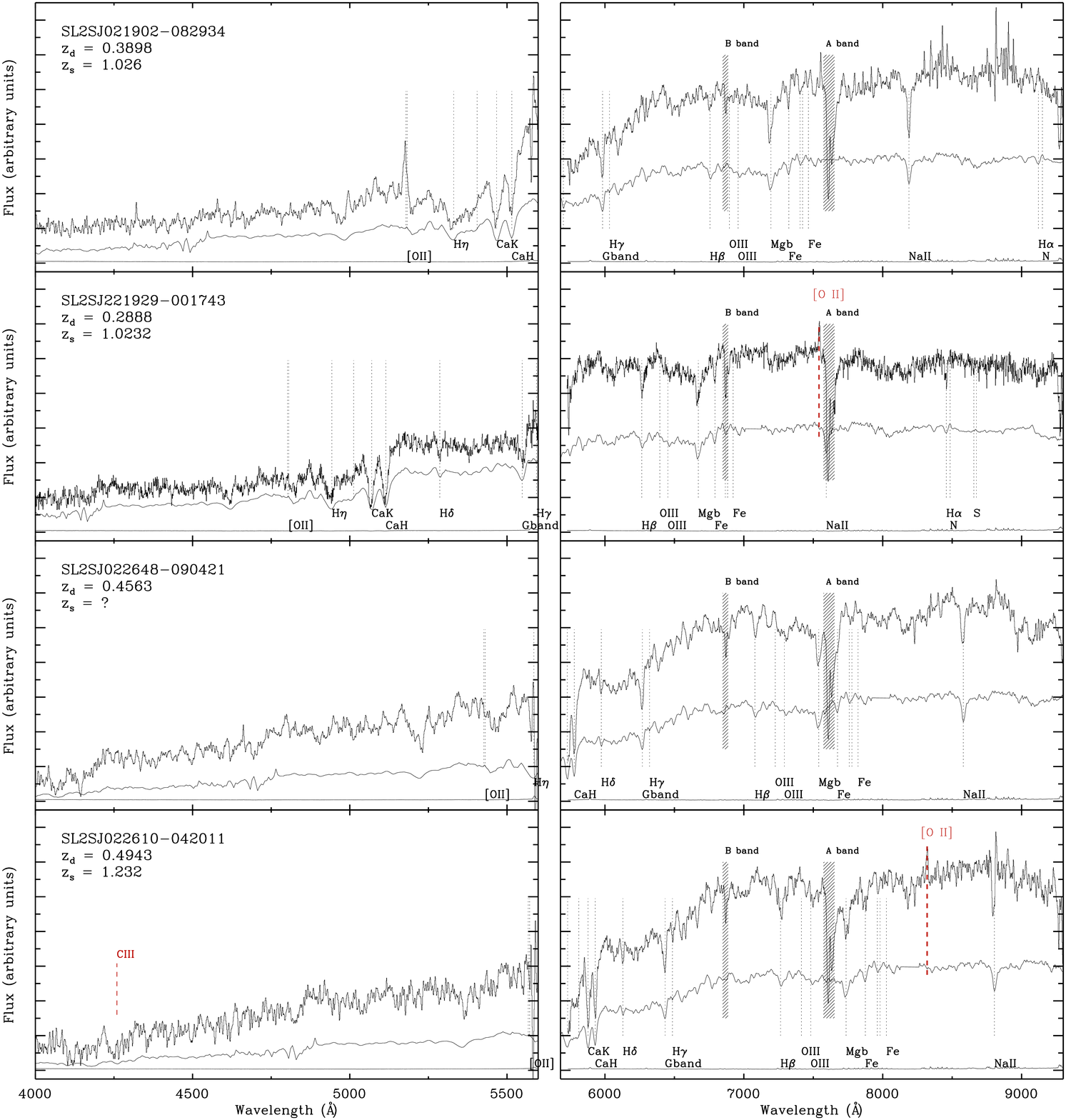}\\
\caption{{\it continued.}}
\end{figure*}
%%%%%%%%%%%%%%%%%%%%%%%%%%%%%%%%%%

%%%%%%%%%%%%%%%%%%%%%%%%%%%%%%%%%%
\begin{figure*}
\figurenum{\ref{fig:spec1}}
\centering\includegraphics[width=0.9\linewidth]{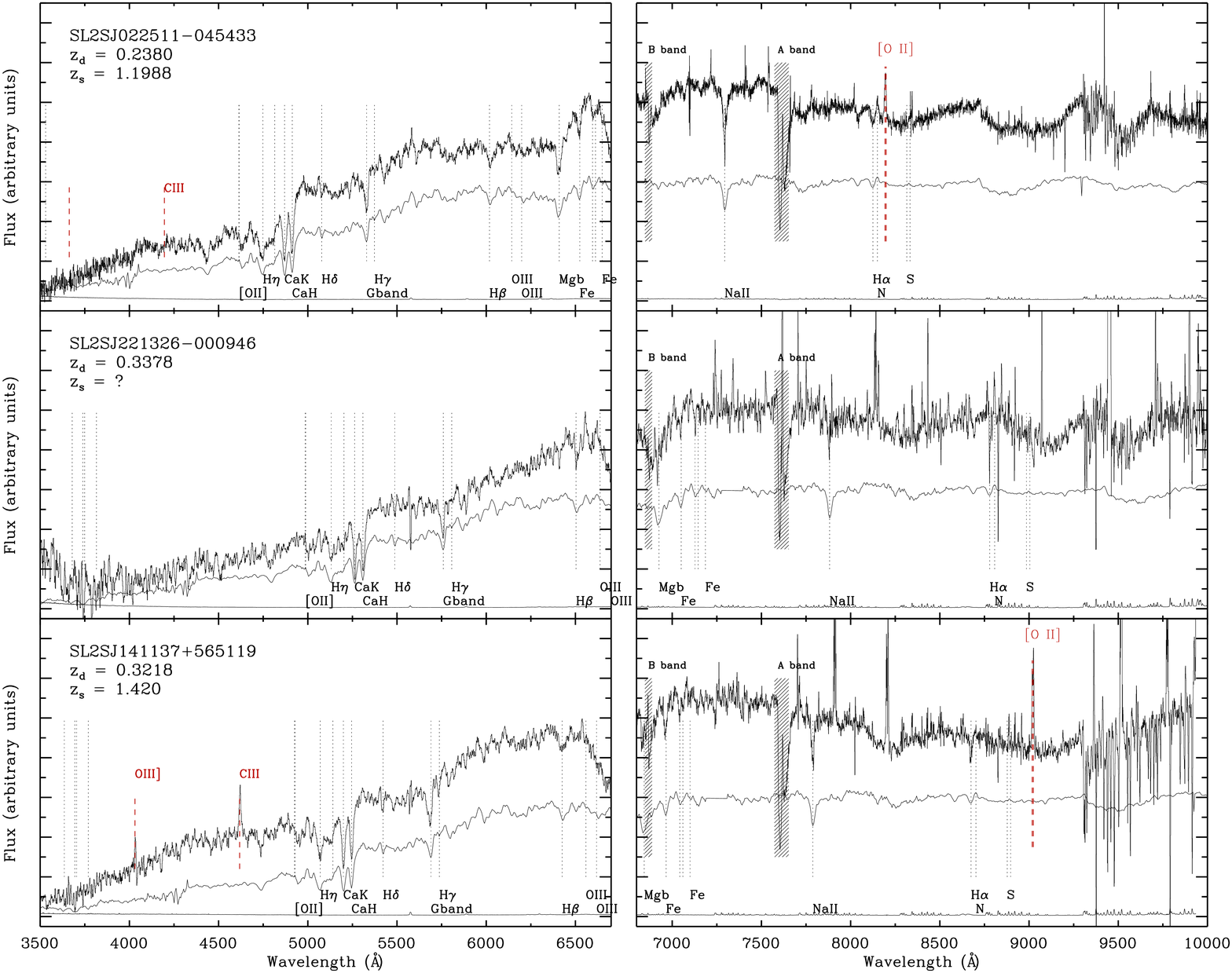}\\
\caption{{\it continued.}}
\end{figure*}
%%%%%%%%%%%%%%%%%%%%%%%%%%%%%%%%%%

Follow up spectroscopy of \Ntot~selected lens candidates was obtained using the
Low Resolution Imaging Spectrograph (LRIS) on the Keck~I telescope over six nights
from 2006 to 2010, with the aim of measuring the deflector and source redshifts
and the velocity dispersion of the deflector. 
%%%%%%%%%%%%%% dre, fix this!
Because of incompleteness in the execution of the \hst snapshot observations
mentioned above, we have spectra of \Nnohst~systems where \hst data is not yet 
available. These lens candidates, however, were very promising on the basis
of their CFHT images alone.
In this section we describe the observations and their analysis in some detail,
outlining our methodology for measuring the lens redshifts, source redshifts,
and lens velocity dispersions.

% - - - - - - - - - - - - - - - - - - - - - - - - - - - - - - - - - - - - - - - 

\subsection{Details of the observations and data reduction}

Observations were made in long slit mode, with the slit centered on the
deflector. The slit orientation was chosen to maximize both the source flux and
the spatial separation of the deflector and arc. 
The total flux is dominated by the deflector, so maximizing the source flux
maximizes the likelihood of measuring the source redshift, $\zs$.  Spatial
separation within the slit also separates the deflector and arc traces in the
2D spectra, allowing us to look for source emission lines directly, as
discussed in Section~\ref{subsec:sourcez}.  
The seeing ranged between 0$\farcs$55 and 1$\farcs$3  and total exposure times
varied from 20 to 45~min, typically with two consecutive 15~min intervals.
Between each exposure, we dithered along the slit by 10$''$ to improve defect
removal and sky subtraction. A summary of each observing run is given in
Table~\ref{table:logs}.

%%%%%%%%%%%%%%%%%%%%%%%%%%%%%%%
\begin{deluxetable}{ccccccc}
	\tabletypesize{\small}
\tablecaption{
\label{table:logs}
Observing logs}
	\tabletypesize{\small}
\tablehead{
run   &     obsdate        & graname &grisname &slit&time &seeing  \\}
\startdata
1     &     20$\,$ Jul 2006 & 831/8200   & 600/4000  & 1$\farcs$5  & 20  & 0$\farcs$84            \\
2a    &     23 Dec 2006 & 831/8200   & 300/5000  & 1$\farcs$5  & 60  & 0$\farcs$60            \\
2b    &     23 Dec 2006 & 831/8200   & 300/5000  & 1$\farcs$5  & 40  & 0$\farcs$80            \\
3a    &     13 Sep 2007 & 400/8500   & 400/3400  & 1$\farcs$0  & 45  &  0$\farcs$73\\
3b    &     13 Sep 2007 & 400/8500   & 400/3400  & 1$\farcs$0  & 30  &  0$\farcs$75\\
4a    &     14 Sep 2007 & 400/8500   & 400/3400  & 1$\farcs$0  & 30  &  0$\farcs$55 \\
4b    &     14 Sep 2007 & 600/7500   & 600/4000  & 0$\farcs$7  & 30  &  0$\farcs$55 \\
4c    &     14 Sep 2007 & 831/8200   & 600/4000  & 1$\farcs$0  & 90  &  0$\farcs$55 \\
5     &     $\,\,$9  Sep 2009 & 600/7500   & 300/5000  & 1$\farcs$0  & 30  &  1$\farcs$0  \\
6     &     14 Jan 2010 & 600/7500   & 300/5000  & 1$\farcs$0  & 45  &  1$\farcs$3 \\
\enddata
\tablecomments{
The exposure time is given in minutes. 
The plate scale for both the red and blue chips is 
0$\farcs$135/pixel for 
observations from 2009 and after. For observing runs before 2009, 
the red chip had  plate scale of 0$\farcs$211/pixel.}
\end{deluxetable}

The later setups took greater advantage of the blue sensitivity of LRIS. We
found that the most effective method to measure all three quantities was using
a 680 dichroic, so that the deflector redshift and velocity dispersion could be
measured in the blue. The 680 dichroic allowed us to choose a large wavelength
coverage for the red, which was useful in detecting \OII~out to $z>1.1$. 

The data was reduced using a Python pipeline (developed by MWA). On the red
side, the night sky lines were used to determine the wavelength solution, 
while standard arclamps were used on the blue side. 
Two different extraction windows were chosen: one wide extraction window to
ensure that light from the source was also included in the final spectrum (half
widths between 5 and 12 pixels from the central trace) and another narrower
extraction window to increase the signal to noise ratio, which is important for
measuring velocity dispersions. 
The reduced spectra are shown in Figure~\ref{fig:spec1}.

% - - - - - - - - - - - - - - - - - - - - - - - - - - - - - - - - - - - - - - - 

\subsection{Measuring deflector redshifts}

Deflector redshifts were measured using the centroids and known rest frame
wavelengths of prominent absorption features. 
In the majority of cases, a minimum of the \CaII~H and K lines were used,
however other absorption features were also used when available. The additional
absorption features used were: H$\eta$~(3835\AA), G~band (4305\AA),
H$\gamma$~(4341\AA), \Hb~(4861\AA), \Mgb~($5175$\AA), \NaII~($5892$\AA) and
H$\alpha$~(6563\AA). 
Typically, between three and five absorption features were centroided to measure
the redshift. The measured redshifts are listed in Table~\ref{table:measured}.

% - - - - - - - - - - - - - - - - - - - - - - - - - - - - - - - - - - - - - - - 

\subsection{Measuring deflector velocity dispersions} 
\label{subsec:vdisp}

Velocity dispersions were measured by fitting combinations of stellar spectra
over regions with prominent absorption features and high signal to noise.
Linear combinations of stellar spectra were used to fit a model to the data and
calculate a velocity dispersion. A Python based implementation of the
\citet{vandermarel1994} velocity dispersion code, developed by MWA and
described by \citet{Suy++09}, was used.  We use a set of templates from the
INDO-US stellar library containing spectra for a set of seven K and G giants
with a variety of temperatures and spectra.  K and G giants were used because
they provide a good description of the spectra of our deflectors, as expected
for massive ellipticals.

The value of the velocity dispersion, $\sigma$ for each deflector was
determined by finding a consistent value over several spectral regions and
features.  If the mean signal to noise (S/N) ratio per rest-frame angstrom  was
\gsim$10$ in the 4000--5000\AA~range, then consistent and reliable $\sigma$
values could be measured. Of the \Ntot~SL2S lenses, 12 spectra had sufficient
S/N to measure $\sigma$.

In general, the rest frame 4000--5000\AA~range was used, as the G-band absorption
feature is often uncontaminated by atmospheric absorption, and contains no sharp
changes in the continuum and the CCD efficiency is also good over this range. 
The average signal to noise ratio per angstrom in the 4000--5000\AA~range for 
the 12 objects with measured velocity dispersions is 24.8.

\begin{figure*}

\centering\begin{minipage}{0.9\linewidth}
\centering\includegraphics[width=0.32\linewidth]{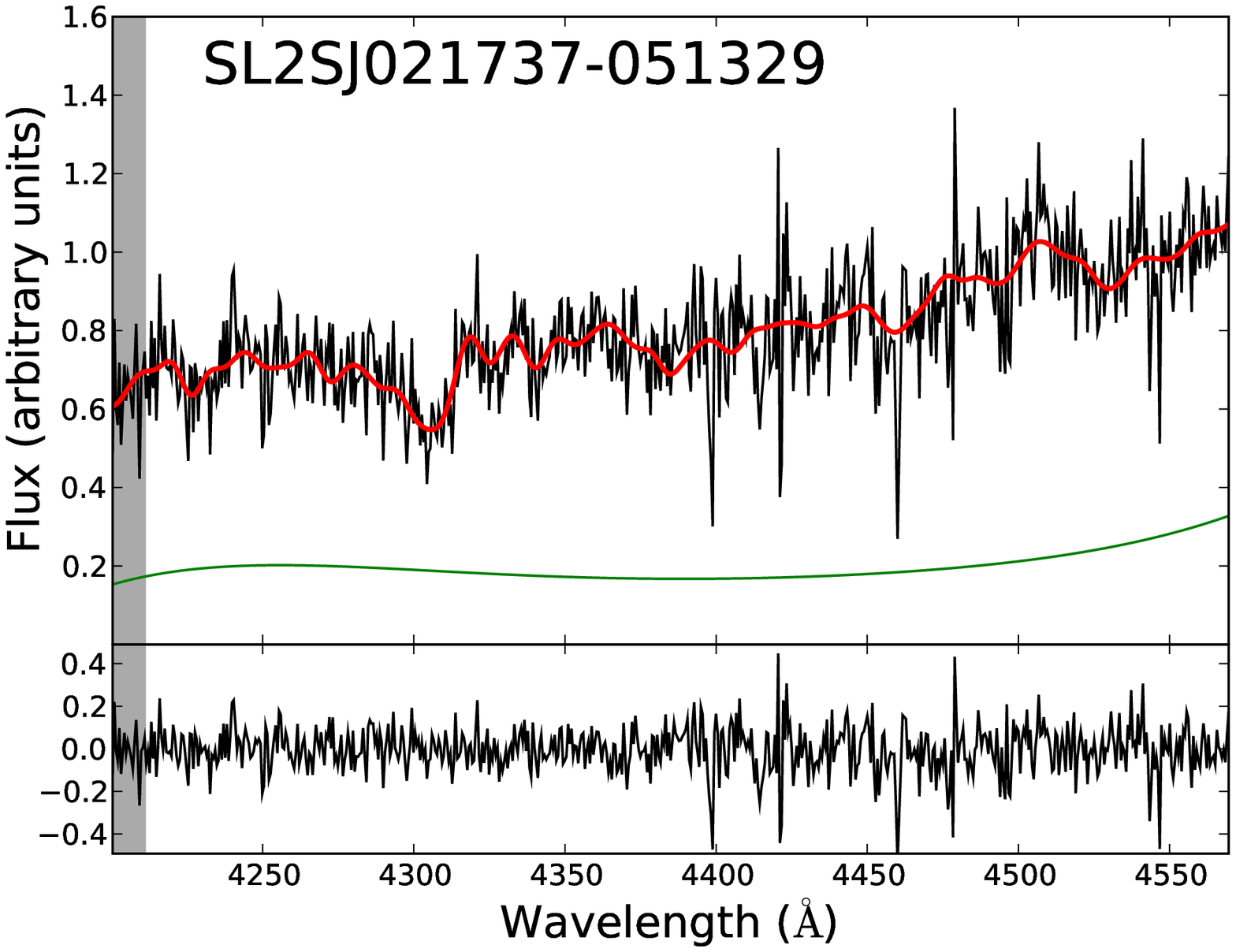}
\centering\includegraphics[width=0.32\linewidth]{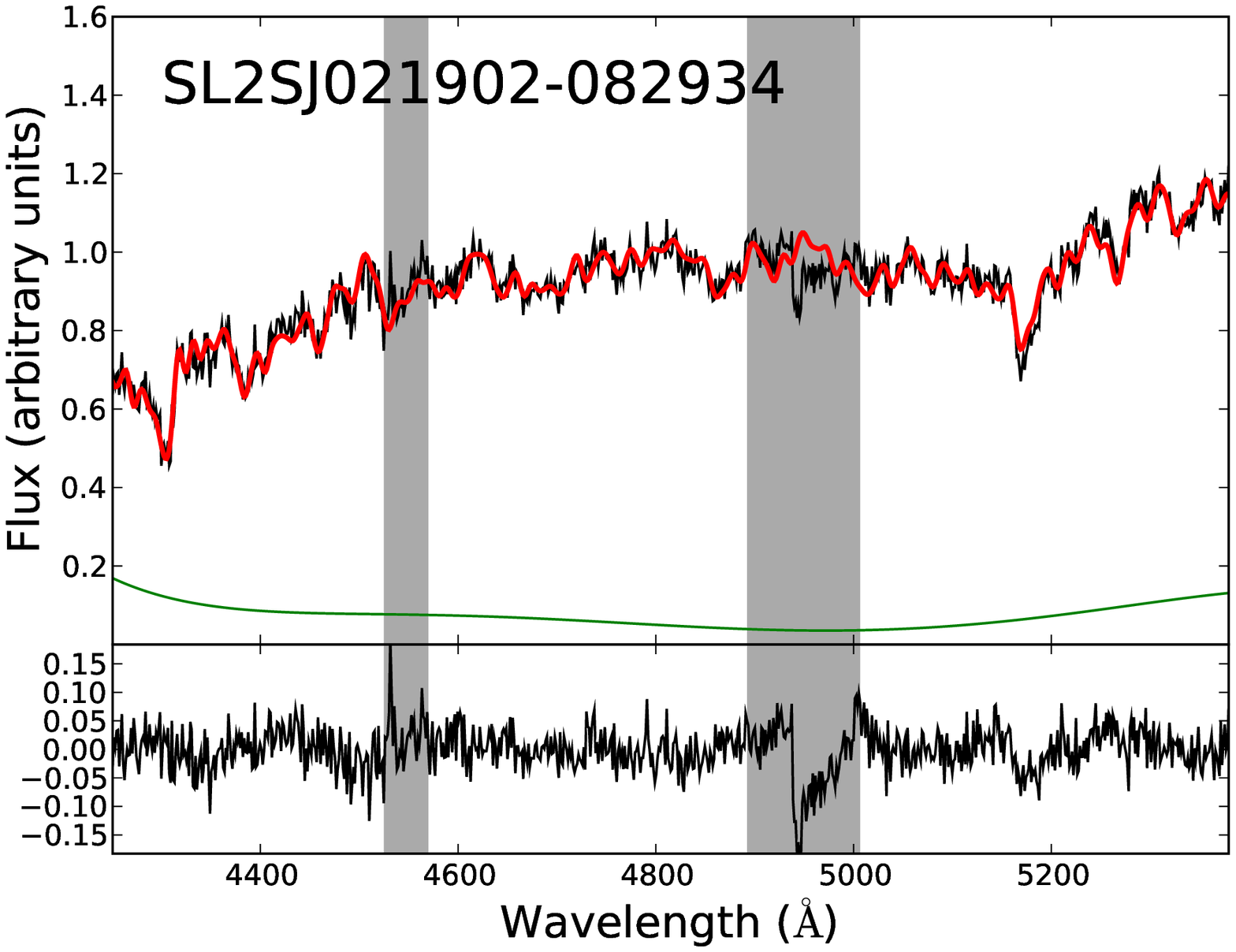}
\centering\includegraphics[width=0.32\linewidth]{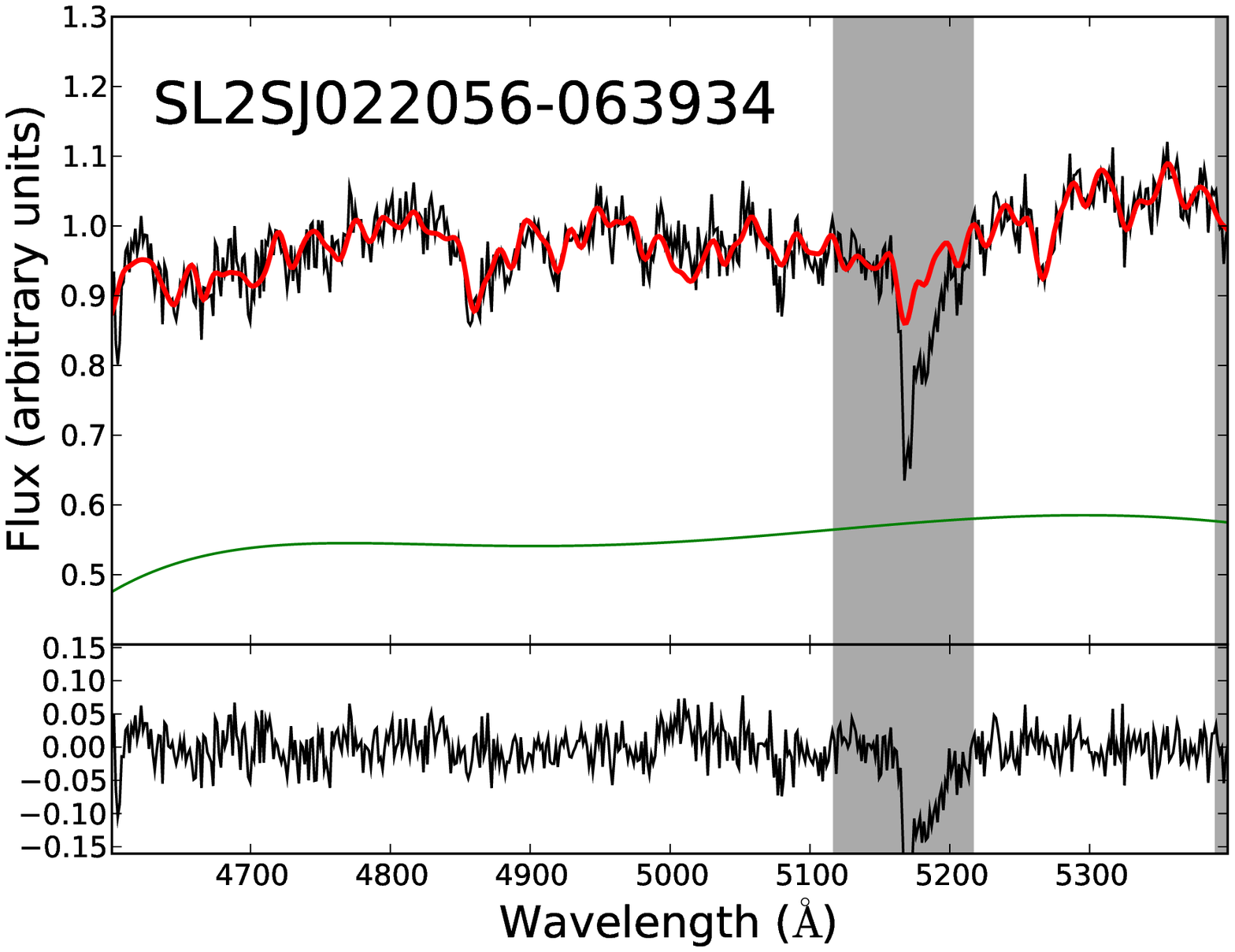}\hfill\\
\centering\includegraphics[width=0.32\linewidth]{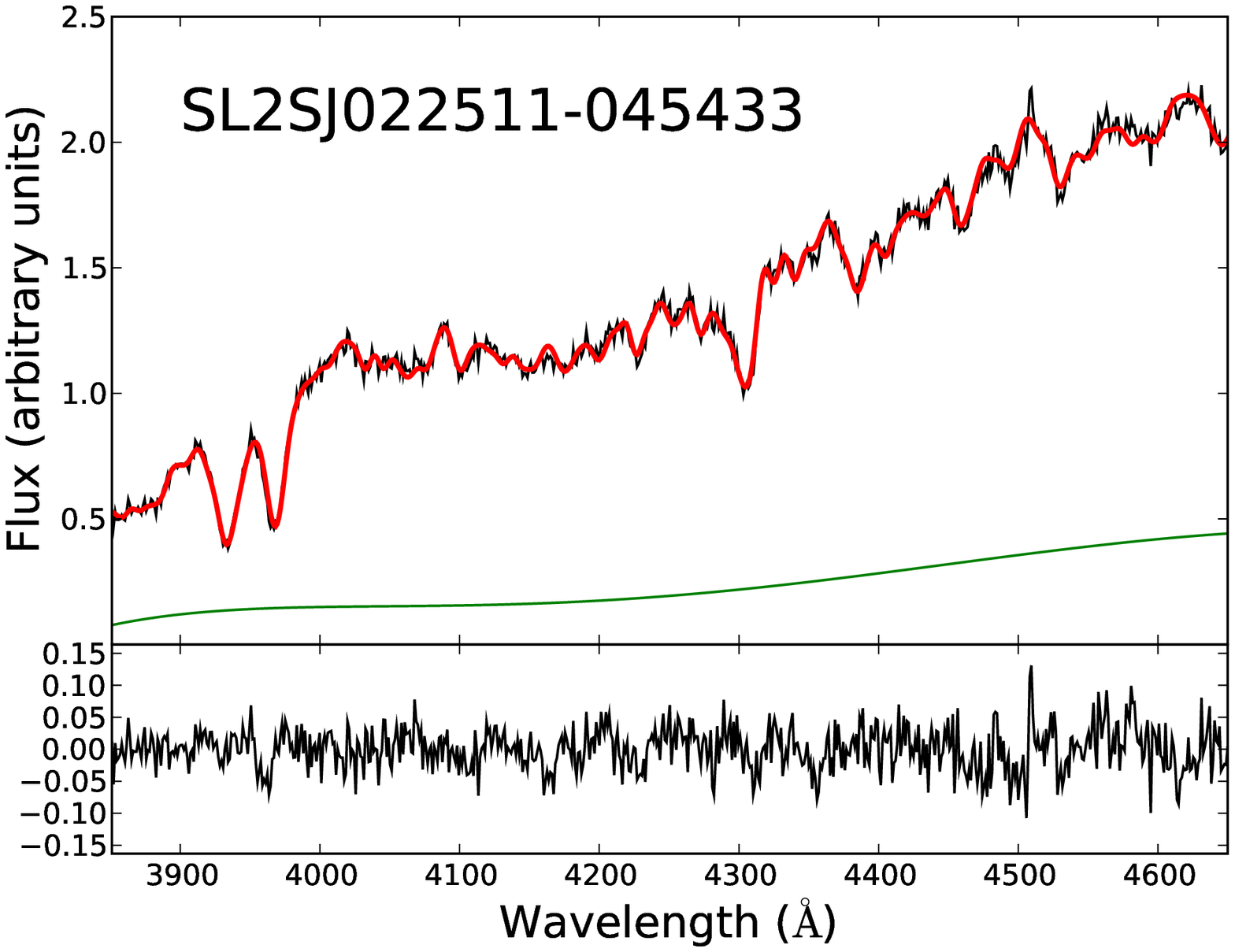}
\centering\includegraphics[width=0.32\linewidth]{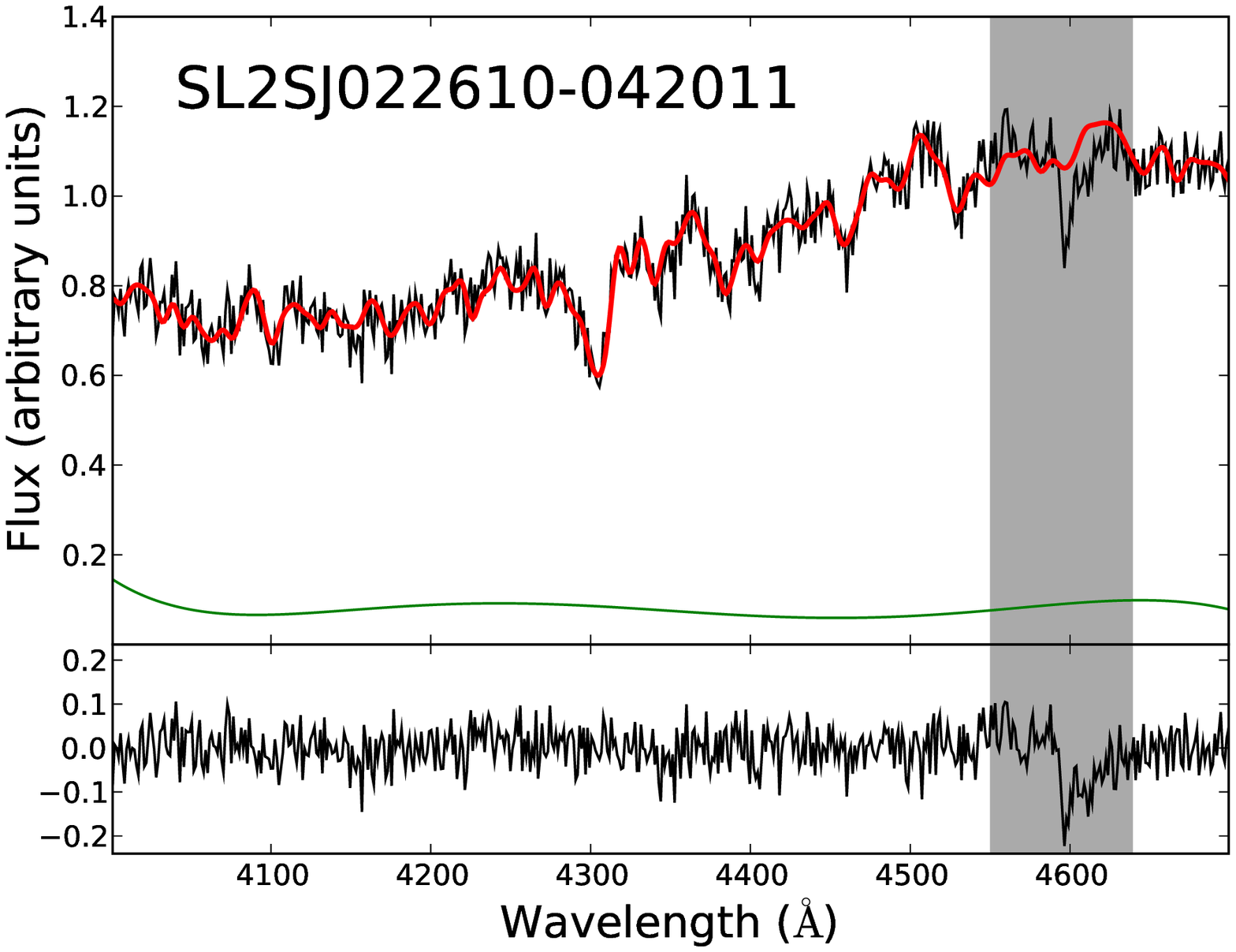}
\centering\includegraphics[width=0.32\linewidth]{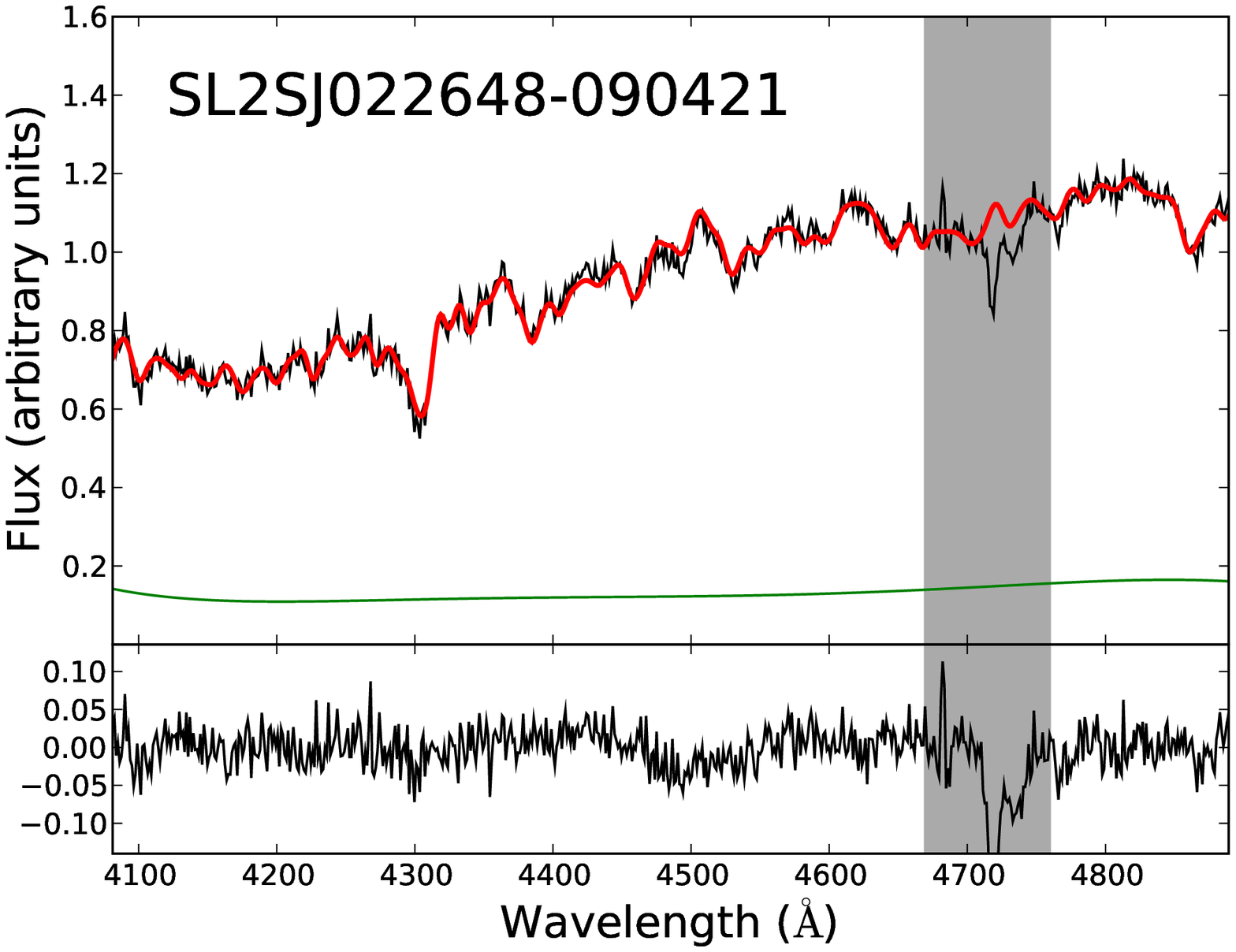}\hfill\\
\centering\includegraphics[width=0.32\linewidth]{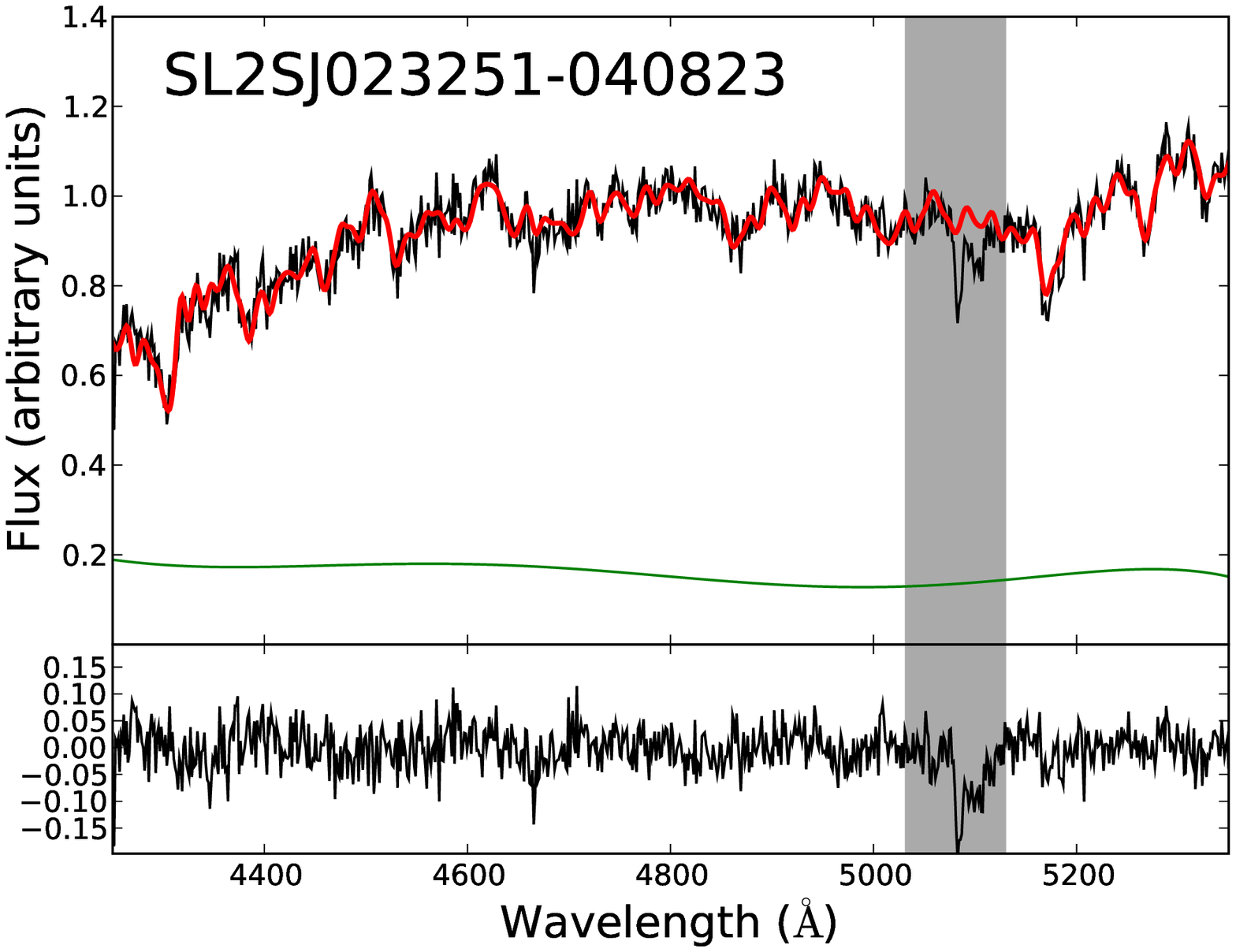}
\centering\includegraphics[width=0.32\linewidth]{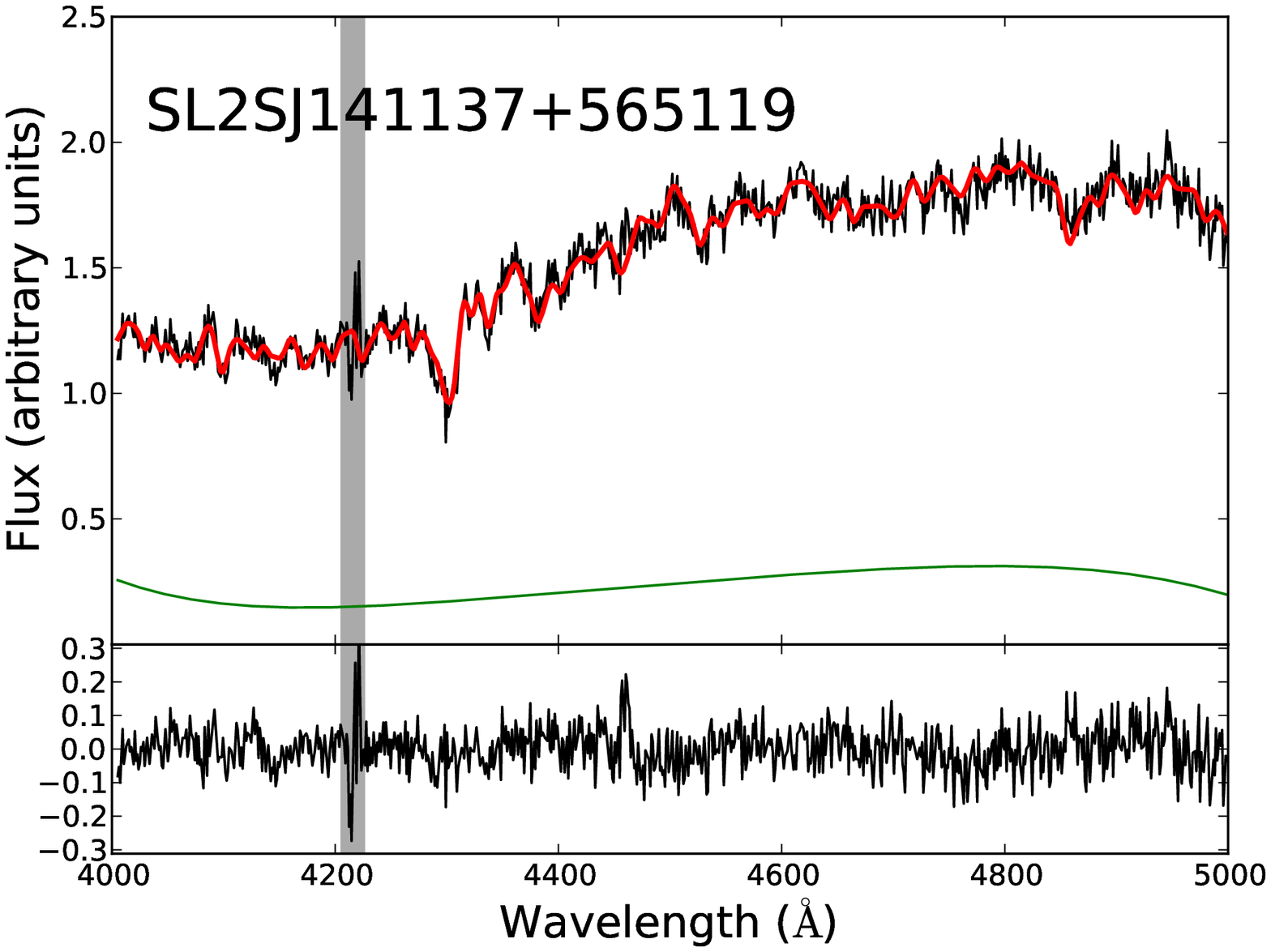}
\centering\includegraphics[width=0.32\linewidth]{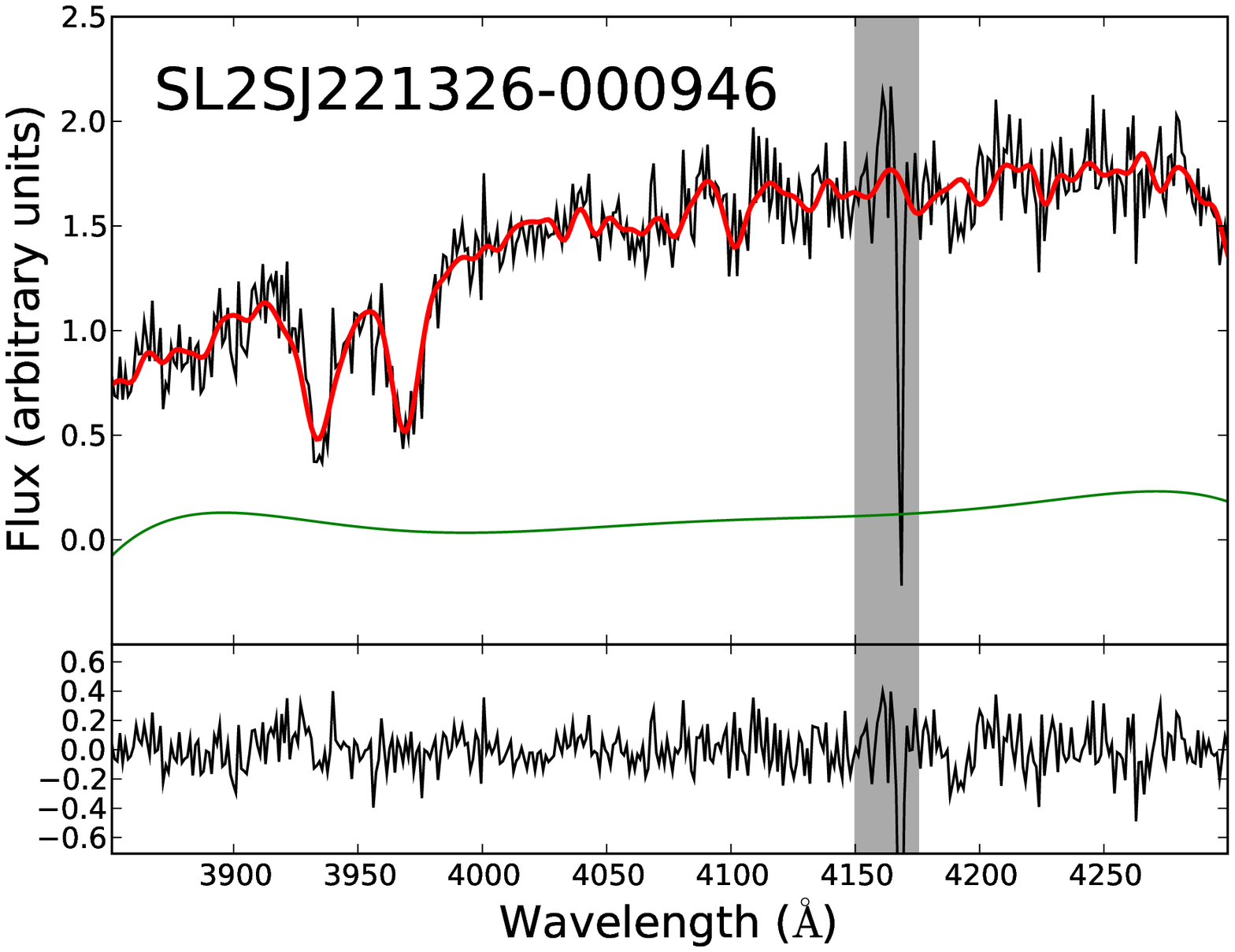}\hfill\\
\centering\includegraphics[width=0.32\linewidth]{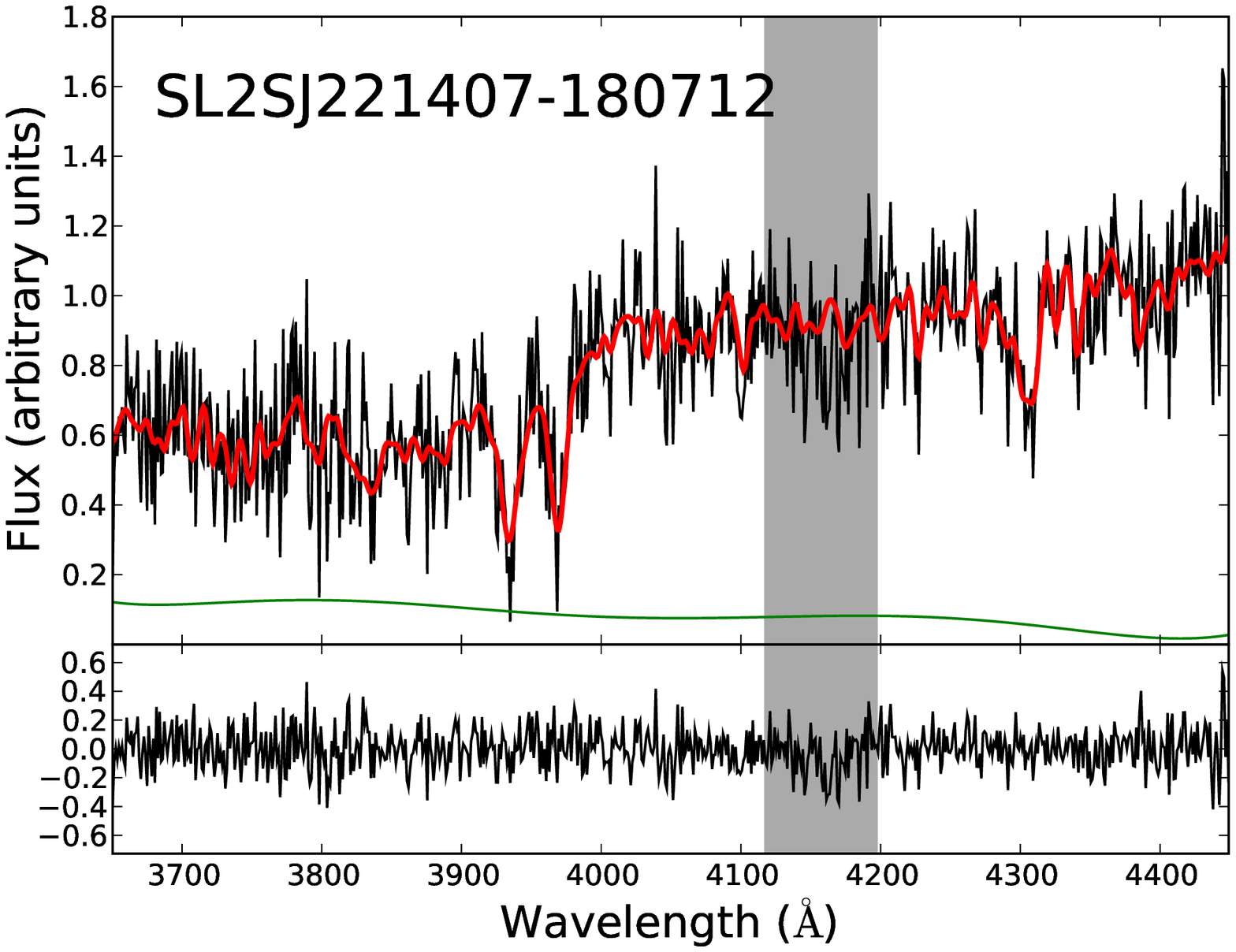}
\centering\includegraphics[width=0.32\linewidth]{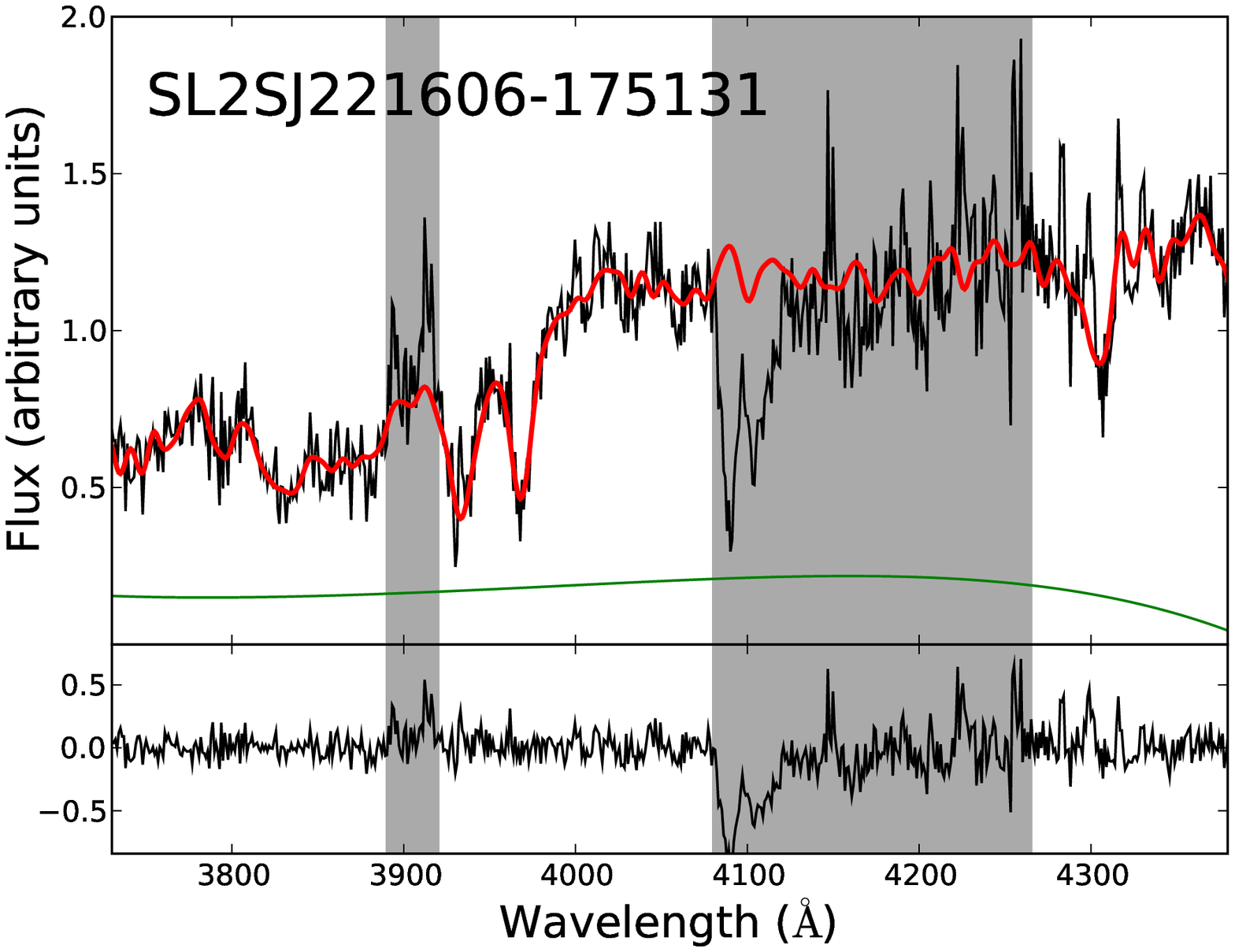}
\centering\includegraphics[width=0.32\linewidth]{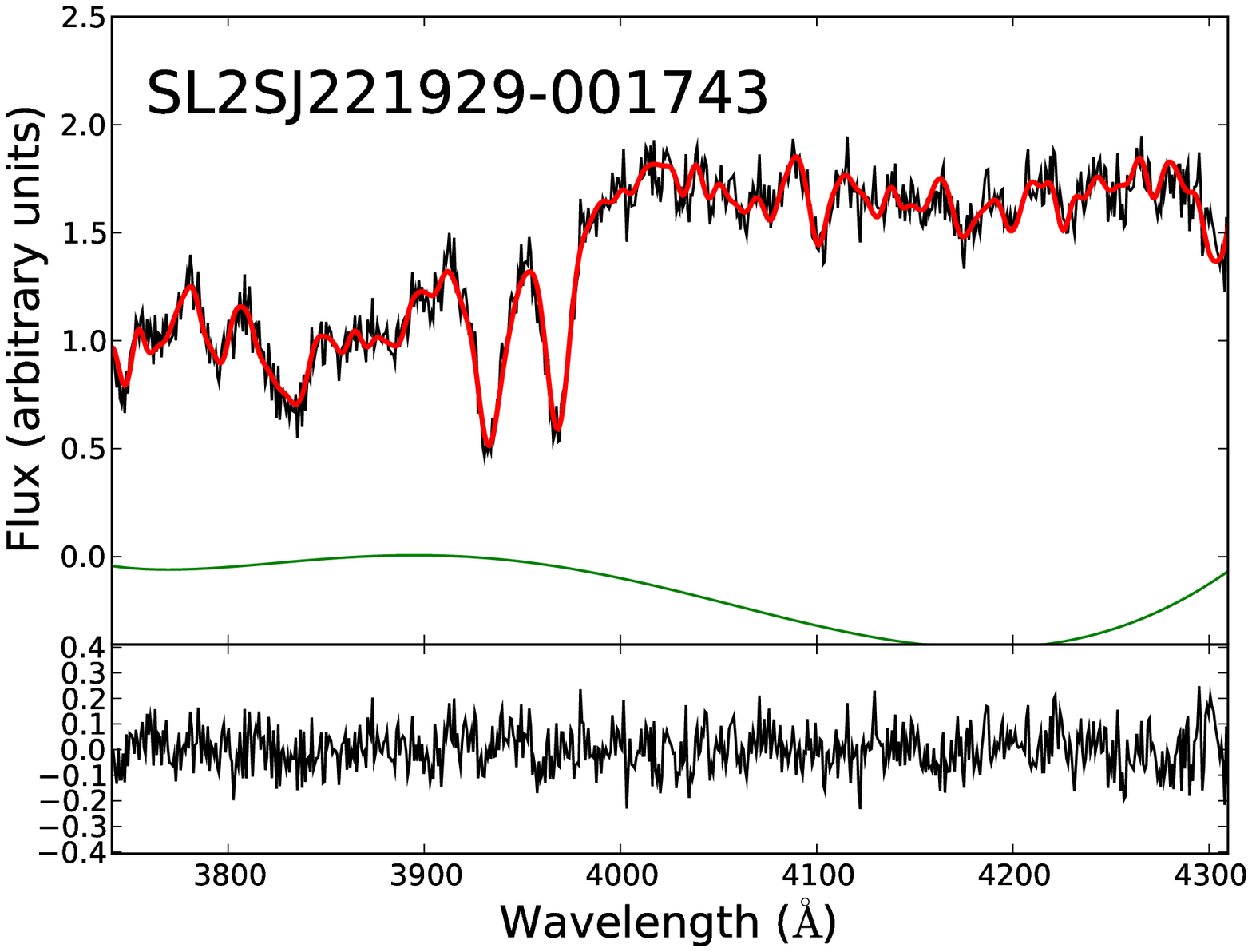}
\hfill
\end{minipage}\hfill
\caption{\label{fig:sigma} Measuring the stellar velocity dispersions of the
lens galaxies, by accurate absorption line fitting.  For each object, we show
the relevant portion of the LRIS spectrum (black line),
compared to  a model generated from all 9 INDO-US templates and a sixth order
polynomial  continuum (in red, with the continuum alone shown in green). The
grey shaded wavelength ranges were not included in the fits.  The lower
sub-panels show the fit residuals in each case.}
\end{figure*}

Other regions that were used to fit the stellar templates were: the
5000--6000\AA~range (matching the \Mgb~and \NaII~absorption features) and the
3500--4000\AA~range was also used, but only where the depth and continuum fit
to the spectra were good.
Generally, three regions of the spectrum that produced good fits and consistent
$\sigma$ values were used to calculate the final $\sigma$ and its associated
uncertainty.  For each lens, one of the models generated to measure $\sigma$ is
shown in Figure~\ref{fig:sigma}. 
Regions where atmospheric absorption was a problem, or the templates did not
produce a good fit were masked out, as shown in the grey regions of
Figure~\ref{fig:sigma}.
The results of the velocity dispersion measurements and the mean S/N per
angstrom in the rest frame 4000--5000\AA~range for each object are listed in
Table~\ref{table:measured}.  The mean measured $\sigma$ for the SL2S sample is
250~km~s$^{-1}$.  The measured velocity dispersions were then corrected to a
uniform physical aperture using the slit width, the size of the extraction
window and the empirical power-law relation of \citet{jorgensen1995}.  The
corrected velocity dispersion, $\sigma_{e2}$, measures the velocity dispersion
at $\Reff/2$, as used by \citet{Bol++08b}. 
Inside the effective radius, the relation is well described by a power law:
\begin{equation} \sigma_{\rm e2} = \sigma_{\rm ap} \left(\frac{\Reff}{2\,r_{\rm
ap}}\right)^{-0.04}, \end{equation} where $2r_{\rm ap} \approx 2(xy/\pi)^{1/2}$
and where $x$ and $y$ are the width and length of the rectangular aperture.

%%%%%%%%%%%%%%%%%%%%%%%%%%%%%%%%%%
%%%%%%%%%%%%%%%%%%%%%%%%%%%%%%%%%%

% - - - - - - - - - - - - - - - - - - - - - - - - - - - - - - - - - - - - - - - 

\subsection{Measuring source redshifts}
\label{subsec:sourcez}
Source redshifts were measured for five objects.  Two objects,
SL2SJ021737-051329 and SL2SJ141137+565119 ($z_s$$\,=\,$1.847 and 1.420,
respectively) have multiple source emission lines that were centroided to
measure the redshift. As can be seen in Figure~\ref{fig:spec1},
SL2SJ021737-051329 has narrow emission lines typical of Type~II AGN
(\CIV~$\lambda$1549, \HeII~$\lambda$1640, \OIII~$\lambda$1666 and
\CIII~$\lambda$1909). SL2SJ141137+565119 shows a clear splitting of the
\OII~doublet as well as strong \CIII~and \OIII~emission. 

The three remaining source redshifts were identified using
\OII~$\lambda\lambda$3726.1, 3728.8~only. SL2SJ022610-042011 and
SL2SJ022511-045433 ($z_s$$\,=\,$1.232 and 1.1988, respectively) show a clear
splitting of the doublet in the 2D spectra.  
For the objects where there was no clear splitting of the \OII~doublet, the
redshift measurement was more difficult.  We used two methods to search for
source emission lines. Firstly, we looked for emission lines in the 2D spectra
at the expected position of the source trace. Secondly, we looked at the
residuals from both a simple fit to a standard elliptical galaxy template and
also the residuals from the velocity dispersion measurement, discussed in
Section~\ref{subsec:vdisp}. 
Despite that three source redshifts were identified using \OII~alone, no other
strong emission features are observed blue-ward of this feature, corroborating
the identification, as we would expect to see other lines if the detected
feature were redder (\eg \OIII, H$\alpha$ or H$\beta$).

%%%%%%%%%%%%%%%%%%%%%%%%%%%%%%%%%%
%trim=l b r t
%%%%%%%%%%%%%%%%%%%%%%%%%%%%%%%%%%
\begin{figure*}
\centering
\includegraphics[width=0.20\textwidth]{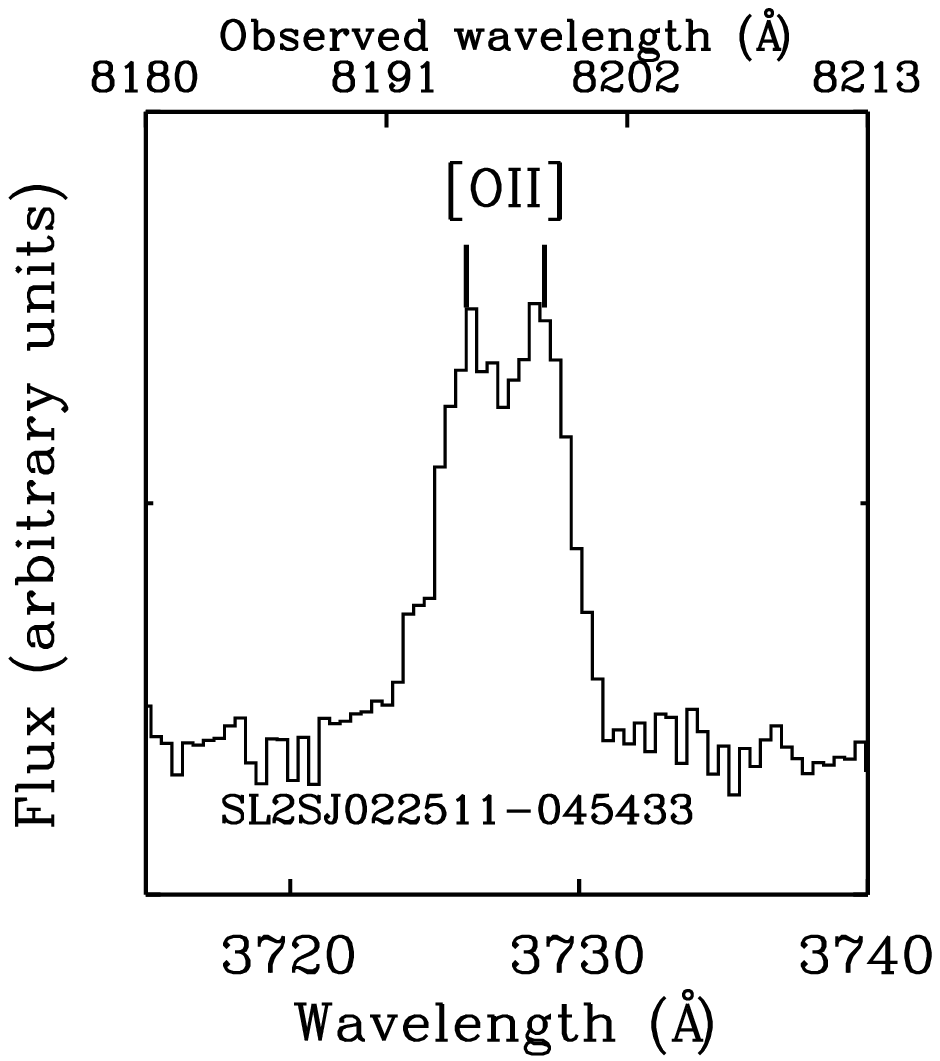}
\includegraphics[width=0.20\textwidth]{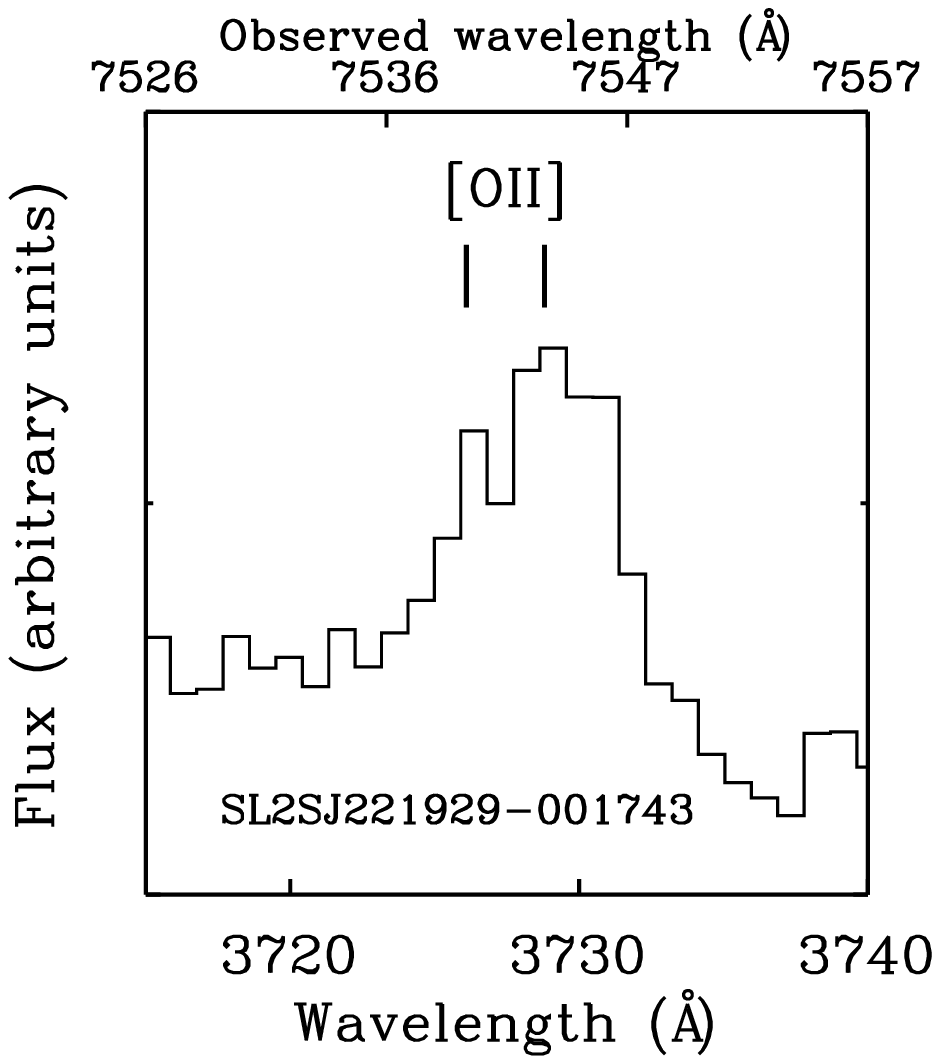}
\includegraphics[width=0.20\textwidth]{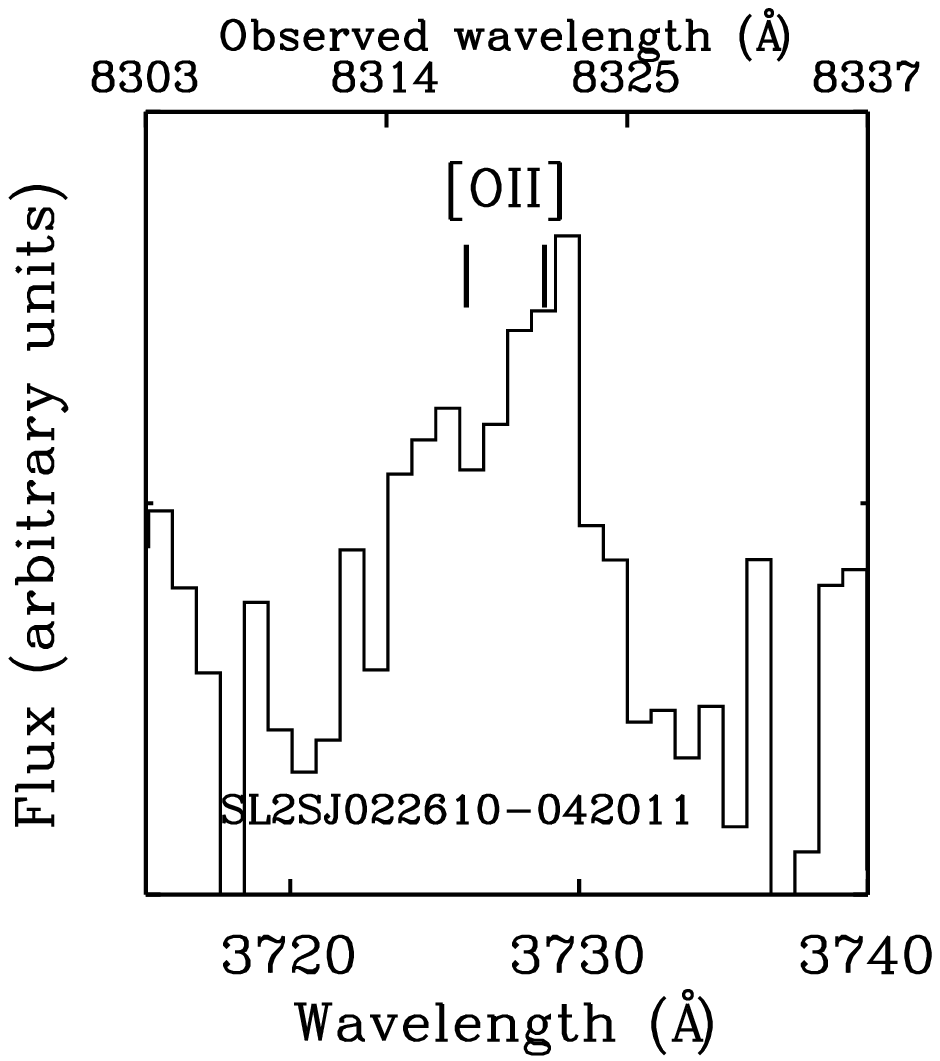}
\includegraphics[width=0.20\textwidth]{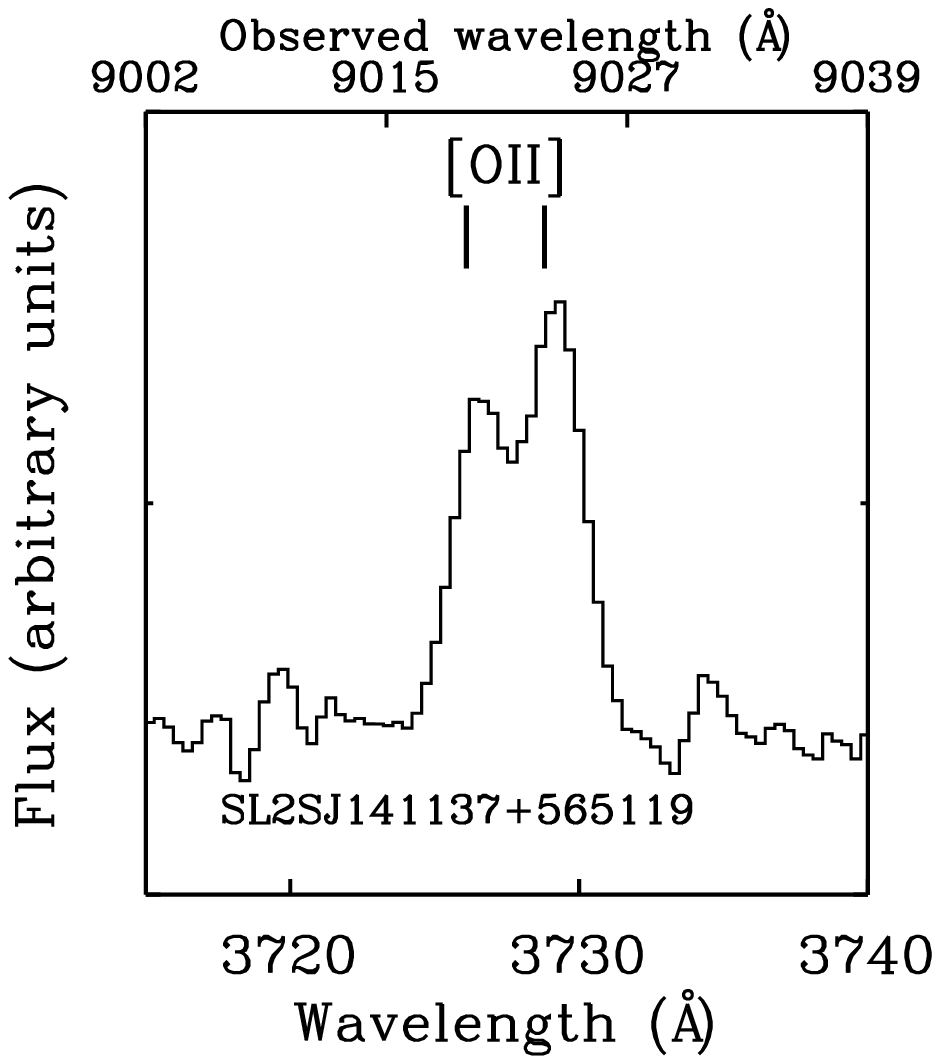}\\
\centering
\caption{\label{figure:2d} The \OII~doublet is shown for each object
with a measured source redshift and observed \OII~doublet.  
}
\end{figure*}
%%%%%%%%%%%%%%%%%%%%%%%%%%%%%%%%%%

The 2D spectrum of SL2SJ022511-045433 shows a slight offset between the
\OII~emission of the arc and the arclet. We estimate the difference in
redshifts to be $\Delta z=0.0005$, corresponding to a relative motion of
$\sim$150~km/s. 
This is most likely due to kinematic structure of the source. Consistent with
this interpretation, the lens model indicates the presence of two separate
peaks in the surface brightness distribution of the lens, \ie possibly a pair
of source galaxies (see Paper~I for details).

% - - - - - - - - - - - - - - - - - - - - - - - - - - - - - - - - - - - - - - --

\subsection{Systems with no source emission lines}

For the remaining 11 sources, for which no source emission lines could be detected,
we must estimate the source redshift from whatever information we have. We
first consider the the \hst and CFHTLS photometry.  We did not attempt to infer
photometric redshifts, since the disentanglement of lens and source colors in
the low-resolution, multi-filter CFHTLS data was deemed likely to lead to
significant systematic uncertainty. Instead, we conservatively used the
redshift distribution of the faint galaxies in the COSMOS survey to provide a
broad probability density function (PDF) for~$\zs$.

\citet{Lea++07} model this distribution using the following functional form:
\begin{eqnarray}
  \pr(\zs|m) &\propto& \zs^2 \,\exp{\left[ 
                           -\left(\frac{z}{z_0(m)}\right)^{1.5}\right]} \\
   z_0(m) &=& \frac{(0.18\,m - 3.3)}{1.412}.                        
\end{eqnarray}
Here $m$ is the AB magnitude of the source in the F814W filter. 
In most cases we
have either a F606W magnitude from the \hst image modeling, or a $g$-band
magnitude from the CFHTLS image modeling: we assume that the sources, as faint
blue galaxies, have spectral energy distributions consistent with flat in the
AB magnitude system, and hence substitute our blue unlensed
source magnitudes (output from the lens inversions) directly. This
approximate transformation introduces a small additional uncertainty which we
neglect relative to the intrinsic PDF width.

This model source redshift distribution was used by, among others,
\citet{Gav++07} and \citet{Lag++09} when estimating the redshift distribution
of background sources.  We note that this approach is qualitatively different
from that taken by \citet{Gav++07} and \citet{Lag++09}: in their weak lensing
studies they did not have a single well-defined source magnitude, but instead
integrated over the number counts down to the magnitude limit. Whereas we are
able to use the small amount of information that the source brightness
provides. 
We neglect the small uncertainty in $z_0(m)$ and the
photometric uncertainty in $m$, and truncate the density to zero at $\zs \leq
\zd$. We then draw samples from $\pr(\zs|m)$ that we can then transform into
distances, physical masses and so on.  This gives the broader PDF,~$\pr(\zs)$.

While the source magnitude provides a very rough photometric redshift, we can
also use the lens geometry to give a similarly rough {\it geometric} redshift.
This involves multiplying the COSMOS prior PDF $\pr(\zs|m)$ by an additional
distribution describing our prior knowledge of the lens strength, as follows.
In practice we do this by importance sampling the COSMOS prior \citep[see
\eg][for descriptions of how this process works]{lewis2002,Suy++10}.  Given
that the velocity dispersion of the dark matter is approximately equal to the
central stellar velocity dispersion, and the apparently universal (approximate)
isothermal profile of lens galaxies \citep[see \eg][]{kochanek1994,Koo++06}, we
can down weight some predicted $z_{\rm s,pdf}$ values based on the velocity
dispersions they predict.  The total deflector mass is modeled as a Singular
Isothermal Ellipsoid (SIE), with a velocity dispersion
\begin{equation} \sigmasie = 186.2\,{\rm km}\,{\rm s}^{-1}\,\times
\sqrt{\frac{\REin}{\rm arcsec} \frac{\Dds}{\Ds}} \end{equation}
where $\Dds$ and $\Ds$ are the angular diameter distances between the deflector
and source and observer and source, respectively.
The calculated $\sigmasie$ for each $\zspdf$ was used to give a
$\pr(\sigmasie)$.  A joint prior, $\pr(\zs|m)\pr(\sigmasie)$, using the
additional information from $\sigmasie$ was then calculated and used to tighten
the constraint on $\zs$. 

%%%%%%%%%%%%%%%%%%%%%%%%%%%%%%%%%%
\begin{figure}
\centering
\includegraphics[width=0.9\linewidth]{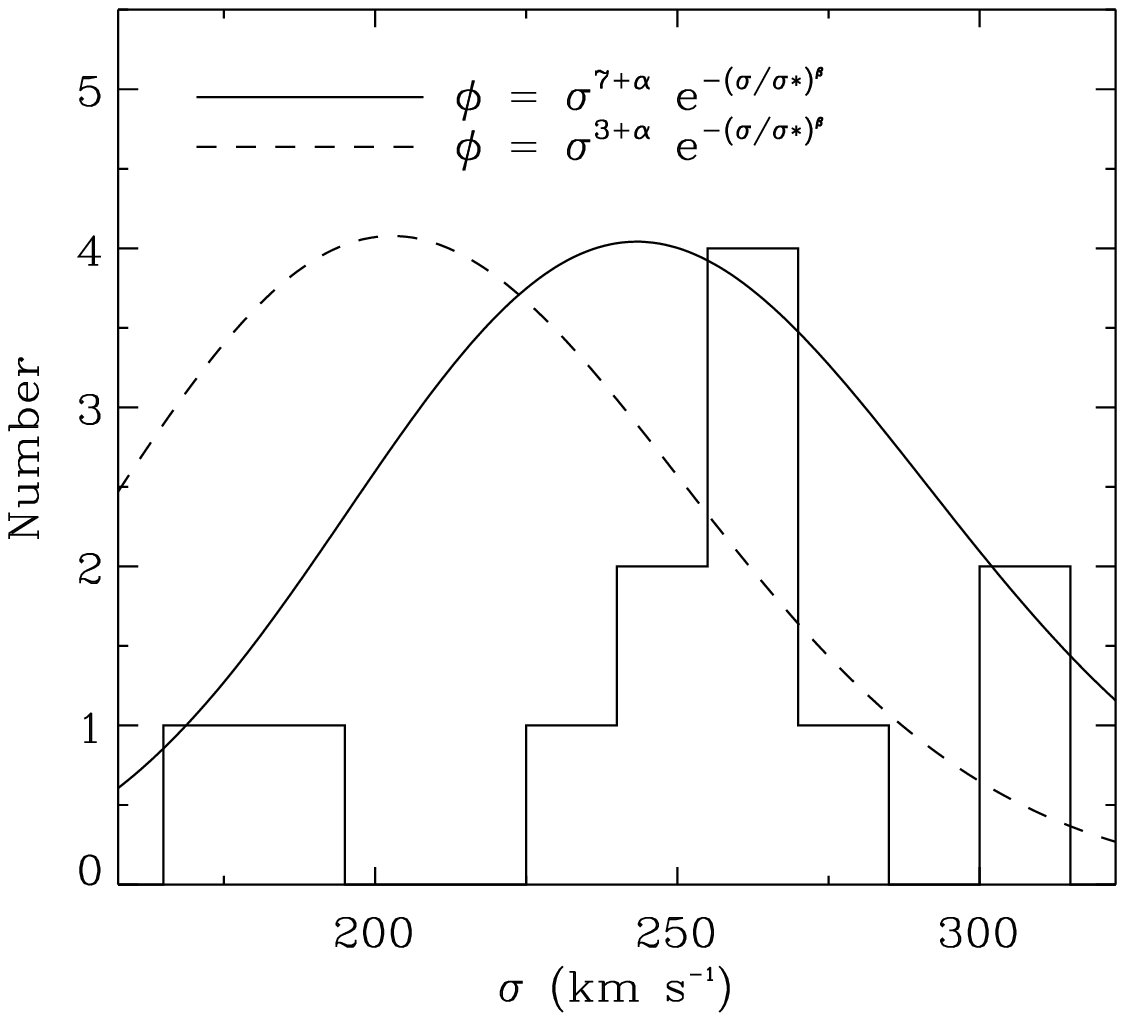}\\
\caption{\label{figure:sheth} The distribution of measured velocity dispersions
 is shown with the \citeauthor{sheth2003} fitting
function overlayed. The solid and dashed curves show the fitting function
multiplied by $\sigma^8$ (lensing and luminosity selection function) and
$\sigma^4$ (only lensing), respectively. 
The fitting functions were normalized to the unit area of the histogram. 
Note that the mean of the SL2S $\sigma$ distribution is in good agreement 
with the peak of the solid curve.
}
\end{figure} 
%%%%%%%%%%%%%%%%%%%%%%%%%%%%%%%%%%

%%%%%%%%%%%%%%%%%%%%%%%%%%%%%%%%%%
\begin{figure*}
\centering\begin{minipage}{0.9\linewidth}
\centering\includegraphics[width=0.9\textwidth]{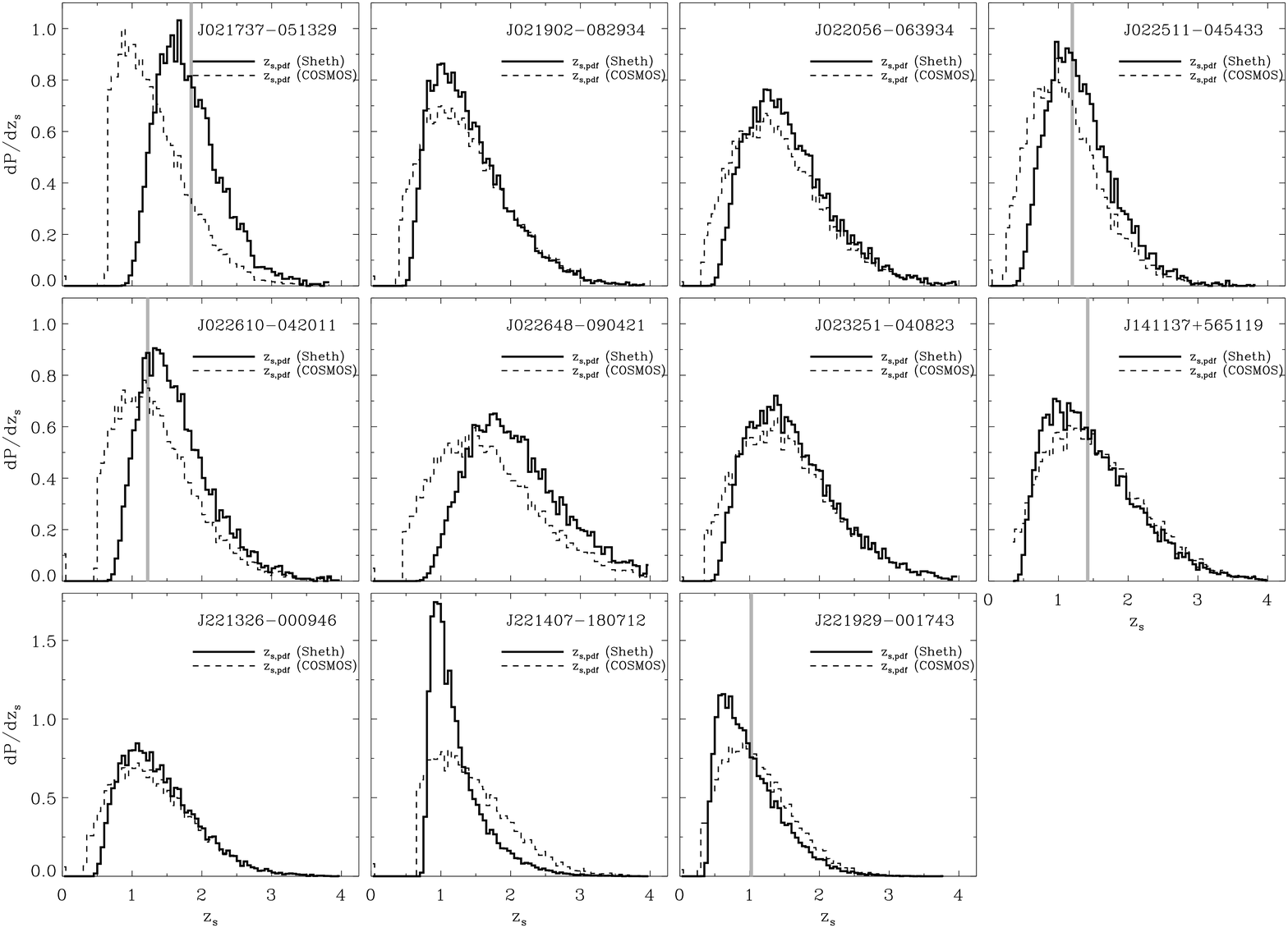}
\end{minipage}\hfill
\caption{\label{fig:zspdf_sheth} The $\zspdf$ distributions are shown for the final sample of lenses. 
The dashed histogram shows the source redshift inferred from the COSMOS distribution.  The thick 
histogram shows the final $\zspdf$, where the inferred source redshifts were weighted 
by the \citeauthor{sheth2003} fitting function (multiplied by $\sigma^8$), as discussed in the text. 
For lenses with a measured $\zs$, the grey vertical line shows the measured source redshift. 
Note that the PDFs have been normalized to unit area and that y-axis on the bottom row is on a different scale. }
\end{figure*}
%%%%%%%%%%%%%%%%%%%%%%%%%%%%%%%%%%

To calculate $\pr(\sigmasie)$, we assumed that the SL2S lens galaxies are at
the high mass end of the velocity function of bright galaxies --
high mass, because we know they are acting as gravitational lenses.
To describe the velocity function of the SL2S lenses, the velocity dispersion
fitting function determined by \citet{sheth2003} was used. The parameters in
the fitting function were determined using measurements of a large sample of
early-type galaxies drawn from the SDSS database. 
The fitting function was multiplied by $\sigma^4$ to mimic the lensing
selection function \citep[see \eg][]{Aug++10}. 
However, the lenses, particularly those with measured $\sigma$, are also
luminosity selected because SDSS is a flux-selected sample and more luminous
objects are drawn from a large volume \citep{HydeBernardi2009,Aug++10}. To
account for the luminosity selection, the \citeauthor{sheth2003} fitting
function was heuristically multiplied by $\sigma^4$.  Figure~\ref{figure:sheth}
shows the distribution of the measured SL2S velocity dispersions with the
\citeauthor{sheth2003} fitting function, multiplied by two different selection
functions, overlayed. 
The dashed curve shows the fitting function multiplied by $\sigma^4$ (mimicking
the lensing selection function), while the solid curve shows the fitting
function multiplied by both the lensing and luminosity selection functions.
Despite the small number of objects, the mean $\sigma$ of the SL2S deflectors
($250\pm39$~km~s$^{-1}$) is in very good agreement with the peak of the solid curve
($243\pm48$~km~s$^{-1}$). 
The prior adopted is then the \citeauthor{sheth2003} fitting function
multiplied by $\sigma^8$;
the resulting distribution was then normalized and used to compute importances
to weight the $\zspdf$ values drawn from the COSMOS prior. We note that final
result is relatively insensitive to the precise choice of the exponent, since
the dominant effect is the exponential cutoff, which eliminates unrealistically high
stellar velocity dispersions. 

Our final prior PDF for the velocity dispersions of the SL2S galaxies is then
given by: 
\begin{equation} 
\pr(\sigmasie) \propto
\sigma^{\alpha+7}\,e^{-\left(\frac{\sigma}{\sigma_*}\right)^{\beta}},
\end{equation}
where the best fit parameters determined by \citet{sheth2003} for early-type
galaxies are: $\sigma_*$=88.8$\pm$17.7~km~s$^{-1}$, $\alpha$=6.5$\pm$1.0 and
$\beta$=1.93$\pm$0.22.  To account for the scatter in $f_{\rm SIE} =
\sigmasie/\sigma_*$, the function was conservatively smeared by $20\%$.
This distribution is quite broad: the assumption of an approximately isothermal
mass profile does not significantly bias our results when inferring the profile
slopes of individual lenses, since we include plausible scatter on $\sigma$.
Distributions of $\zspdf$ for the final sample of lenses are shown in
Figure~\ref{fig:zspdf_sheth}.  The dashed line shows $\zspdf$ distributions
from the COSMOS prior without weightings and the solid line indicates the final
$\zspdf$ for each of the lens.  

The primary effect of $\pr(\sigmasie)$ is to disfavor lower redshift solutions
to the lens equation, that are offered by the COSMOS redshift distribution with
its relatively low $\zs$ peak, which would indicate unrealistically large
stellar velocity dispersions, well above 400~km~s$^{-1}$.  For each 
non-spectroscopic case, we give the median of each lens system's source
redshift PDF in Table~\ref{table:measured}, along with the 16$^{\rm th}$ and
84$^{\rm th}$ percentiles as indicators of the uncertainty.  
The method was verified by comparing the 5 spectroscopically measured redshifts to
those inferred via this procedure, using the same 5 systems.  The grey vertical
lines in Figure~\ref{fig:zspdf_sheth} show the measured source redshifts, which
are in good agreement with $\zspdf$ after weighting by the \citet{sheth2003}
fitting function and the selection function.

% - - - - - - - - - - - - - - - - - - - - - - - - - - - - - - - - - - - - - - - 

% PJM: Main table should go here, after all spec measurements have been 
% described. Needs to go at end to be landscape though (emulateapj bug)

% - - - - - - - - - - - - - - - - - - - - - - - - - - - - - - - - - - - - - - - 

\section{Photometric measurements}\label{sect:mstar}

We now turn to our photometric data, in order to measure sizes and stellar
masses of our lens galaxies. The latter are required in order for us to probe
the dark matter fractions in the core regions of these objects, and are
obtained by stellar population analysis of the CFHTLS photometry. 

% - - - - - - - - - - - - - - - - - - - - - - - - - - - - - - - - - - - - - - - 

\subsection{Model size and magnitude estimation}

We fit an elliptically symmetric de Vaucouleurs profile \citep{dev48} surface
brightness distributions to each of the lens galaxies in our sample, 
to measure the total magnitudes and effective radii. For each lens, we used
the \galfit software package to fit for the position, effective (half-light)
radius, ellipticity, orientation angle and total magnitude.

When available, we also used the \hst data that generally
provide a better estimate of the deflector effective radius. In particular,
the light profile in the reddest available filter should be a better tracer
of the underlying stellar mass distribution profile. Otherwise we used the
average effective radius of the CFHT r, i and z bands and the internal
filter-to-filter r.m.s. scatter to get a precise estimate of the effective
radius and its associated measurement error.
Under the assumption of a spheroidal distribution of stars, with a 
single stellar population and no color gradients, the scatter across the filter
set in each morphological parameter gives a simple estimate of the uncertainty
of each parameter, and in particular on $\Reff$. In most cases, the typical error
on $\Reff$ is of order 5\%. Note also that $\Reff$ is expressed
along the geometric mean axis, that is the radius of the circle enclosing the
same area as the elliptical isophote enclosing half the light.

To build up a picture of each lens' spectral energy distribution (SED), with
the deflector photometry devoid of light coming from the lensed background
source, we follow a two-step iterative process to mask out the lensed images:
first we fit models to the images with no masking, and then use the residual
image to manually mask out the lensed features, before refitting the model.
Apparent magnitudes are then corrected for Galactic dust extinction
\citep{schlegel98}. Errors are dominated by systematic photometric zero point
uncertainties, dust correction errors and variations of \galfit results
depending on the masking strategy. All together these amount to systematic
errors of $\sim$0.05 magnitudes in each CFHT bands.
The resulting effective radii and magnitudes are given in
Table~\ref{table:measured}.

% - - - - - - - - - - - - - - - - - - - - - - - - - - - - - - - - - - - - - - - 

\subsection{SED fitting methodology}\label{sec:stellarmass}

Estimates of the stellar masses were calculated using the CFHT \galfit model
magnitudes and composite stellar population models using a code developed by
\citet{Aug++09}.  The code employs a Bayesian exploration of the stellar
populations of galaxies using composite stellar population models produced by
\citet{bruz2003}.  The code takes a set of photometric data for each object (g,
r, i, z) magnitudes and their uncertainties from the CFHT filters and the
measured redshift. 
%colours (u was not used since the zero point is not well understood). 
%
The free parameters of the star formation history are: the time when star
formation began, the time scale of the exponential decay of the star formation
rate, internal reddening due to dust extinction, and metallicity. The parameter
space was then explored using a Markov Chain Monte Carlo (MCMC) routine,
allowing a full determination of the posterior probability distribution
function for each parameter; see \citet{Aug++09} for a comprehensive
description of the method. We employ uniform priors for all of the model
parameters (including the metallicity) due to the absence of any
well-calibrated priors for the stellar population model parameters of massive
galaxies at the redshifts of these lenses.
%SFR ~ Exp[ t/tau ]

The code was run for both \citet{Salpeter1955} and \citet{Chabrier2003} initial
mass functions (IMFs). The masses derived from the \citeauthor{Salpeter1955}
IMF were a factor of $\sim$1.7 greater than the masses derived using the
\citeauthor{Chabrier2003} IMF. 
In a recent study of 56 gravitational lenses identified by the SLACS survey,
\citet{Tre++10} found that for massive early-type galaxies, the
\citeauthor{Salpeter1955} IMF provides stellar masses in approximate agreement
with those inferred by lensing and dynamical models, whereas the Chabrier IMF
gives (on average) an underestimate. 
However, the IMF normalization may be mass dependent being more similar to a  
\citeauthor{Chabrier2003} IMF for the lower mass systems in the SLACS sample, with
$\sigma$$\,\sim\,$200~km~s$^{-1}$ (\citealt{Barnabe2010, Aug++10}; see also
\citealt{Cap++06,Cap++09}). 
For simplicity, we adopt the \citeauthor{Salpeter1955} IMF normalization,
although they can easily be converted to other normalizations, by multiplying
by an appropriate factor \citep[\eg][]{Aug++09}. Total stellar masses for a
\citeauthor{Salpeter1955} IMF are given in Table~\ref{table:inferred}.

%LSD stellar masses
LSD stellar masses were also calculated using the method described above with
photometry from \citet{K+T02,K+T03} and \citet{T+K04}. This method could only be applied
to 4 of the 5 LSD lenses, as photometry in multiple bands is required. Using
this method we inferred total stellar masses (in units of 10$^{11}$$\msun$) of:
5.01$\pm$2.06, 2.24$\pm$0.58, 4.26$\pm$0.95 and 2.49$\pm$0.82 for H1417+526,
H1543+535, MG2016+112 and 0047-281, respectively.

% - - - - - - - - - - - - - - - - - - - - - - - - - - - - - - - - - - - - - - - 

\section{Lens modeling}\label{sect:modelling}

The modeling of our high resolution \hst images is described in detail in
\paperI. Here, we briefly outline the procedure followed and describe the
relevant output parameters that we use in our joint analysis. Following
standard image reduction, the bright lens galaxies were subtracted from postage
stamp images of the lenses, iteratively masking the lensed arcs.  The remaining
residual image was fitted using a flexibly parametrized source (an elliptical
exponential profile with free position, ellipticity, orientation, size and
flux) traced through a simple elliptically symmetric lens potential. We model
each lens mass distribution using a Singular Isothermal Ellipsoid (SIE)
model~\citep{KSB94}, with centroid fixed at the centroid of the \galfit model
lens light distribution. The lens has three free parameters: Einstein
radius~$\REin$, orientation~$\phi$ and axis ratio~$q$. In all cases the lens
redshift is known spectroscopically, allowing us to work with Einstein radius
in~kpc.

The parameters of each lens were inferred by exploring their posterior PDF
using a Markov Chain Monte Carlo (MCMC) sampler. While the samples do
characterize the posterior PDF as required, the simplicity of the lens model
and the idealized source light distribution  mean that, in practice, the
posterior width does not provide  a good estimate of the uncertainty on, for
example, the Einstein radius. Instead, we use independent reconstructions in
several bands (which we have for a subset of the lenses) to estimate the
uncertainties \citep[see \eg][]{Mar++07}, which are found to be about 5\% on
$\REin$.

The properties of the lens models are discussed in detail in \paperI; here we
are concerned only with the combination of the lensing mass with the dynamical
mass from our spectroscopic measurements. With this in mind, the definition of
elliptical radius in the lens modeling was chosen to preserve the enclosed mass
with circular apertures, adopting the same definition as the SLACS survey
\citep{Koo++06,Bol++08b,Aug++09}. The mass enclosed within the Einstein
radius,~$\MEin$, is essentially independent of the profile of the
lens density profile: although we used an SIE model to infer $\MEin$, we can
treat it as a profile-independent, circularly symmetric enclosed mass and
then combine it with a dynamical spherical mass estimate in order to infer the
total density profile slope (see Section~\ref{section:gammald} below). For more
discussion of the accuracy of this assumption see \citet{Aug++09}.

Three systems J140614+520252, J220629+005728 and J221606-175131, were
observed with LRIS before attempting to perform a lens model. But subsequently
we were not able to find a satisfactory lens model and we failed in measuring 
their Einstein radius essentially because \hst imaging is not available for
them yet. Deflector photometry along with effective radius and redshift are
nevertheless reported in Table~\ref{table:measured}. A final conclusion on
their lensing nature is left for future work (see also \paperI\ for further
discussion).

% - - - - - - - - - - - - - - - - - - - - - - - - - - - - - - - - - - - - - - - 

\section{Characterization of the SL2S sample}

In this section we explore the basic properties of the SL2S lens galaxies, and
compare them with two previously studied samples of lenses.  The SLACS sample is a
lower redshift sample, $0.08<\zd<0.51$, of 85 lenses from the Sloan Digital Sky
Survey \citep{Bol++06,Bol++08a}. The LSD survey measured the internal
kinematics of 5 early-type lens galaxies over a large range of redshifts, 
$0.48<\zd<1.00$, and masses \citep{T+K04}. 

\subsection{The redshift distributions of SL2S lenses}

%%%%%%%%%%%%%%%%%%%%%
\begin{figure}
\centering
\includegraphics[width=0.9\linewidth]{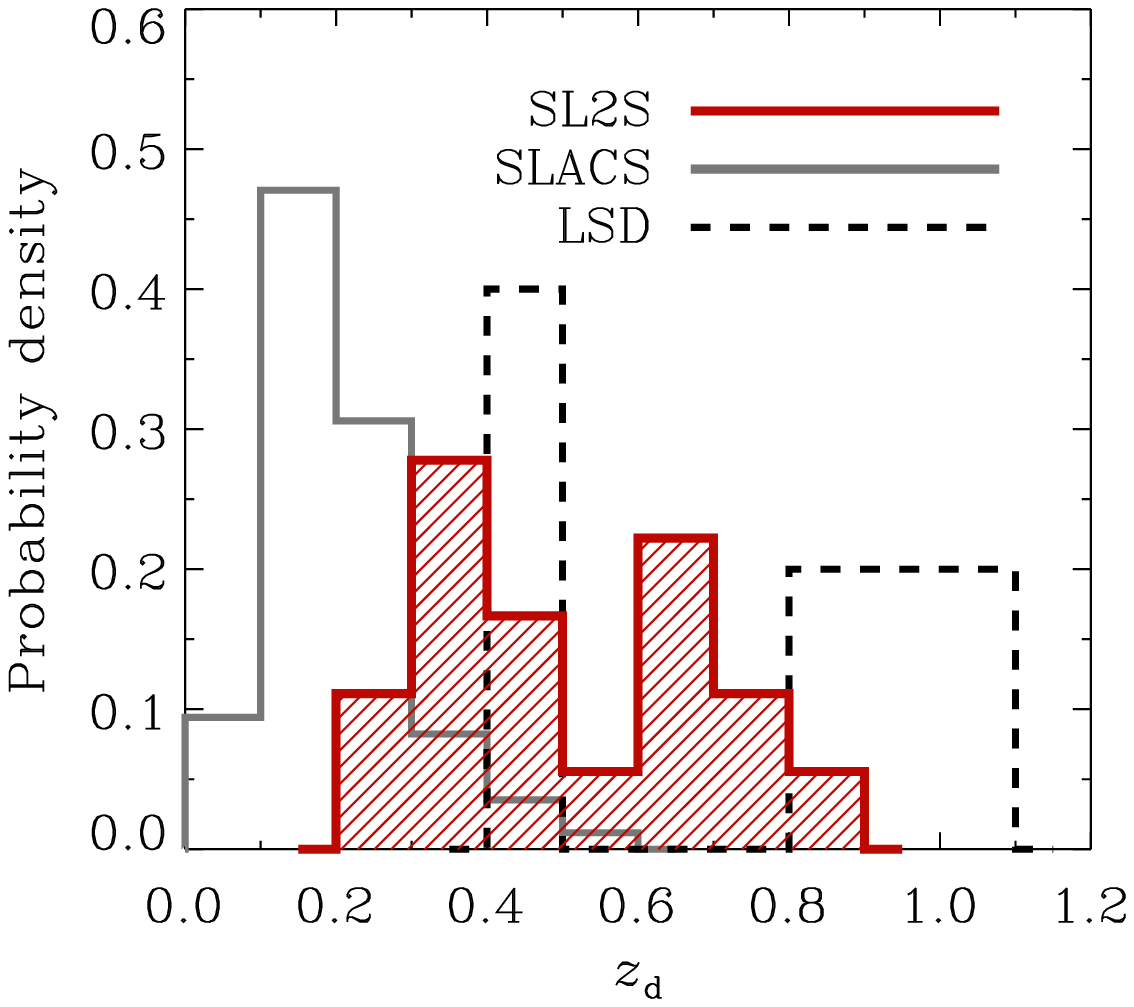}\\
\caption{\label{figure:redshift} Redshift distributions of the SL2S,  
SLACS and LSD surveys. All three distributions have been normalized to one.}
\end{figure}
%I got the SLACS data from IX (Auger)
%%%%%%%%%%%%%%%%%%%%%

It is instructive to compare the distribution of lens redshifts in the SL2S
sample, described in the preceding section, with those of the lower and higher
redshift reference samples, SLACS and LSD, respectively. In
Figure~\ref{figure:redshift}, we show histograms of deflector redshifts for the SL2S
(red), SLACS (grey) and LSD (black) samples. We see that, as anticipated, the
SL2S lenses lie at higher redshift than the SLACS sample. The median
and 68\% confidence intervals are $\zd$=\zdmedian$^{+\zdupper}_{-\zdlower}$,
$\zs$=\zsmedian$^{+\zsupper}_{-\zslower}$ for the SL2S sample,
$\zd$=\zdslacsmedian$^{+\zdslacsupper}_{-\zdslacslower}$
$\zs$=\zsslacsmedian$^{+\zsslacsupper}_{-\zsslacslower}$ for SLACS and
$\zd$=\zdlsdmedian$^{+\zdslacsupper}_{-\zdslacslower}$
$\zs$=\zslsdmedian$^{+\zslsdupper}_{-\zslsdlower}$ for LSD.

% - - - - - - - - - - - - - - - - - - - - - - - - - - - - - - - - - - - - - - -

\subsection{Comparison to previous lens samples}\label{sect:prevsamples}

\begin{figure}
\centering\includegraphics[width=0.9\linewidth]{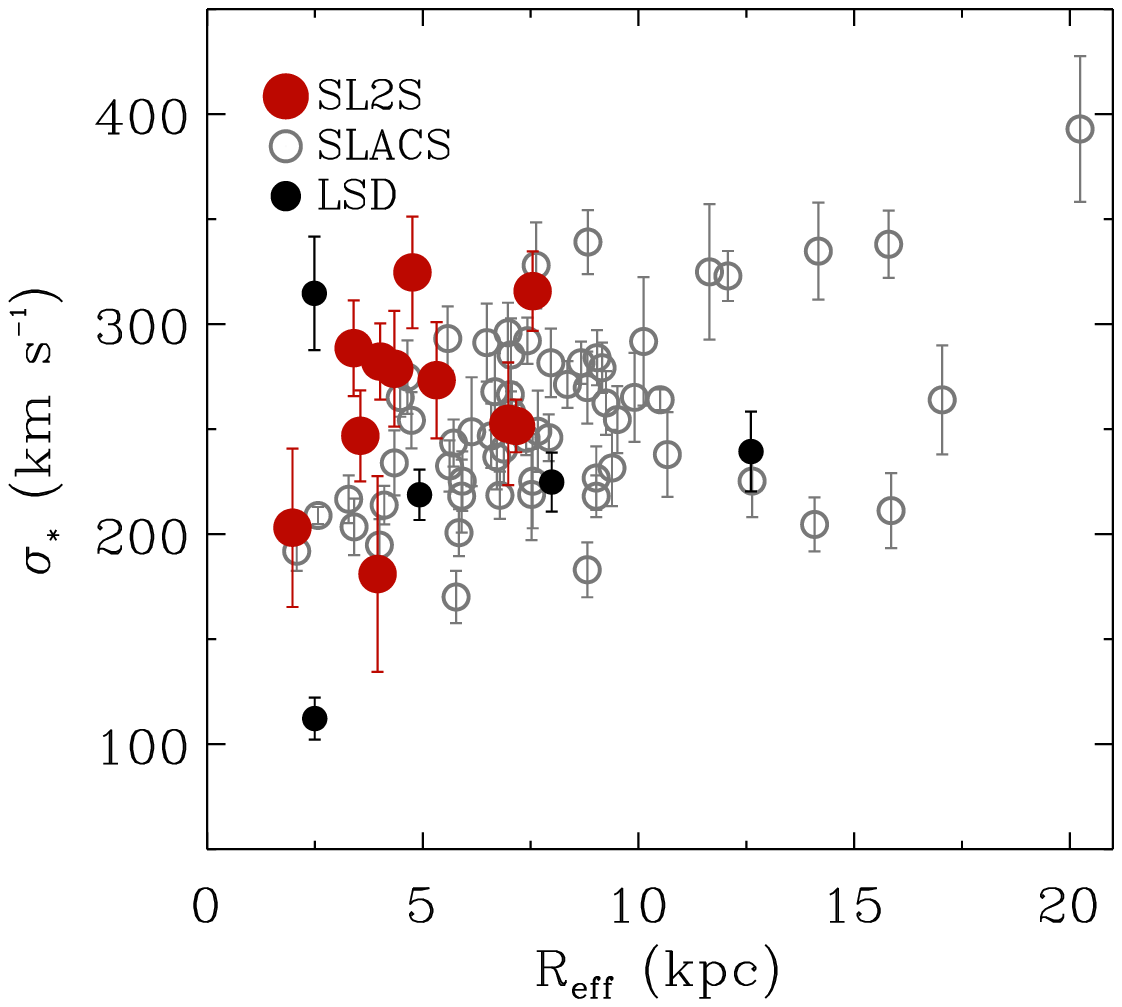}\\
\caption{\label{figure:vdisp_reff} Stellar velocity dispersions are
plotted against the effective radius in the V~band. The SL2S, SLACS
\citep{Aug++09} and LSD \citep{T+K04} data points are shown in red, grey and
black, respectively. Note that $\Reff$ is measured at the intermediate axis 
and the stellar velocity dispersions have been corrected to $\sigma_{e2}$. } 
\end{figure}

The measured stellar velocity dispersions and effective radii provide us with a
means to compare the SL2S, SLACS and LSD samples, to assess whether we are studying
the same {\it types} of galaxies.  
In Figure~\ref{figure:vdisp_reff}, we plot
our spectroscopic velocity dispersions, $\sigma$ against the effective radii,
$\Reff$ inferred from the CFHTLS data above, 
% (where $\Reff$ is measured at the intermediate axis) 
in order to compare the types  of galaxies selected in the SL2S, SLACS and LSD
samples.  
The stellar velocity dispersions are consistent between the samples, while the
effective radii of the SLACS galaxies are slightly higher.  Note that the
stellar velocity dispersions have been normalized to a standard aperture,
$\sigmae2$ for all samples. 

%%%%%%%%%%%%%%%%%%%%%%%%%%%%
%%%%%%%%%%%%%%%%%%%%%%%%%%%%

To compare the samples quantitatively, we applied the
2D Kolmogorov-Smirnov test in the $\sigma - \Reff$ plane.
We found the KS statistic to be 0.28, and the significance level to be 91\%.
Given this high significance, we conclude that we are drawing
from the same population of galaxies, despite the marginal tendency towards
larger sizes in the SLACS sample.

%%%%%%%%%%%%%%%%%%%%%%%%%%%%
\begin{figure}
\centering\includegraphics[width=0.9\linewidth]{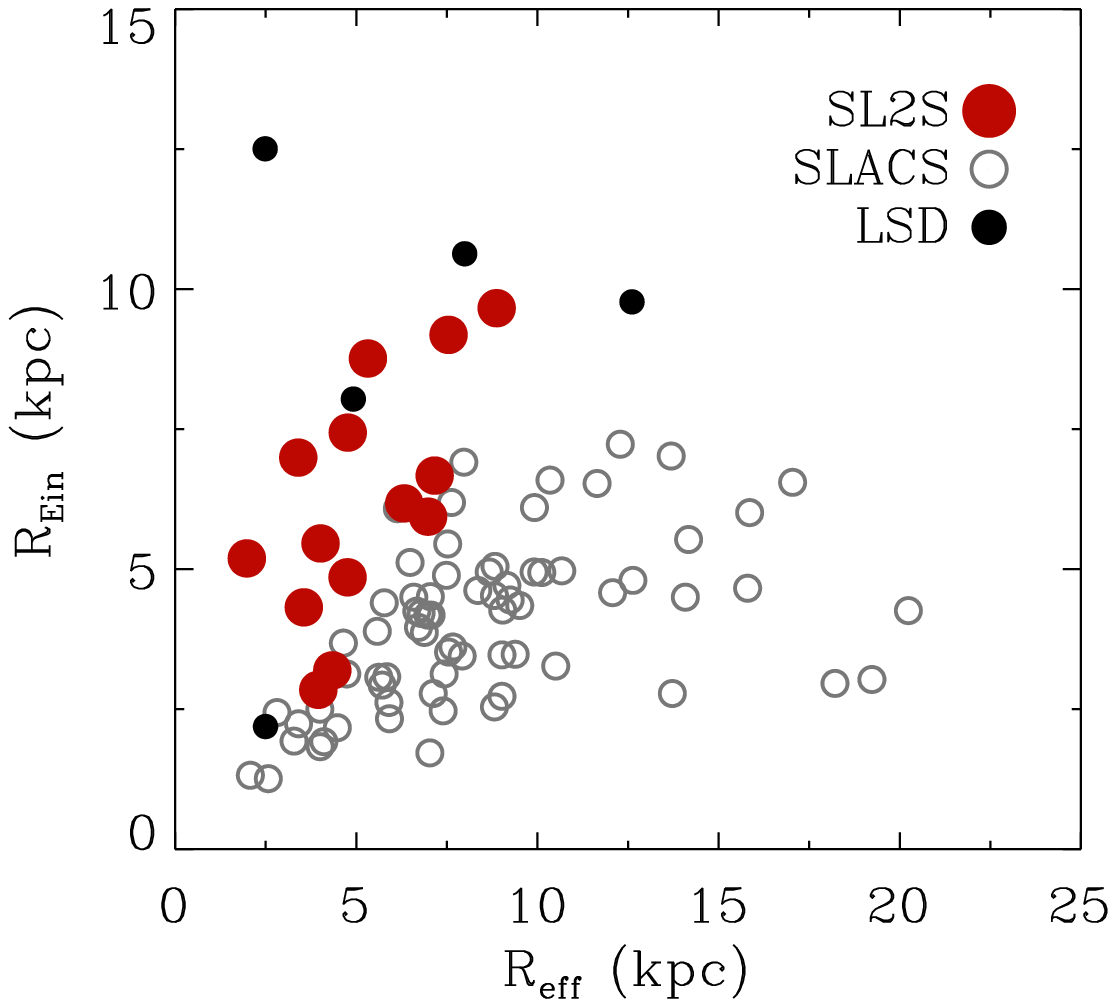}\\
\caption{\label{figure:reffrein} The effective radius is plotted against the Einstein radius. 
The SL2S, SLACS and LSD points are shown in red, grey and black, respectively. 
Typical uncertainties on $\REin$ and $\Reff$ are 5\% and 10\%, respectively. }
\end{figure}
%%%%%%%%%%%%%%%%%%%%%%%%%%%%

Figure~\ref{figure:reffrein} shows the distribution of $\Reff$ and $\REin$ for
the SL2S, SLACS and LSD samples. The higher redshift samples (SL2S and LSD),
have much higher physical $\REin$ values while the effective radii are slightly
smaller, as discussed above. The difference in $\REin$ is partly due to the
method used to select the lens candidates: resolvable SL2S ring radii
of~0.5-2~arcsec correspond to larger physical radii than those of the SLACS
sample. 
SLACS lenses were pre-selected to be bright SDSS target galaxies. This
selection method ensures that they lie at relatively low redshift, which in
turn means that the characterizing angular size of Einstein rings translates to
a smaller physical Einstein radii for the SLACS lenses than the SL2S lenses.
Furthermore, the source redshifts are on average significantly higher for the
LSD and SL2S samples, implying larger Einstein Radii for the same mass
distribution.

We conclude that we do seem to be measuring similar types of galaxies, with
similar sizes, in a similar mass range: the SLACS and SL2S lenses are
consistent with having been drawn from the same population. However, the
different lens geometry (due to the difference in redshift between the samples)
means that we are sampling the mass density profile at a larger radius, which
is approximately a factor of 2 times greater compared to the SLACS lenses.

% - - - - - - - - - - - - - - - - - - - - - - - - - - - - - - - - - - - - - - -

\section{The internal mass structure of SL2S lensing galaxies}\label{sect:lsd}

In this section we use our new spectroscopic measurements to carry out a joint
lensing and dynamics analysis of the SL2S sample, estimating the total mass
density profile slope and dark matter fraction of each individual lens
galaxy.  We first describe the simple model within which we work, and then
present our inferences of these two key parameters.

% - - - - - - - - - - - - - - - - - - - - - - - - - - - - - - - - - - - - - - - 

\subsection{Power-law density profiles}

We choose to 
work in the context of a spherically-symmetric power law total mass density
profile model \citep[\eg][]{T+K02a,Koo++06,Suy++10}:
\begin{eqnarray}
  \rho_{tot}(r) &\propto& r^{-\gamma'} \\
  \Sigma(R) &\propto& R^{1-\gamma'}.
\label{eq:powerlawprops}  
\end{eqnarray}
As in the lens modeling,  we use capitalized $R$ to denote projected radius. 
We can normalize these power law profiles in terms of $\MEin$, the
robustly-estimated, profile-independent Einstein Mass, from Section~\ref{sect:modelling}
above: our approach is to use the well-measured Einstein radii and Einstein
masses from the lens  models of \paperI, and re-interpret them within the
context of  this spherical power-law model.

The spherical power law model allows us to predict the observed  spectroscopic
velocity dispersion of the tracer stellar population, via the Jeans equation 
\citep[see, \eg][]{T+K04,Koo++06,Aug++10,Suy++10}. 
Normalizing the surface density and integrating to the
(observable) Einstein radius~$\REin$, we find that
\begin{eqnarray}
  \Sigma(R) &=& \frac{(3-\gamma')}{2}\, \Sigmacrit \left( \frac{R}{\REin} \right)^{1-\gamma'} \\
  \rho_{tot}(r) &=& \frac{(3-\gamma')}{2\pi^{1/2}} 
                     \frac{\Gamma(\frac{\gamma'}{2})}{\Gamma(\frac{\gamma'-1}{2})}
                     \frac{\Sigmacrit}{\REin}
                     \left( \frac{r}{\REin} \right)^{-\gamma'}. 
\label{eq:powerlaweqs}  
\end{eqnarray}
Since $\MEin = \pi \REin^2 \Sigmacrit$,
we can choose the two parameters of our model to be $\MEin$ and $\gamma'$. 
In the next subsection we outline how the stellar velocity
dispersion is predicted from these parameters.

% - - - - - - - - - - - - - - - - - - - - - - - - - - - - - - - - - - - - - - - 

\subsection{Stellar dynamics analysis}

As described by \citet{Suy++10} and \citet{Aug++10}, we predict the stellar
velocity dispersion at the appropriate aperture radius using the spherical
Jeans equation. Qualitatively we use the power law profile to compute the total
spherical mass enclosed within radius $r$, and assume the tracer stars follow a
Hernquist distribution with scale radius related to the measured effective
radius, and then integrate to predict the stellar velocity dispersion,
$\sigma(\MEin,\gamma')$. We direct the reader to Section~2.3 of \citet{Suy++10}
for all relevant equations.  We assume isotropic orbits for all of the
systems, which is approximately found to be the case from a more detailed
analysis of the resolved kinematics of several SLACS lenses \citep{Bar++09}.

% - - - - - - - - - - - - - - - - - - - - - - - - - - - - - - - - - - - - - - - 

\subsection{Joint analysis}

We propagate the uncertainties in the Einstein mass (which can be significant,
for systems with no spectroscopic source redshift), and the spectroscopic
velocity dispersion, in the following way. The output of the lens modeling
procedure for each lens is the posterior probability distribution for the
Einstein mass given the \hst image data~$\pixels$,
$\pr(\MEin|\pixels,\zd,\zs)$, characterized as a set of sample $\MEin$ values
-- this can be viewed as the {\it prior} PD for $\MEin$ for the joint
analysis. The analysis of that lens' spectrum yields the likelihood, 
$\pr(\sigma^{\rm obs}|\MEin,\gamma')$, which we assume to be Gaussian in
$\sigma^{\rm obs}$, with mean equal to the Jeans
prediction~$\sigma(\MEin,\gamma')$ from the previous subsection. 

The posterior PDF that we seek for each lens is then
\begin{eqnarray}
 \pr(\MEin,\gamma'|\sigma^{\rm obs},\pixels,\zd,\zs) 
 & \propto & \pr(\sigma^{\rm obs}|\MEin,\gamma') \nonumber\\ 
 & & \cdot \pr(\MEin|\pixels,\zd,\zs) \nonumber\\
 & & \cdot \pr(\gamma').
\end{eqnarray}
Since we are only working in the context of a single model, we do not compute
the evidence to normalize the right-hand side of this equation; all we need
are samples drawn from the posterior PDF $\pr(\MEin,\gamma'|\sigma^{\rm obs})$.

For the prior on the slope~$\gamma'$ we assume a uniform distribution with
limits of $-1.2$ and $-2.8$, which we find to be broad enough to enclose all
the likelihood (and are in fact close to the mathematical limits required  for
the normalizability of the profile). We draw an equal number of sample
$\gamma'$~values from this uniform distribution to match the $\MEin$ samples,
and then  use the likelihood evaluated at each 2-dimensional sample position
as weights: for each sample $\{\MEin,\gamma'\}$ we solve the Jeans equation to
calculate $\sigma(\MEin,\gamma')$, and then evaluate the Gaussian function 
$\pr(\sigma^{\rm obs}|\MEin,\gamma')$. (For more details on importance
sampling, see the appendix of \citeauthor{Suy++10}~\citeyear{Suy++10}.) This
set of weighted samples can then be used to compute integrals over the
posterior, confidence limits, histograms to represent the marginalized
distributions and so on. We present these inferences in the following two
sections, and summarize our numerical results in Table~\ref{table:inferred}.

% - - - - - - - - - - - - - - - - - - - - - - - - - - - - - - - - - - - - - - - 

\subsection{The density profile slope~$\gamma'$ from lensing and stellar
dynamics}\label{section:gammald}

\begin{figure}
\centering\includegraphics[width=0.9\linewidth]{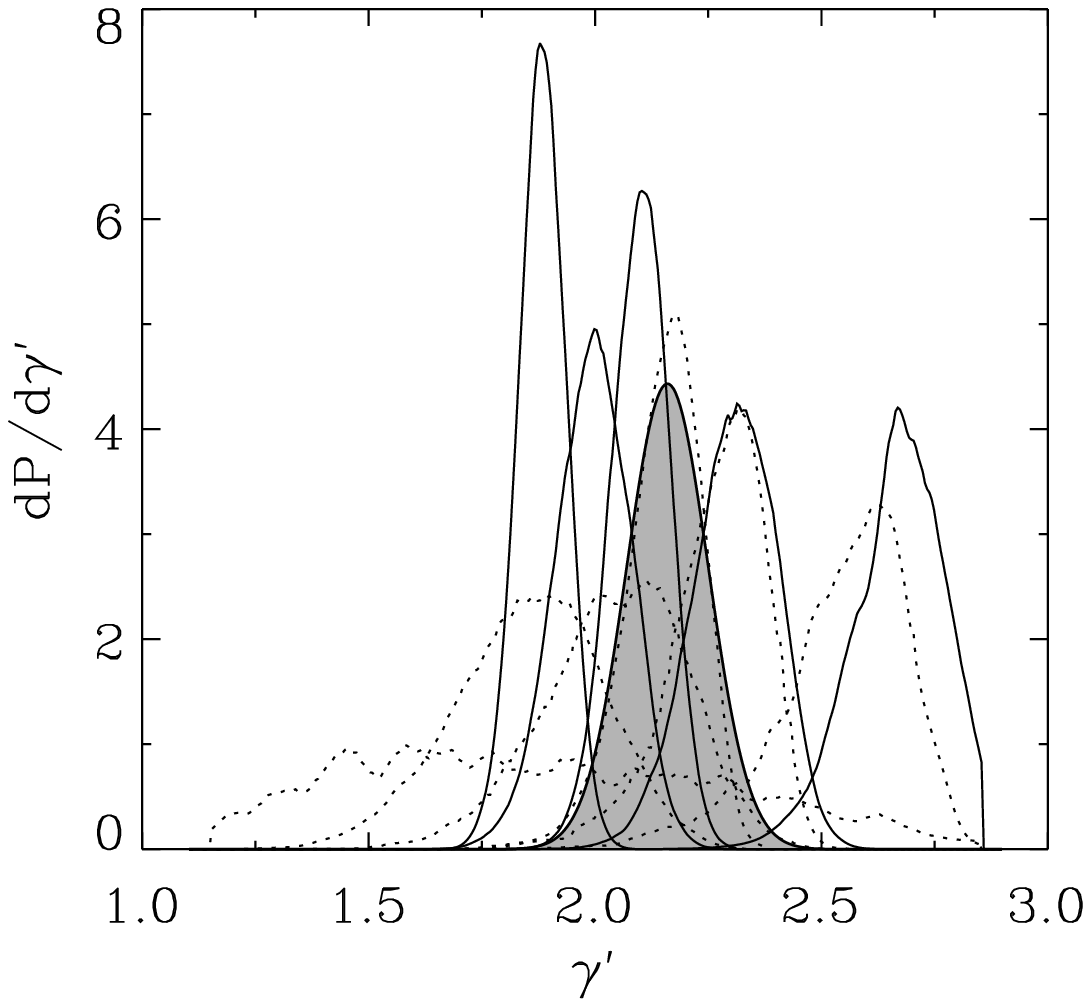}\\
\caption{\label{fig:gammadist} Posterior probability distributions for
$\gamma'$ using a uniform prior. The solid and dashed curves show $\gamma'$ 
distributions for individual lenses with and without measured source redshifts, respectively. 
The shaded region indicates the posterior PDF for the mean of the Gaussian
distribution from which the sample was inferred
to have been drawn: 
$\langle\gamma'\rangle$$\,=\,$$\meangammap^{+\errplusmeangammap}_{-\errminusmeangammap}$. 
Note that the distributions have been normalized to unit area. }
\end{figure}

In total we have \Ndyn~SL2S lenses with both lensing and dynamical mass
estimates. In Figure~\ref{fig:gammadist} we show the posterior probability
distributions for the logarithmic density profile slope~$\gamma'$, resulting
from the joint lensing and dynamics  analysis. We see that the mean of the
population lies %(shaded curve, from the product of the individual PDFs) lies 
at~$\langle\gamma'\rangle$$\,=\,$$\meangammap^{+\errplusmeangammap}_{-\errminusmeangammap}$, 
and is shown by a shaded region in Figure~\ref{fig:gammadist}.   
The intrinsic (Gaussian) scatter of the sample, inferred assuming
Gaussian errors on the individual $\gamma'$~values,  
is~$S_{\gamma'}$$\,=\,$$\scattergammap^{+\errplusscattergammap}_{-\errminusscattergammap}$.  
For each of the 11 lenses in the final sample, the median $\gamma'$ (with 
16$\th$ and 84$\th$ percentiles) is given in Table~\ref{table:inferred}

%%%%%%%%%%%%%%%%%%%%
%%%%%%%%%%%%%%%%%%%%%

%%%%%%%%%%%%%%%%%%%%%
\begin{figure}
\centering\includegraphics[width=0.9\linewidth]{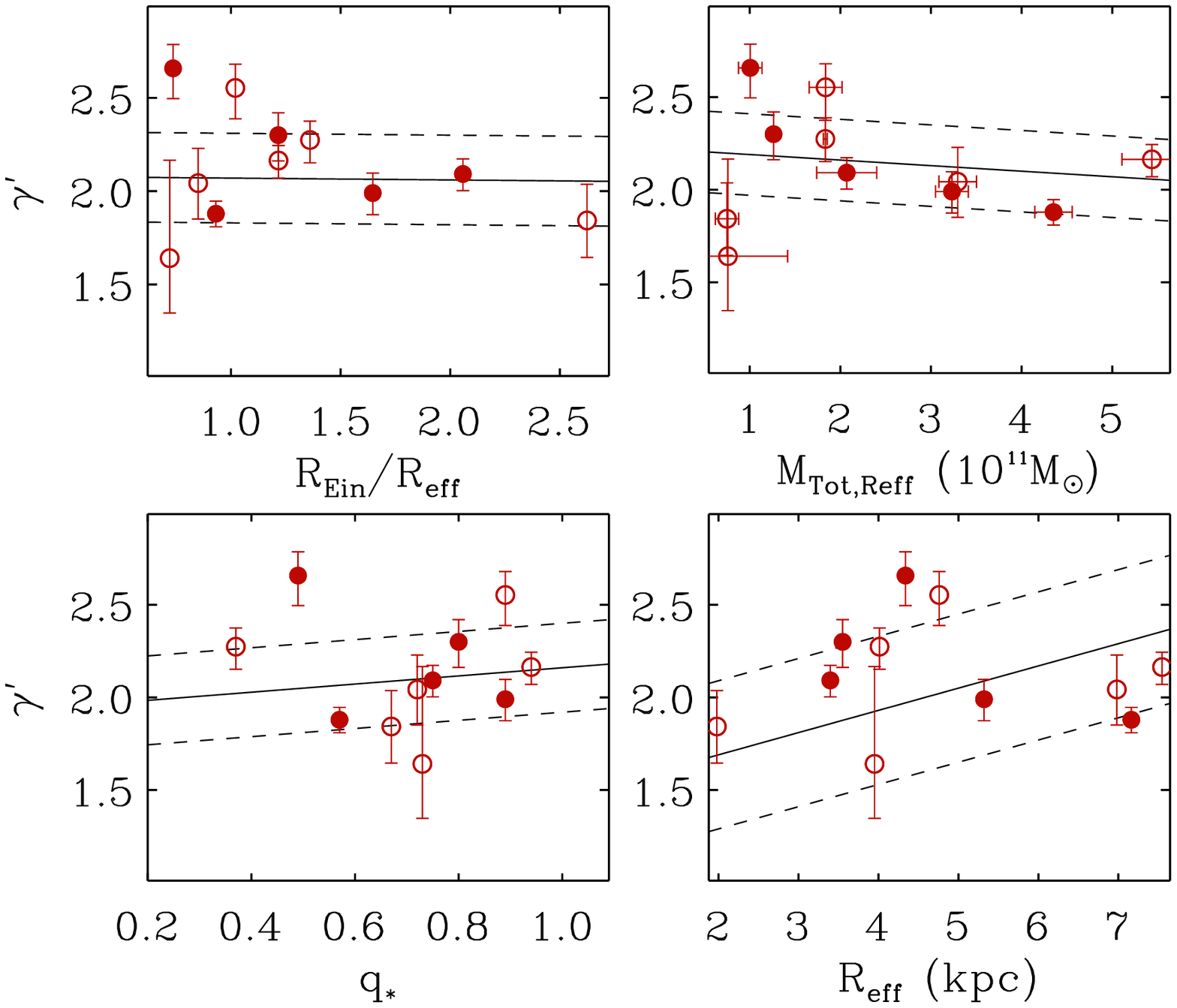}\\
\caption{\label{fig:gammaexplore} The slope of the density profile is plotted against 
Einstein radius, projected mass within the Einstein radius, stellar ellipticity and  
effective radius.  
The error bars show the 16$\th$ and 84$\th$ percentiles. 
The solid line shows the linear best fit to the data and the dashed lines indicate the scatter. 
} 
\end{figure}
%%%%%%%%%%%%%%%%%%%%%

We explore how $\gamma'$ varies with: $\REin/\Reff$, total mass within $\REin$, 
axis ratio 
and effective radius of the lenses in Figure~\ref{fig:gammaexplore}.  
In each panel, the solid line shows the linear best fit to the data and the 
dashed lines show the scatter.  
The gradients of the fits, including 16$\th$ and 84$\th$ percentiles are: 
\begin{eqnarray*}
d\gamma'/d(\REin/\Reff) &=& -0.01^{+0.21}_{-0.11} \\
d\gamma'/dM_{\Reff} &=& -0.03^{+0.17}_{-0.05} \\
d\gamma'/dq_* &=& 0.22^{+0.29}_{-0.17} \\
d\gamma'/d\Reff &=& 0.12^{+0.14}_{-0.07}, 
%d\gamma'/d &=& ^{+}_{-} 
\end{eqnarray*}
with $M_{\Reff}$ in units of $10^{11}\msun$, and $\Reff$ in kpc. 
The large uncertainties on the gradients indicate negligible trends between 
$\gamma'$ and these variables.

As discussed in Section~\ref{sect:discuss}, cosmic evolution of $\gamma'$ could
be mimicked by a dependence on $\REin/\Reff$.  However, a negligible
correlation between $\gamma'$ and $\REin/\Reff$ suggests that this is unlikely. 

We find no correlation between $\gamma'$ and either $\Reff$ or the projected
mass inside $\Reff$, indicating that $\gamma'$ is independent of galaxy size
and mass.

% - - - - - - - - - - - - - - - - - - - - - - - - - - - - - - - - - - - - - - - 

\subsection{The dark matter fraction in the SL2S lenses}\label{section:fdm}

We now turn to our second key parameter: the dark matter mass fraction, $\fdm$
in the cores of lens galaxies. Following the SLACS analyses of
\citet{Koo++06} and \citet{Aug++09}, we use $\Reff/2$ as the fiducial
aperture  radius for dark matter fraction estimation.  The stellar masses
calculated in Section~\ref{sec:stellarmass} are total  stellar masses. These must be
corrected to the mass within our fiducial aperture by integrating  (under the
assumption that stellar surface density follows surface brightness) the de
Vaucouleurs profile, which is given by:
\begin{equation} 
I(r) = I_{\rm eff}\, {\rm e}^{-7.67 \left[\left(\frac{r}{\Reff}\right)^{1/4} - 1\right] }.
\end{equation}
The fraction of mass enclosed within $\Reff$/2 is therefore given by:
\begin{equation}
\frac{M_*(\Reff/2)}{M_{*,\rm total}} = 
  \frac{\int_0^{\Reff/2} I(r)\, r\, dr}{\int_0^{\infty} I(r)\, r\, dr} = 0.320.%%%%%%%%%%0.319828
\label{eq:mstarextrapolate}
\end{equation}
We obtained total masses within the
same radius by integrating the power law total surface density profile:
\begin{equation}
\frac{M(\Reff/2)}{\MEin} = \left( \frac{\Reff/2}{\REin} \right)^{3-\gamma'}.
\label{eq:mtotalextrapolate}
\end{equation}
Total masses within the Einstein radius and total stellar masses are 
given in Table~\ref{table:inferred}.
The dark matter fraction is then defined by: 
\begin{equation}
\fdm = 1 - \frac{M_*(\Reff/2)}{M(\Reff/2)}.
\label{eq:fdm}
\end{equation}

We do not just have a single number for each of $M_*(\Reff/2)$ and 
$M(\Reff/2)$; rather, we have probability distributions for each. We combine
these to form the posterior PDF for $\fdm$, by applying the formulae in 
Equations~\ref{eq:mstarextrapolate},~\ref{eq:mtotalextrapolate}
and~\ref{eq:fdm} to each sample $\{M_*,M\}$ drawn from the  product of
$\pr(M_*)$ (from the SED modeling) and $\pr(M)$ (from the lensing plus
dynamics joint analysis) -- since these PDFs are independent,
the members of each $\{M_*,M\}$ pair can be drawn randomly from each
individual ensemble. 
The resulting PDF for $\fdm$ (visualized as a histogram of $\fdm$
samples) can have significant probability at $\fdm < 0$, due to the
uncertainty in each mass estimate, and the fact that our model only enforces
positivity for the total and stellar mass distributions, not their
difference. For consistency with the SLACS analysis, we allow negative values
of $\fdm$  (equivalent to $M_* > M$).  We found that 
truncating the PDFs at $\fdm \geq 0$ for the SL2S lenses did not significantly 
affect the numerical results.

%%%%%%%%%%%%%%%%%%%%%%%%%%
\begin{figure}
\centering\includegraphics[width=0.9\linewidth]{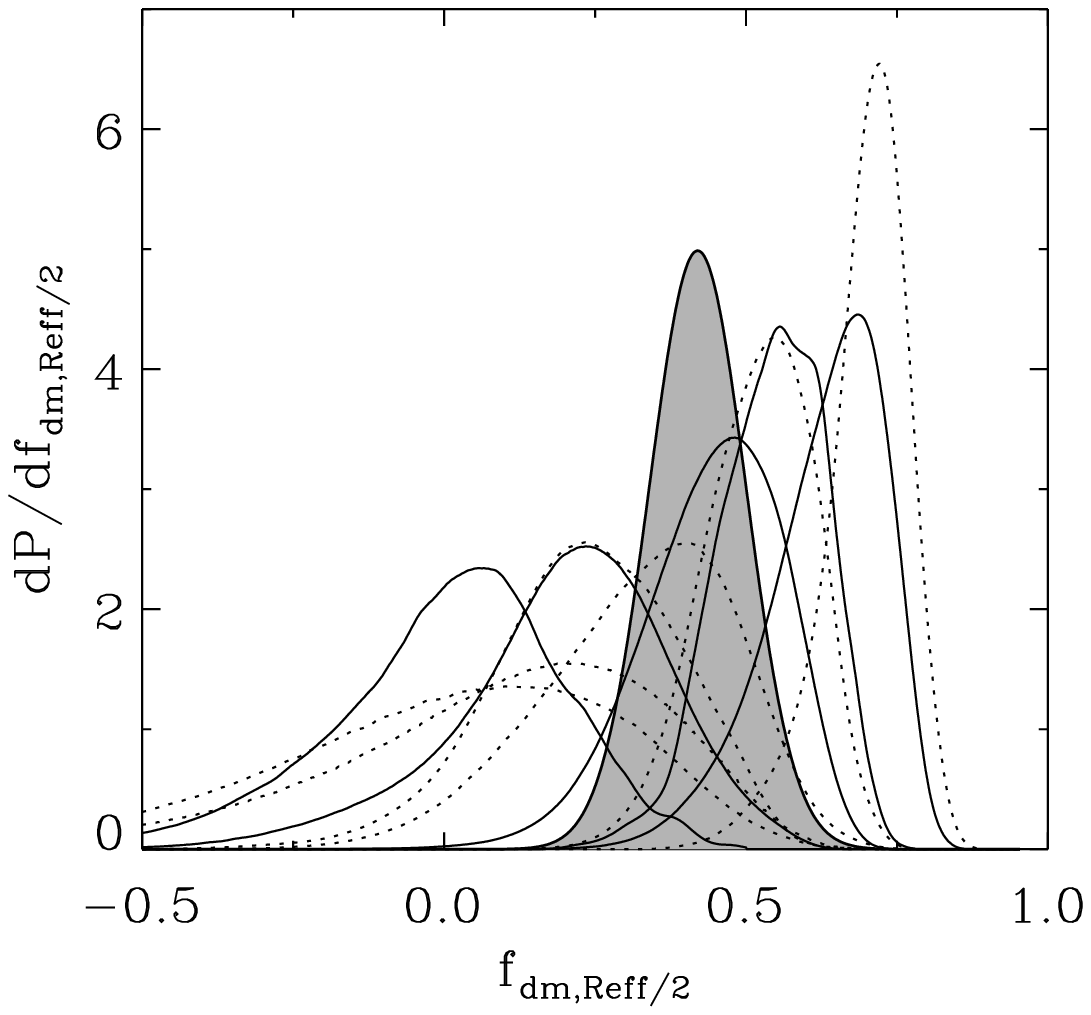}
\caption{\label{fig:fdmdist} Posterior probability distributions for the
dark matter fraction in the SL2S lenses. The dark matter distributions of lenses 
with measured source redshifts are shown in a solid line. The dashed lines show 
the distributions for lenses with no measured source redshift. 
Note that the distributions have been normalized to unit area. 
The shaded region indicates the posterior PDF for the mean of the Gaussian
distribution from which the sample was inferred
to have been drawn: $\langle\fdm\rangle = 
\meanfdm^{+\errplusmeanfdm}_{-\errminusmeanfdm}$. }
\end{figure}
%%%%%%%%%%%%%%%%%%%%%%%%%%

In Figure~\ref{fig:fdmdist} we show the resulting dark matter fraction
posterior probability distribution for each SL2S lens, assuming a
\citeauthor{Salpeter1955} IMF. The median and
16$\th$ and 84$\th$ percentiles of each dark matter fraction distribution are
given in Table~\ref{table:inferred}.  We see that the mean of the population
lies at~$\meanfdm^{+\errplusmeanfdm}_{+\errminusmeanfdm}$, as shown by the
shaded region.  
This distribution has an intrinsic
width~$\scatterfdm^{+\errplusscatterfdm}_{-\errminusscatterfdm}$.
Our results are sensitive to the choice of a universal
\citeauthor{Salpeter1955} IMF in the stellar populations analysis. 
If we instead assert a universal \citeauthor{Chabrier2003} IMF, 
we find that the mean dark matter fraction
is~$\meanfdmchab^{+\errplusmeanfdmchab}_{+\errminusmeanfdmchab}$,
while the scatter is essentially unchanged
($\scatterfdmchab^{+\errplusscatterfdmchab}_{-\errminusscatterfdmchab}$).
As expected, a \citeauthor{Chabrier2003} IMF predicts a
higher dark matter fraction, as shown in Table~\ref{table:inferred}.

%%%%%%%%%%%%%%%%%%%%%%%%%%
\begin{figure}
\centering\includegraphics[width=0.9\linewidth]{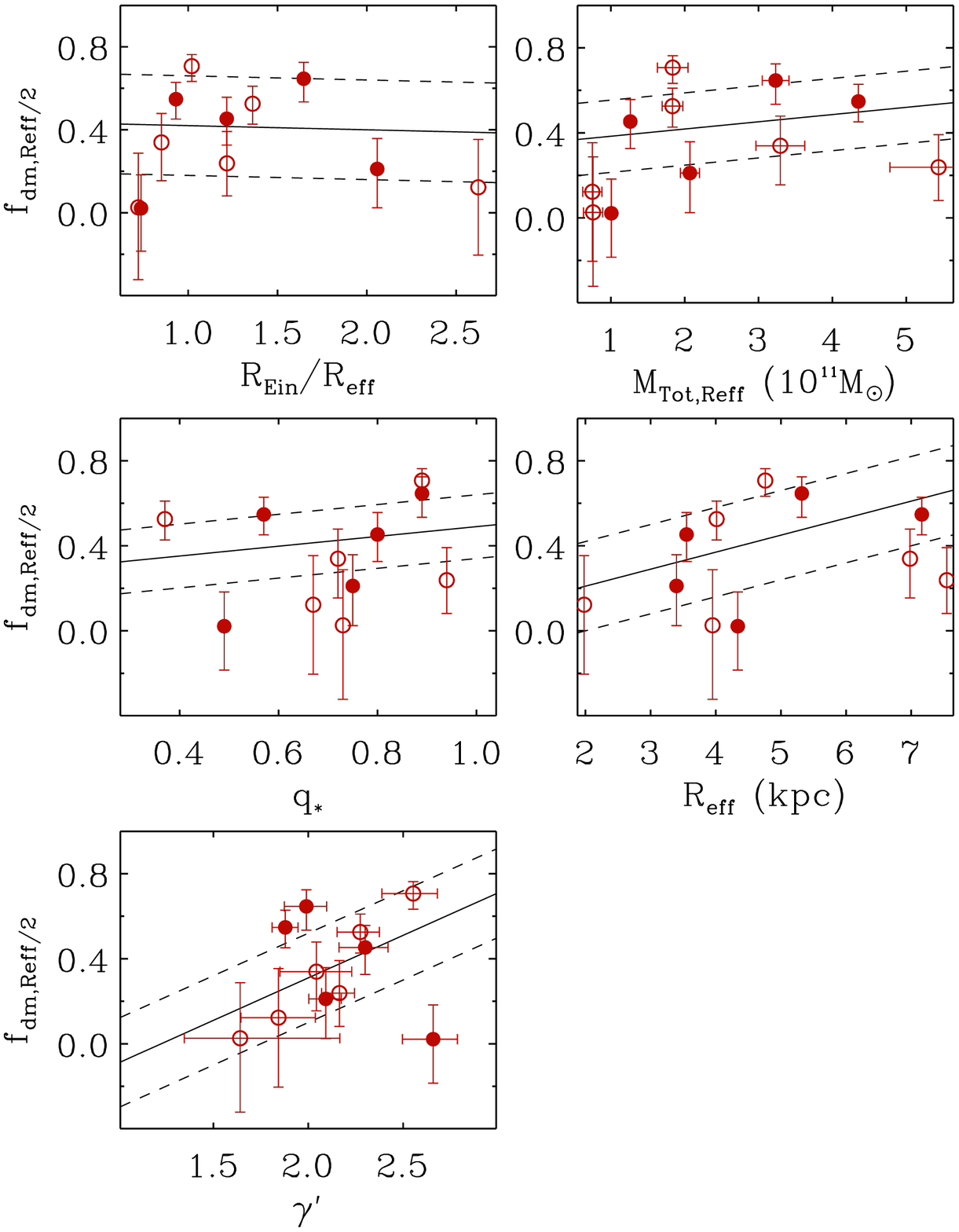}\\
\caption{\label{fig:fdmexplore} Dark matter fraction plotted against  
$\REin/\Reff$, the total mass within $\Reff$, axis ratio, 
total mass density profile slope, and $\Reff$.
The error bars show the 16$\th$ and 84$\th$ percentiles. 
The solid line shows the linear best fit to the data and the dashed lines indicate the scatter. 
}
\end{figure}
%%%%%%%%%%%%%%%%%%%%%%%%%%

In Figure~\ref{fig:fdmexplore} we show how the SL2S lens dark matter fractions
vary with: $\REin/\Reff$, the total mass within $\Reff$, axis ratio, 
 density profile slope, and $\Reff$.  In each panel, the solid
line shows the linear best fit to the data and the dashed lines show the
scatter. 
The gradients of the fits, including 16$\th$ and 84$\th$ percentiles are: 
\begin{eqnarray*}
d\fdm/d(\REin/\Reff) &=& -0.02^{+0.14}_{-0.12} \\
d\fdm/dM_{\Reff} 		 &=& 0.034^{+0.048}_{-0.045} \\
d\fdm/dq_* 					 &=& 0.23^{+0.29}_{-0.32} \\
d\fdm/d\Reff 				 &=& 0.08^{+0.10}_{-0.08} \\
d\fdm /d\gamma' 		 &=& 0.40^{+0.31}_{-0.12},
%d\gamma'/d &=& ^{+}_{-} 
\end{eqnarray*}
with $M_{\Reff}$ in units of $10^{11}\msun$, $\Reff$ in kpc and $\fdm$ is the dark 
matter within $\Reff/2$.  
There is evidence for a correlation between $\fdm$ and $\gamma'$.  
For all other variables, the large uncertainties on the gradients indicate that 
there is only a negligible trend with $\fdm$.

%-------------------------------------------------------------------------------

\section{Cosmic evolution}\label{sec:evolution}

We now investigate evolution in the properties of massive galaxies using our
new lens sample. We focus on the two key quantities studied in
Section~\ref{sect:lsd}: the density profile slope $\gamma'$, and the dark
matter fraction~$\fdm$. 

The low redshift reference measurements of $\gamma'$ and $\fdm$ come from the
SLACS analysis, which found density profiles very close to isothermal inside
one effective radius in a sample of 63 SLACS strong-lens early-type galaxies:
$\langle \gamma'_{\rm SLACS}\rangle =2.078\pm0.027$, with a scatter of~$0.16$
\citep{Aug++10}.  Likewise \citet{Aug++09} inferred a mean $\fdm$ within
$\Reff/2$ for 85 SLACS lenses of 0.3 with a scatter of 0.2, using a
\citeauthor{Salpeter1955} IMF. These results are broadly consistent with our
findings for SL2S systems.  We can therefore anticipate only a mild cosmic evolution
of these quantities.

\begin{figure}
\centering\includegraphics[width=0.9\linewidth]{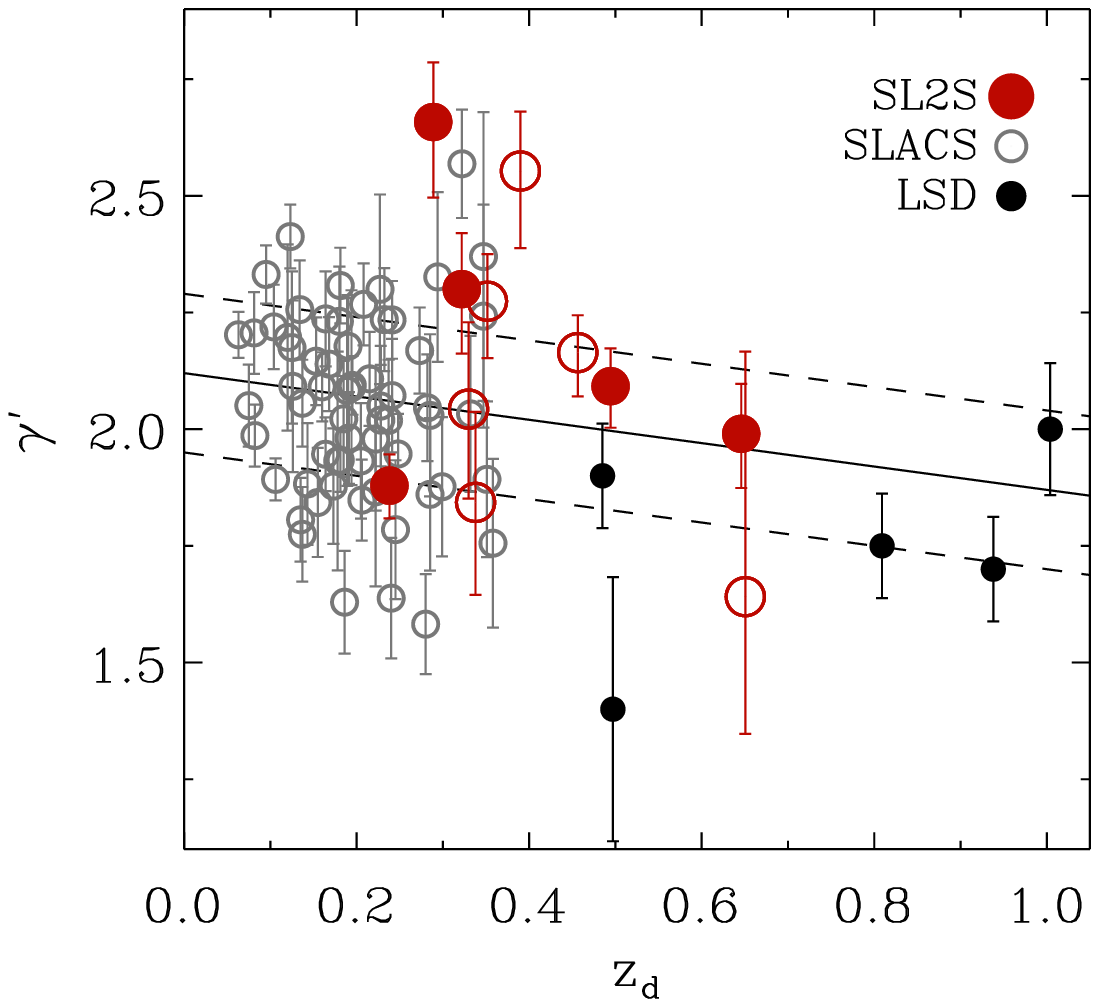}
\caption{\label{fig:gvsz} Cosmic evolution of total mass density 
slope, $\gamma'$. The SLACS and LSD values were taken from: \citep{Aug++10} 
and \citep{T+K02a,K+T03,T+K04}, respectively. 
The error bars show the 16$\th$ and 84$\th$ percentiles. 
The best fit to the data is shown by the solid line and the scatter is shown by the dashed lines. }
% gamma prime vs z_d LSD data from \citep{T+K04}, SLACS data, Not from8
\end{figure}

In Figure~\ref{fig:gvsz}, we show how the density profile slope
$\gamma'$ varies with redshift, using the SL2S, SLACS and LSD samples
together to cover the redshift range 0.05 to 1. 
We quantify this statement by fitting the $\gamma'(\zd)$ data with a linear
relation in the mean slope, still including Gaussian scatter about that
relation:
\begin{equation}
  \langle\gamma'\rangle(\zd) = \langle\gamma'_0\rangle 
    + \frac{\partial\langle\gamma'\rangle}{\partial\zd}\, \zd \pm S_{\gamma'}.
\end{equation}
For the SL2S data alone, we 
find~$\langle\gamma'_0\rangle$$\,=\,$$\meanvgammap^{+\errplusmeanvgammap}_{-\errminusmeanvgammap}$, 
$\partial\langle\gamma'\rangle/\partial\zd$$\,=\,$$\gradvgammap^{+\errplusgradvgammap}_{-\errminusgradvgammap}$ 
for the gradient and, in this evolving~$\gamma'$ case, the scatter is 
$S_{\gamma'}$$\,=\,$$\scattervgammap^{+\errplusscattervgammap}_{-\errminusscattervgammap}$. 
When we include the SLACS and LSD data points, we 
find~$\langle\gamma'_0\rangle$$\,=\,$$\meanallvgammap^{+\errplusmeanallvgammap}_{-\errminusmeanallvgammap}$,  
$\partial\langle\gamma'\rangle/\partial\zd$$\,=\,$$\gradallvgammap^{+\errplusgradallvgammap}_{-\errminusgradallvgammap}$,
and~$S_{\gamma'}$$\,=\,$$\scatterallvgammap^{+\errplusscatterallvgammap}_{-\errminusscatterallvgammap}$. 

These results are inconsistent with no evolution in the total density profile
slope of massive lens galaxies since $z\simeq1$: the probability of the linear
gradient in $\langle\gamma'\rangle$ being positive is just 2\%. The lens data
suggest (at approximately the 2-$\sigma$ level) that the mean total density
profile of massive galaxies has become slightly steeper over cosmic time.

%%%%%%%%%%%%%%%%%%%%%%%%
%%%%%%%%%%%%%%%%%%%%%%%%

%%%%%%%%%%%%%%%%%%%%%%%%
\begin{figure}
\centering\includegraphics[width=0.9\linewidth]{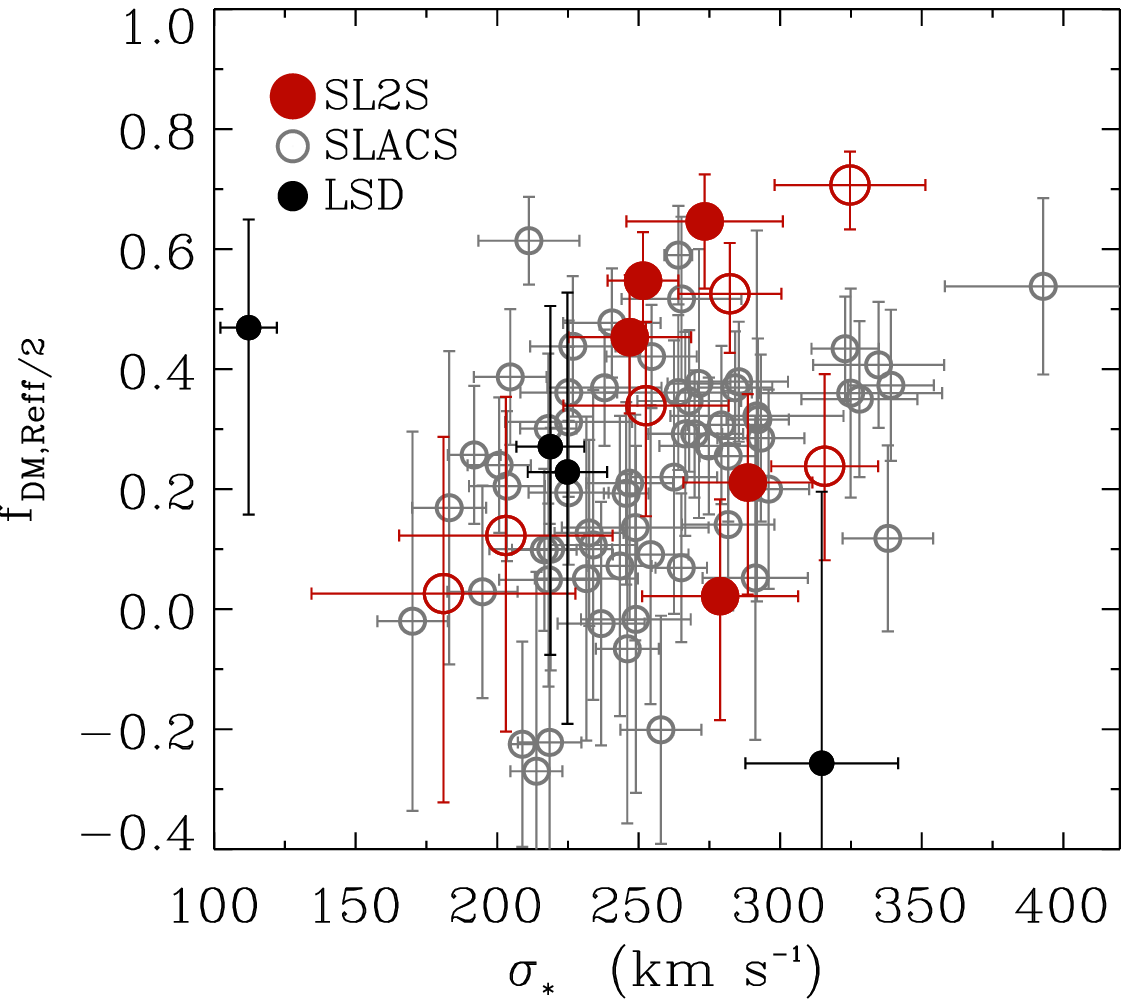}\\
\caption{\label{fig:fvsv} The dark matter fraction within $\Reff/2$ 
as a function of measured velocity dispersion. 
The error bars show the 16$\th$ and 84$\th$ percentiles. 
For all samples, the velocity dispersion was normalized to a standard aperture, $\Reff/2$.}
\end{figure}
%%%%%%%%%%%%%%%%%%%%%%%

%%%%%%%%%%%%%%%%%%%%%%%%
\begin{figure}
\centering\includegraphics[width=0.9\linewidth]{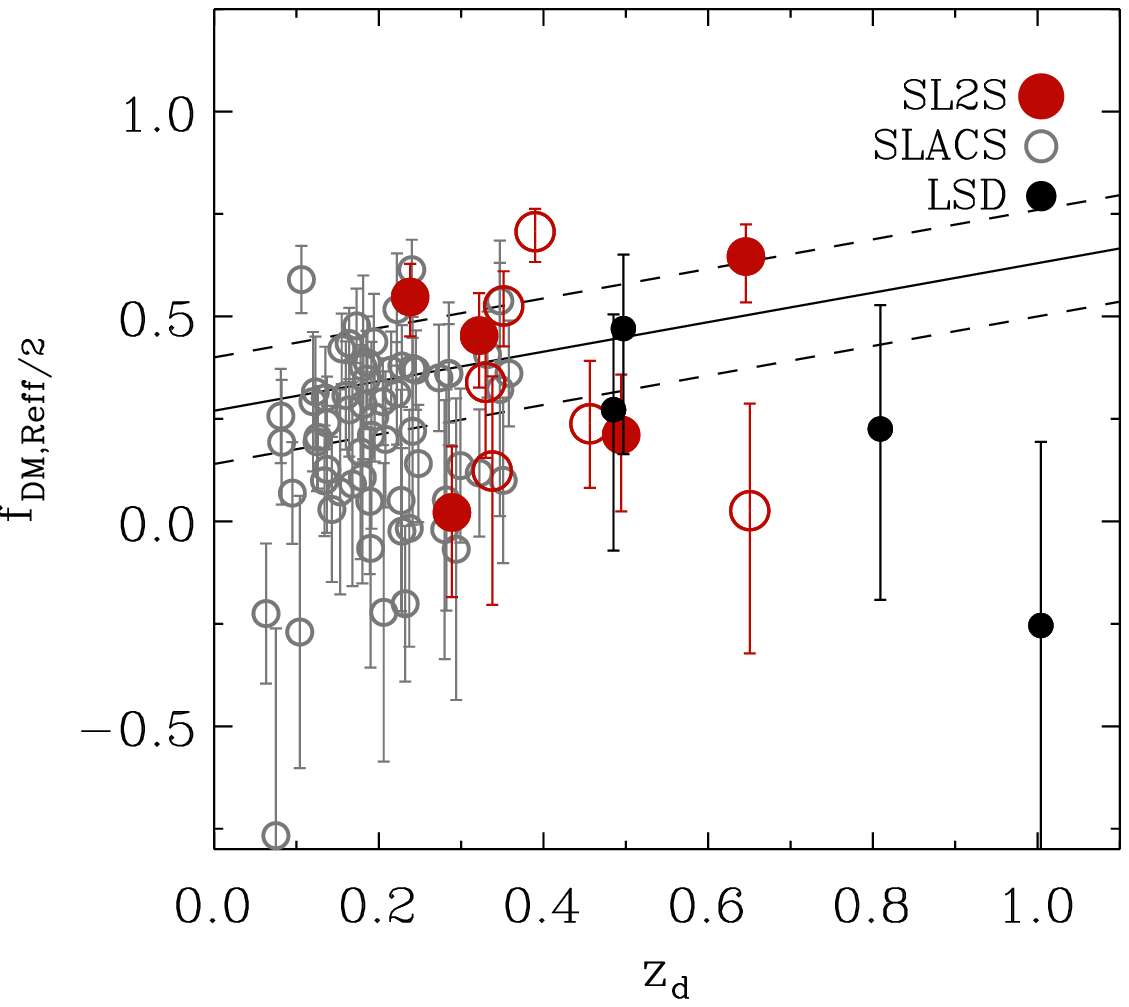}\\
\caption{\label{fig:fvsz} Evolution of dark matter fraction with redshift for a 
\citeauthor{Salpeter1955} IMF. 
The error bars show the 16$\th$ and 84$\th$ percentiles. 
The solid and dashed lines show the best fit to the data and the scatter, respectively.}
\end{figure}
%%%%%%%%%%%%%%%%%%%%%%%%

Figure~\ref{fig:fvsv} shows the dark matter fraction as a function of the
measured stellar velocity dispersion, $\sigmae2$ for the SL2S, SLACS and LSD
samples. Together, these samples cover a large rage in stellar velocity
dispersion, however, no significant trend with central dark matter fraction was
found. 
In Figure~\ref{fig:fvsz}, we show how the dark matter fraction varies with
redshift, using the SL2S, SLACS and LSD samples together to cover the redshift
range 0.05 to 1.0, and focusing on the \citeauthor{Salpeter1955} IMF.  (The
Chabrier IMF assignment only affects the overall normalization of the stellar
masses and not their evolution.) Only 4 of the 5 LSD lenses were included,
since photometry in multiple bands was required to infer stellar masses  (given
in Section~\ref{sec:stellarmass}).  We find that the mean dark matter fraction
has not evolved strongly with cosmic time, however, there is marginal evidence
for some change in the population.  Again, we quantify this statement by
fitting the $\fdm(\zd)$ data with a linear relation in the mean slope, still
including Gaussian scatter about that relation:
\begin{equation} 
\langle\fdm\rangle(\zd) = \langle\fdmbar\rangle +
\frac{\partial\langle\fdm\rangle}{\partial\zd}\, \zd \pm S_{\fdm}.
\end{equation}
For the SL2S data alone, we
find~$\langle\fdmbar\rangle$$\,=\,$$\meanvfdm^{+\errplusmeanvfdm}_{-\errminusmeanvfdm}$,
$\partial\langle\fdm\rangle/\partial\zd$$\,=\,$$\gradvfdm^{+\errplusgradvfdm}_{-\errminusgradvfdm}$
for the gradient and
$S_{\fdm}$$\,=\,$$\scattervfdm^{+\errplusscattervfdm}_{-\errminusscattervfdm}$
for the scatter.  Including the SLACS and LSD data points, we
find~$\langle\fdmbar\rangle$$\,=\,$$\meanallvfdm^{+\errplusmeanallvfdm}_{-\errminusmeanallvfdm}$,
$\partial\langle\fdm\rangle/\partial\zd$$\,=\,$$\gradallvfdm^{+\errplusgradallvfdm}_{-\errminusgradallvfdm}$,
and~$S_{\fdm}$$\,=\,$$\scatterallvfdm^{+\errplusscatterallvfdm}_{-\errminusscatterallvfdm}$.
The probability of the linear gradient in $\langle\fdm\rangle$ being positive
is 98\%: the lens data suggest (again at approximately the 2-$\sigma$ level)
that the mean projected dark matter fraction in massive galaxies, within half
their effective radius, has decreased slightly over cosmic time.

%-------------------------------------------------------------------------------

\section{Discussion}\label{sect:discuss}

By combining the inferred total mass density slopes of the SL2S, SLACS
and LSD samples, we found a tantalizing suggestion that
$\gamma'$ has become slightly steeper over cosmic time. 

%%%%%%%%%%%%%%%%%%%%%%%%
\begin{figure}
\centering\includegraphics[width=0.9\linewidth]{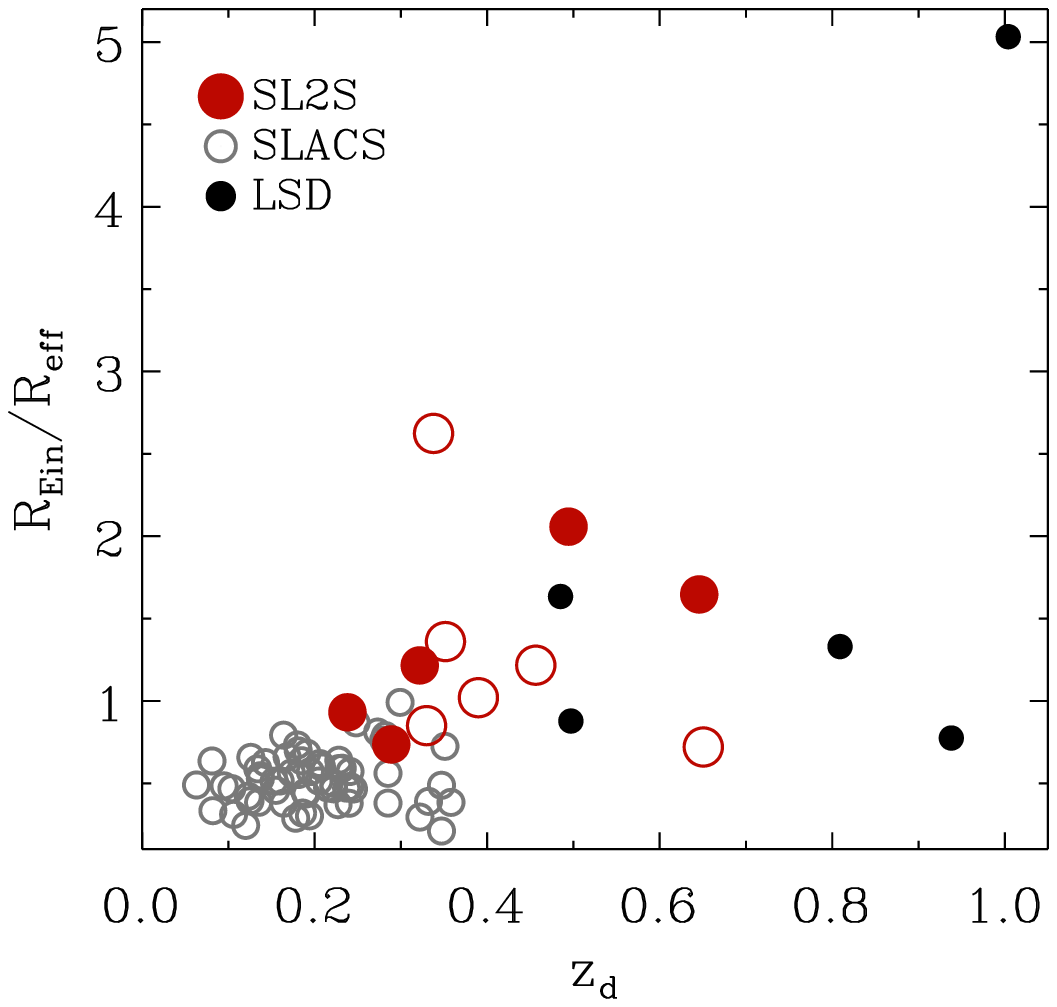}
\caption{\label{fig:reinreff_zd} Evolution of $\REin/\Reff$ with redshift.  The
ratio of $\REin/\Reff$ tends to increase with redshift. } \end{figure}
%%%%%%%%%%%%%%%%%%%%%%%

Before interpreting this result, and its implications for our
understanding of the formation and evolution of early-type galaxies, it
is important to address a potential source of systematic error. As
shown in Figure~\ref{fig:reinreff_zd}, the ratio of $\REin/\Reff$
increases with redshift, mainly because the physical size of $\REin$ increases 
with redshift (as discussed in Section~\ref{sect:prevsamples}), but also because 
of the increasing source redshift.
Since the slope of the total mass density is determined by measuring
the total mass at two different radii, one of which is the Einstein
radius, SL2S lenses sample $\gamma'$ at larger radii than SLACS lenses. 
Therefore, a trend in $\REin/\Reff$ with
redshift could mimic the inferred evolution of $\gamma'$. 
However, for this effect to mimic the evolution of $\gamma'$, the
density profile would have to become shallower and then steeper with
increasing radius, which is unlikely given the total mass density
profile of local early-type galaxies.
We also note that $\gamma'$ and $\fdm$ vary differently with cosmic
time, which is unexpected if the evolution being mimicked by a trend
in $\REin/\Reff$ with redshift. As discussed in
Sections~\ref{section:gammald} and~\ref{section:fdm},
Figures~\ref{fig:gammaexplore} and~\ref{fig:fdmexplore} show that both
$\gamma'$ and $\fdm$ show a negligible trend with $\REin/\Reff$,
suggesting that the evolution of $\gamma'$ is not an artifact due to a
dependence on $\REin/\Reff$.

For these reasons we conclude that true evolution of the average mass
density profile is the most likely interpretation of our findings,
although additional measurements of $\gamma'$ at different radii are
needed to conclusively rule out alternate interpretations.

It is important to stress that both the mean value and evolution of $\gamma'$
are completely consistent with previous results, $\langle\gamma'_{0,\rm
SLACS+LSD}\rangle$$\,=$$\,$$2.10\pm 0.07$ and
$\partial\langle\gamma'\rangle/\partial\zd$$\,=\,$$-0.23\pm0.16$
\citep{Koo++06}. It is only by virtue of our larger sample that we have
been able to reduce the error bars and find marginal evidence for evolution.
If this evidence for evolution is confirmed by larger samples, it would
indicate that growth of massive galaxies since $z\sim1$ has not occurred
through dry (dissipationless) mergers alone, since they preserve $\gamma'$
\citep[see \eg][]{NTB09}. In dissipative merging events, $\gamma'$ increases as
a result of baryons cooling and sinking towards the center. Therefore,
dissipative processes would be required to contribute, at least partially, to
the evolution of early-type galaxies since $z\sim1$. 
Although they cannot be the dominant process, because of tight limits on recent
star formation in massive early-type galaxies since $z\sim1$
\citep[e.g.,][]{Tre++05}, our result seems to suggest that they cannot be
completely neglected either. This is consistent with evidence for a
``frosting'' of recent stars found in detailed studies of the stellar
populations of massive galaxies \citep[e.g.,][]{Tra++00a}.

Another notable feature of the SL2S $\gamma'$ distribution is its large
intrinsic scatter. The intrinsic scatter in the SLACS sample is just
0.16$\pm$0.02 \citep{Aug++10}, compared to
$\scattergammap^{+\errplusscattergammap}_{-\errminusscattergammap}$ for SL2S.
From the LSD lenses, \citet{T+K04} also found that the intrinsic scatter in
$\gamma'$ was larger at $z=1$. This may indicate an overall trend toward more
complete dynamical relaxation over Gyr timescales, and perhaps a reduced
contribution of external convergence, due to line of sight structure in nearby
systems, where the Einstein radii are smaller. Again, a larger sample size is
required to confirm this interpretation. 
Finally, we find that the dark matter fraction within half of the effective
radius has decreased slightly with cosmic time. This is again consistent with
some contribution from dissipational processes in early-type galaxy formation
and evolution, where baryons move to the central region as they cool. However,
it is important to keep in mind residual uncertainties in estimating the
stellar mass, which may be redshift dependent, since the average age of the
stellar populations is a function of cosmic time. Cosmic evolution of the IMF
would be another source of ambiguity in interpreting these results
\citep{vdo08,Tre++10}. 

To conclude, we have developed a new method to estimate source
redshifts, in the absence of spectroscopic measurements. The method is
completely general, does not require accurate multiband photometry of
the sources, and allows us to infer $\gamma'$ and $\fdm$ with errors
that are just 2-3 times as large as if we had spectroscopic
redshifts. Although spectroscopic redshifts are in general preferable,
this may be a good strategy for exploiting future surveys of thousands
of strong lenses, where wholesale spectroscopy of the complete sample
may not be practical.

%-------------------------------------------------------------------------------

\section{Summary and Conclusions}\label{sect:concl}

New spectroscopic measurements with deflector redshifts and velocity
dispersions were presented for 11 lenses.  These spectroscopic measurements
were combined with lens models and photometry, described in \paperI, to infer
the total density slope and dark matter fraction of each of the 11 SL2S
galaxies in the final sample. 

The main results are summarized below:
\begin{enumerate}
\item The SL2S sample has a median deflector redshift, $\zd$$\,=\,$$\zdmedian$,
  source redshift, $\zs$$\,=\,$$\zsmedian$ and velocity dispersion, 
  $\sigmae2$$\,=\,$$\sigmamedian$~km~s$^{-1}$. 
\item The SL2S, SLACS and LSD lenses are the same types of galaxies, however, 
  the physical size of $\REin$ is generally larger for higher redshift deflectors. 
\item We developed a new method to estimate the source redshift probability
  distribution function for lenses with no spectroscopic $\zs$.
  This lack of accurate source redshift produces uncertainties on $\gamma'$
  and $\fdm$ that are only a factor 2-3 greater.
  Uncertainties would eventually further decrease with better multiband source
  photometry, and this would allow large samples of high 
  redshift lenses to be analyzed in the next generation of cosmological surveys, where
  spectroscopy of all the systems may not be affordable.
\item The average total density slope measured from the SL2S sample alone is: 
  $\langle\gamma'_0\rangle$$\,=\,$$\meangammap^{+\errplusmeangammap}_{-\errminusmeangammap}$, 
  with a scatter of
  $\scattergammap^{+\errplusscattergammap}_{-\errminusscattergammap}$. 
\item Combining the SL2S $\gamma'$ measurements with previous analyses from
  SLACS and LSD, we find, 
  $\langle\gamma'_0\rangle$$\,=\,$$\meanallvgammap^{+\errplusmeanallvgammap}_{-\errminusmeanallvgammap}$,  
  $\partial\langle\gamma'\rangle/\partial\zd$$\,=\,$$\gradallvgammap^{+\errplusgradallvgammap}_{-\errminusgradallvgammap}$,
  and~$S_{\gamma'}$$\,=\,$$\scatterallvgammap^{+\errplusscatterallvgammap}_{-\errminusscatterallvgammap}$. 
  This suggests (at approximately the 2-$\sigma$ level) that the mean total density 
  profile of massive galaxies has become slightly steeper over cosmic time.
\item Stellar masses were estimated using CFHT photometry, and   
  enabled us to disentangle the total and stellar masses. 
\item   From this we inferred the dark matter fraction within 
  half the effective radius:  
  $\langle\fdmbar\rangle$$\,=\,$$\meanvfdm^{+\errplusmeanvfdm}_{-\errminusmeanvfdm}$, 
  with a scatter
	$S_{\fdm}$$\,=\,$$\scattervfdm^{+\errplusscattervfdm}_{-\errminusscattervfdm}$
  for the SL2S sample, using a \citeauthor{Salpeter1955} IMF.
\item The combined dark matter fractions from all three samples 
	suggest that the dark matter fraction within $\Reff/2$ has decreased slightly 
	since $z\sim1$.  
	We find the mean dark matter fraction within $\Reff/2$, gradient of evolution 
	over cosmic time and scatter to be: 
	$\langle\fdmbar\rangle$$\,=\,$$\meanallvfdm^{+\errplusmeanallvfdm}_{-\errminusmeanallvfdm}$,  
	$\partial\langle\fdm\rangle/\partial\zd$$\,=\,$$\gradallvfdm^{+\errplusgradallvfdm}_{-\errminusgradallvfdm}$,
	and $S_{\fdm}$$\,=\,$$\scatterallvfdm^{+\errplusscatterallvfdm}_{-\errminusscatterallvfdm}$, respectively. 
	
\end{enumerate}

%-------------------------------------------------------------------------------

\acknowledgments

%%%  
We thank our friends of the SLACS and SL2S collaborations for many
useful and insightful discussions over the course of the past years.

AJR acknowledges the support of an Australian Postgraduate Award.
RG and FB acknowledge support from the Centre National des Etudes
Spatiales (CNES).
PJM was given support by the TABASGO and Kavli foundations in the
form of two research fellowships. 
TT acknowledges support from the NSF through CAREER award NSF-0642621,
and from the Packard Foundation through a Packard Research Fellowship.
Based on observations obtained with MegaPrime/MegaCam, a joint project
of CFHT and CEA/DAPNIA, at the Canada-France-Hawaii Telescope (CFHT)
which is operated by the National Research Council (NRC) of Canada,
the Institut National des Sciences de l'Univers of the Centre National
de la Recherche Scientifique (CNRS) of France, and the University of
Hawaii.  This work is based in part on data products produced at
TERAPIX and the Canadian Astronomy Data Centre as part of the
Canada-France-Hawaii Telescope Legacy Survey, a collaborative project
of NRC and CNRS.
This research is supported by NASA through Hubble Space Telescope
programs GO-10876, GO-11289, GO-11588 and in part by the National
Science Foundation under Grant No. PHY99-07949, and is based on
observations made with the NASA/ESA Hubble Space Telescope and
obtained at the Space Telescope Science Institute, which is operated
by the Association of Universities for Research in Astronomy, Inc.,
under NASA contract NAS 5-26555, and at the W.M. Keck Observatory,
which is operated as a scientific partnership among the California
Institute of Technology, the University of California and the National
Aeronautics and Space Administration. The Observatory was made
possible by the generous financial support of the W.M. Keck
Foundation. The authors wish to recognize and acknowledge the very
significant cultural role and reverence that the summit of Mauna Kea
has always had within the indigenous Hawaiian community.  We are most
fortunate to have the opportunity to conduct observations from this
mountain.

%-------------------------------------------------------------------------------

\bibliographystyle{apj}

\begin{thebibliography}{87}
\expandafter\ifx\csname natexlab\endcsname\relax\def\natexlab#1{#1}\fi

\bibitem[{{Auger} {et~al.}(2009){Auger}, {Treu}, {Bolton}, {Gavazzi},
  {Koopmans}, {Marshall}, {Bundy}, \& {Moustakas}}]{Aug++09}
{Auger}, M.~W., {Treu}, T., {Bolton}, A.~S., {Gavazzi}, R., {Koopmans},
  L.~V.~E., {Marshall}, P.~J., {Bundy}, K., \& {Moustakas}, L.~A. 2009, \apj,
  705, 1099

\bibitem[{{Auger} {et~al.}(2010){Auger}, {Treu}, {Bolton}, {Gavazzi},
  {Koopmans}, {Marshall}, \& {Burles}}]{Aug++10}
{Auger}, M.~W., {Treu}, T., {Bolton}, A.~S., {Gavazzi}, R., {Koopmans},
  L.~V.~E., {Marshall}, P.~J., \& {Burles}, S. 2010, \apj

\bibitem[{{Barnabe} {et~al.}(2010){Barnabe}, {Auger}, {Treu}, {Koopmans},
  {Bolton}, {Czoske}, \& {Gavazzi}}]{Barnabe2010}
{Barnabe}, M., {Auger}, M.~W., {Treu}, T., {Koopmans}, L., {Bolton}, A.~S.,
  {Czoske}, O., \& {Gavazzi}, R. 2010, \mnras, 955

\bibitem[{{Barnab{\`e}} {et~al.}(2009){Barnab{\`e}}, {Czoske}, {Koopmans},
  {Treu}, {Bolton}, \& {Gavazzi}}]{Bar++09}
{Barnab{\`e}}, M., {Czoske}, O., {Koopmans}, L.~V.~E., {Treu}, T., {Bolton},
  A.~S., \& {Gavazzi}, R. 2009, \mnras, 399, 21

\bibitem[{{Bernardi} {et~al.}(2005){Bernardi}, {Sheth}, {Nichol}, {Schneider},
  \& {Brinkmann}}]{Ber++05}
{Bernardi}, M., {Sheth}, R.~K., {Nichol}, R.~C., {Schneider}, D.~P., \&
  {Brinkmann}, J. 2005, \aj, 129, 61

\bibitem[{{Bertin} \& {Stiavelli}(1993)}]{B+S93}
{Bertin}, G., \& {Stiavelli}, M. 1993, Reports on Progress in Physics, 56, 493

\bibitem[{{Bolton} {et~al.}(2008{\natexlab{a}}){Bolton}, {Burles}, {Koopmans},
  {Treu}, {Gavazzi}, {Moustakas}, {Wayth}, \& {Schlegel}}]{Bol++08a}
{Bolton}, A.~S., {Burles}, S., {Koopmans}, L.~V.~E., {Treu}, T., {Gavazzi}, R.,
  {Moustakas}, L.~A., {Wayth}, R., \& {Schlegel}, D.~J. 2008{\natexlab{a}},
  \apj, 682, 964

\bibitem[{{Bolton} {et~al.}(2006){Bolton}, {Burles}, {Koopmans}, {Treu}, \&
  {Moustakas}}]{Bol++06}
{Bolton}, A.~S., {Burles}, S., {Koopmans}, L.~V.~E., {Treu}, T., \&
  {Moustakas}, L.~A. 2006, \apj, 638, 703

\bibitem[{{Bolton} {et~al.}(2008{\natexlab{b}}){Bolton}, {Treu}, {Koopmans},
  {Gavazzi}, {Moustakas}, {Burles}, {Schlegel}, \& {Wayth}}]{Bol++08b}
{Bolton}, A.~S., {Treu}, T., {Koopmans}, L.~V.~E., {Gavazzi}, R., {Moustakas},
  L.~A., {Burles}, S., {Schlegel}, D.~J., \& {Wayth}, R. 2008{\natexlab{b}},
  \apj, 684, 248

\bibitem[{{Bruzual} \& {Charlot}(2003)}]{bruz2003}
{Bruzual}, G., \& {Charlot}, S. 2003, \mnras, 344, 1000

\bibitem[{{Bundy} {et~al.}(2005){Bundy}, {Ellis}, \& {Conselice}}]{BEC05}
{Bundy}, K., {Ellis}, R.~S., \& {Conselice}, C.~J. 2005, \apj, 625, 621

\bibitem[{{Bundy} {et~al.}(2007){Bundy}, {Treu}, \& {Ellis}}]{BTE07}
{Bundy}, K., {Treu}, T., \& {Ellis}, R.~S. 2007, \apjl, 665, L5

\bibitem[{{Cabanac} {et~al.}(2007){Cabanac}, {Alard}, {Dantel-Fort}, {Fort},
  {Gavazzi}, {Gomez}, {Kneib}, {Le F{\`e}vre}, {Mellier}, {Pello}, {Soucail},
  {Sygnet}, \& {Valls-Gabaud}}]{Cab++07}
{Cabanac}, R.~A., {Alard}, C., {Dantel-Fort}, M., {Fort}, B., {Gavazzi}, R.,
  {Gomez}, P., {Kneib}, J.~P., {Le F{\`e}vre}, O., {Mellier}, Y., {Pello}, R.,
  {Soucail}, G., {Sygnet}, J.~F., \& {Valls-Gabaud}, D. 2007, \aap, 461, 813

\bibitem[{{Cappellari} {et~al.}(2006){Cappellari}, {Bacon}, {Bureau}, {Damen},
  {Davies}, {de Zeeuw}, {Emsellem}, {Falc{\'o}n-Barroso}, {Krajnovi{\'c}},
  {Kuntschner}, {McDermid}, {Peletier}, {Sarzi}, {van den Bosch}, \& {van de
  Ven}}]{Cap++06}
{Cappellari}, M., {Bacon}, R., {Bureau}, M., {Damen}, M.~C., {Davies}, R.~L.,
  {de Zeeuw}, P.~T., {Emsellem}, E., {Falc{\'o}n-Barroso}, J., {Krajnovi{\'c}},
  D., {Kuntschner}, H., {McDermid}, R.~M., {Peletier}, R.~F., {Sarzi}, M., {van
  den Bosch}, R.~C.~E., \& {van de Ven}, G. 2006, \mnras, 366, 1126

\bibitem[{{Cappellari} {et~al.}(2009){Cappellari}, {di Serego Alighieri},
  {Cimatti}, {Daddi}, {Renzini}, {Kurk}, {Cassata}, {Dickinson},
  {Franceschini}, {Mignoli}, {Pozzetti}, {Rodighiero}, {Rosati}, \&
  {Zamorani}}]{Cap++09}
{Cappellari}, M., {di Serego Alighieri}, S., {Cimatti}, A., {Daddi}, E.,
  {Renzini}, A., {Kurk}, J.~D., {Cassata}, P., {Dickinson}, M., {Franceschini},
  A., {Mignoli}, M., {Pozzetti}, L., {Rodighiero}, G., {Rosati}, P., \&
  {Zamorani}, G. 2009, \apjl, 704, L34

\bibitem[{{Cardone} \& {Tortora}(2010)}]{Cardone10}
{Cardone}, V.~F., \& {Tortora}, C. 2010, ArXiv e-prints

\bibitem[{{Cardone} {et~al.}(2009){Cardone}, {Tortora}, {Molinaro}, \&
  {Salzano}}]{Cardone09}
{Cardone}, V.~F., {Tortora}, C., {Molinaro}, R., \& {Salzano}, V. 2009, \aap,
  504, 769

\bibitem[{{Cassata} {et~al.}(2010){Cassata}, {Giavalisco}, {Guo}, {Ferguson},
  {Koekemoer}, {Renzini}, {Fontana}, {Salimbeni}, {Dickinson}, {Casertano},
  {Conselice}, {Grogin}, {Lotz}, {Papovich}, {Lucas}, {Straughn}, {Gardner}, \&
  {Moustakas}}]{Cas++09}
{Cassata}, P., {Giavalisco}, M., {Guo}, Y., {Ferguson}, H., {Koekemoer}, A.~M.,
  {Renzini}, A., {Fontana}, A., {Salimbeni}, S., {Dickinson}, M., {Casertano},
  S., {Conselice}, C.~J., {Grogin}, N., {Lotz}, J.~M., {Papovich}, C., {Lucas},
  R.~A., {Straughn}, A., {Gardner}, J.~P., \& {Moustakas}, L. 2010, \apjl, 714,
  L79

\bibitem[{{Chabrier}(2003)}]{Chabrier2003}
{Chabrier}, G. 2003, \pasp, 115, 763

\bibitem[{{Cimatti} {et~al.}(2006){Cimatti}, {Daddi}, \& {Renzini}}]{CDR06}
{Cimatti}, A., {Daddi}, E., \& {Renzini}, A. 2006, \aap, 453, L29

\bibitem[{{Ciotti}(2009)}]{Cio09a}
{Ciotti}, L. 2009, \nat, 460, 333

\bibitem[{{Ciotti} {et~al.}(2007){Ciotti}, {Lanzoni}, \& {Volonteri}}]{CLV07}
{Ciotti}, L., {Lanzoni}, B., \& {Volonteri}, M. 2007, \apj, 658, 65

\bibitem[{{Ciotti} {et~al.}(2009){Ciotti}, {Ostriker}, \& {Proga}}]{COP09}
{Ciotti}, L., {Ostriker}, J.~P., \& {Proga}, D. 2009, \apj, 699, 89

\bibitem[{{Croton} {et~al.}(2006){Croton}, {Springel}, {White}, {De Lucia},
  {Frenk}, {Gao}, {Jenkins}, {Kauffmann}, {Navarro}, \& {Yoshida}}]{Cro++06}
{Croton}, D.~J., {Springel}, V., {White}, S.~D.~M., {De Lucia}, G., {Frenk},
  C.~S., {Gao}, L., {Jenkins}, A., {Kauffmann}, G., {Navarro}, J.~F., \&
  {Yoshida}, N. 2006, \mnras, 365, 11

\bibitem[{{Daddi} {et~al.}(2005){Daddi}, {Renzini}, {Pirzkal}, {Cimatti},
  {Malhotra}, {Stiavelli}, {Xu}, {Pasquali}, {Rhoads}, {Brusa}, {di Serego
  Alighieri}, {Ferguson}, {Koekemoer}, {Moustakas}, {Panagia}, \&
  {Windhorst}}]{Dad++05}
{Daddi}, E., {Renzini}, A., {Pirzkal}, N., {Cimatti}, A., {Malhotra}, S.,
  {Stiavelli}, M., {Xu}, C., {Pasquali}, A., {Rhoads}, J.~E., {Brusa}, M., {di
  Serego Alighieri}, S., {Ferguson}, H.~C., {Koekemoer}, A.~M., {Moustakas},
  L.~A., {Panagia}, N., \& {Windhorst}, R.~A. 2005, \apj, 626, 680

\bibitem[{{de Vaucouleurs}(1948)}]{dev48}
{de Vaucouleurs}, G. 1948, Annales d'Astrophysique, 11, 247

\bibitem[{{Dehnen}(2005)}]{Deh05}
{Dehnen}, W. 2005, \mnras, 360, 892

\bibitem[{{di Serego Alighieri} {et~al.}(2005){di Serego Alighieri}, {Vernet},
  {Cimatti}, {Lanzoni}, {Cassata}, {Ciotti}, {Daddi}, {Mignoli}, {Pignatelli},
  {Pozzetti}, {Renzini}, {Rettura}, \& {Zamorani}}]{diS++05}
{di Serego Alighieri}, S., {Vernet}, J., {Cimatti}, A., {Lanzoni}, B.,
  {Cassata}, P., {Ciotti}, L., {Daddi}, E., {Mignoli}, M., {Pignatelli}, E.,
  {Pozzetti}, L., {Renzini}, A., {Rettura}, A., \& {Zamorani}, G. 2005, \aap,
  442, 125

\bibitem[{{Faure} {et~al.}(2008){Faure}, {Kneib}, {Covone}, {Tasca},
  {Leauthaud}, {Capak}, {Jahnke}, {Smolcic}, {de la Torre}, {Ellis},
  {Finoguenov}, {Koekemoer}, {Le Fevre}, {Massey}, {Mellier}, {Refregier},
  {Rhodes}, {Scoville}, {Schinnerer}, {Taylor}, {Van Waerbeke}, \&
  {Walcher}}]{Fau++08}
{Faure}, C., {Kneib}, J.-P., {Covone}, G., {Tasca}, L., {Leauthaud}, A.,
  {Capak}, P., {Jahnke}, K., {Smolcic}, V., {de la Torre}, S., {Ellis}, R.,
  {Finoguenov}, A., {Koekemoer}, A., {Le Fevre}, O., {Massey}, R., {Mellier},
  Y., {Refregier}, A., {Rhodes}, J., {Scoville}, N., {Schinnerer}, E.,
  {Taylor}, J., {Van Waerbeke}, L., \& {Walcher}, J. 2008, \apjs, 176, 19

\bibitem[{{Gavazzi} {et~al.}(2007){Gavazzi}, {Treu}, {Rhodes}, {Koopmans},
  {Bolton}, {Burles}, {Massey}, \& {Moustakas}}]{Gav++07}
{Gavazzi}, R., {Treu}, T., {Rhodes}, J.~D., {Koopmans}, L.~V.~E., {Bolton},
  A.~S., {Burles}, S., {Massey}, R.~J., \& {Moustakas}, L.~A. 2007, \apj, 667,
  176

\bibitem[{{Graves} {et~al.}(2009){Graves}, {Faber}, \& {Schiavon}}]{GFS09a}
{Graves}, G.~J., {Faber}, S.~M., \& {Schiavon}, R.~P. 2009, \apj, 693, 486

\bibitem[{{Grillo} {et~al.}(2008){Grillo}, {Gobat}, {Rosati}, \&
  {Lombardi}}]{Gri++08b}
{Grillo}, C., {Gobat}, R., {Rosati}, P., \& {Lombardi}, M. 2008, \aap, 477, L25

\bibitem[{{Hopkins} {et~al.}(2010){Hopkins}, {Bundy}, {Hernquist}, {Wuyts}, \&
  {Cox}}]{Hop++10d}
{Hopkins}, P.~F., {Bundy}, K., {Hernquist}, L., {Wuyts}, S., \& {Cox}, T.~J.
  2010, \mnras, 401, 1099

\bibitem[{{Hyde} \& {Bernardi}(2009)}]{HydeBernardi2009}
{Hyde}, J.~B., \& {Bernardi}, M. 2009, \mnras, 396, 1171

\bibitem[{{Jiang} \& {Kochanek}(2007)}]{J+K07}
{Jiang}, G., \& {Kochanek}, C.~S. 2007, \apj, 671, 1568

\bibitem[{{J{\o}rgensen} {et~al.}(1995){J{\o}rgensen}, {Franx}, \&
  {Kjaergaard}}]{jorgensen1995}
{J{\o}rgensen}, I., {Franx}, M., \& {Kjaergaard}, P. 1995, \mnras, 276, 1341

\bibitem[{{Juneau} {et~al.}(2005){Juneau}, {Glazebrook}, {Crampton},
  {McCarthy}, {Savaglio}, {Abraham}, {Carlberg}, {Chen}, {Le Borgne}, {Marzke},
  {Roth}, {J{\o}rgensen}, {Hook}, \& {Murowinski}}]{Jun++05}
{Juneau}, S., {Glazebrook}, K., {Crampton}, D., {McCarthy}, P.~J., {Savaglio},
  S., {Abraham}, R., {Carlberg}, R.~G., {Chen}, H., {Le Borgne}, D., {Marzke},
  R.~O., {Roth}, K., {J{\o}rgensen}, I., {Hook}, I., \& {Murowinski}, R. 2005,
  \apjl, 619, L135

\bibitem[{{Kazantzidis} {et~al.}(2006){Kazantzidis}, {Zentner}, \&
  {Kravtsov}}]{KZK06}
{Kazantzidis}, S., {Zentner}, A.~R., \& {Kravtsov}, A.~V. 2006, \apj, 641, 647

\bibitem[{{Khochfar} \& {Silk}(2006)}]{K+S06}
{Khochfar}, S., \& {Silk}, J. 2006, \apjl, 648, L21

\bibitem[{{Kochanek}(1994)}]{kochanek1994}
{Kochanek}, C.~S. 1994, \apj, 436, 56

\bibitem[{{Koopmans} {et~al.}(2009){Koopmans}, {Bolton}, {Treu}, {Czoske},
  {Auger}, {Barnab{\`e}}, {Vegetti}, {Gavazzi}, {Moustakas}, \&
  {Burles}}]{Koo++09}
{Koopmans}, L.~V.~E., {Bolton}, A., {Treu}, T., {Czoske}, O., {Auger}, M.~W.,
  {Barnab{\`e}}, M., {Vegetti}, S., {Gavazzi}, R., {Moustakas}, L.~A., \&
  {Burles}, S. 2009, \apjl, 703, L51

\bibitem[{{Koopmans} \& {Treu}(2002)}]{K+T02}
{Koopmans}, L.~V.~E., \& {Treu}, T. 2002, \apjl, 568, L5

\bibitem[{{Koopmans} \& {Treu}(2003)}]{K+T03}
---. 2003, \apj, 583, 606

\bibitem[{{Koopmans} {et~al.}(2006){Koopmans}, {Treu}, {Bolton}, {Burles}, \&
  {Moustakas}}]{Koo++06}
{Koopmans}, L.~V.~E., {Treu}, T., {Bolton}, A.~S., {Burles}, S., \&
  {Moustakas}, L.~A. 2006, \apj, 649, 599

\bibitem[{{Kormann} {et~al.}(1994){Kormann}, {Schneider}, \&
  {Bartelmann}}]{KSB94}
{Kormann}, R., {Schneider}, P., \& {Bartelmann}, M. 1994, \aap, 284, 285

\bibitem[{{Lackner} \& {Ostriker}(2010)}]{L+O10}
{Lackner}, C.~N., \& {Ostriker}, J.~P. 2010, \apj, 712, 88

\bibitem[{{Lagattuta} {et~al.}(2010){Lagattuta}, {Fassnacht}, {Auger},
  {Marshall}, {Brada{\v c}}, {Treu}, {Gavazzi}, {Schrabback}, {Faure}, \&
  {Anguita}}]{Lag++09}
{Lagattuta}, D.~J., {Fassnacht}, C.~D., {Auger}, M.~W., {Marshall}, P.~J.,
  {Brada{\v c}}, M., {Treu}, T., {Gavazzi}, R., {Schrabback}, T., {Faure}, C.,
  \& {Anguita}, T. 2010, \apj, 716, 1579

\bibitem[{{Leauthaud} {et~al.}(2007)}]{Lea++07}
{Leauthaud}, A., {et~al.} 2007, \apjs, 172, 219

\bibitem[{Lewis \& Bridle(2002)}]{lewis2002}
Lewis, A., \& Bridle, S. 2002, Phys. Rev. D, 66, 103511

\bibitem[{{Mancini} {et~al.}(2010){Mancini}, {Daddi}, {Renzini}, {Salmi},
  {McCracken}, {Cimatti}, {Onodera}, {Salvato}, {Koekemoer}, {Aussel},
  {Floc'h}, {Willott}, \& {Capak}}]{Man++10}
{Mancini}, C., {Daddi}, E., {Renzini}, A., {Salmi}, F., {McCracken}, H.~J.,
  {Cimatti}, A., {Onodera}, M., {Salvato}, M., {Koekemoer}, A.~M., {Aussel},
  H., {Floc'h}, E.~L., {Willott}, C., \& {Capak}, P. 2010, \mnras, 401, 933

\bibitem[{{Marshall} {et~al.}(2007){Marshall}, {Treu}, {Melbourne}, {Gavazzi},
  {Bundy}, {Ammons}, {Bolton}, {Burles}, {Larkin}, {Le Mignant}, {Koo},
  {Koopmans}, {Max}, {Moustakas}, {Steinbring}, \& {Wright}}]{Mar++07}
{Marshall}, P.~J., {Treu}, T., {Melbourne}, J., {Gavazzi}, R., {Bundy}, K.,
  {Ammons}, S.~M., {Bolton}, A.~S., {Burles}, S., {Larkin}, J.~E., {Le
  Mignant}, D., {Koo}, D.~C., {Koopmans}, L.~V.~E., {Max}, C.~E., {Moustakas},
  L.~A., {Steinbring}, E., \& {Wright}, S.~A. 2007, \apj, 671, 1196

\bibitem[{{Merritt}(1999)}]{Mer99}
{Merritt}, D. 1999, \pasp, 111, 129

\bibitem[{{Meza} {et~al.}(2003){Meza}, {Navarro}, {Steinmetz}, \&
  {Eke}}]{Mez++03}
{Meza}, A., {Navarro}, J.~F., {Steinmetz}, M., \& {Eke}, V.~R. 2003, \apj, 590,
  619

\bibitem[{{Miralda-Escude}(1995)}]{Mir95}
{Miralda-Escude}, J. 1995, \apj, 438, 514

\bibitem[{{Naab} {et~al.}(2009){Naab}, {Johansson}, \& {Ostriker}}]{NJO09}
{Naab}, T., {Johansson}, P.~H., \& {Ostriker}, J.~P. 2009, \apjl, 699, L178

\bibitem[{{Naab} {et~al.}(2007){Naab}, {Johansson}, {Ostriker}, \&
  {Efstathiou}}]{Naa++07}
{Naab}, T., {Johansson}, P.~H., {Ostriker}, J.~P., \& {Efstathiou}, G. 2007,
  \apj, 658, 710

\bibitem[{{Natarajan} \& {Kneib}(1996)}]{N+K96}
{Natarajan}, P., \& {Kneib}, J.-P. 1996, \mnras, 283, 1031

\bibitem[{{Navarro} {et~al.}(2010){Navarro}, {Ludlow}, {Springel}, {Wang},
  {Vogelsberger}, {White}, {Jenkins}, {Frenk}, \& {Helmi}}]{Nav++10}
{Navarro}, J.~F., {Ludlow}, A., {Springel}, V., {Wang}, J., {Vogelsberger}, M.,
  {White}, S.~D.~M., {Jenkins}, A., {Frenk}, C.~S., \& {Helmi}, A. 2010,
  \mnras, 402, 21

\bibitem[{{Newman} {et~al.}(2010){Newman}, {Ellis}, {Treu}, \&
  {Bundy}}]{New++10}
{Newman}, A.~B., {Ellis}, R.~S., {Treu}, T., \& {Bundy}, K. 2010, \apjl, 717,
  L103

\bibitem[{{Nipoti} {et~al.}(2009{\natexlab{a}}){Nipoti}, {Treu}, {Auger}, \&
  {Bolton}}]{Nip++09}
{Nipoti}, C., {Treu}, T., {Auger}, M.~W., \& {Bolton}, A.~S.
  2009{\natexlab{a}}, \apjl, 706, L86

\bibitem[{{Nipoti} {et~al.}(2009{\natexlab{b}}){Nipoti}, {Treu}, \&
  {Bolton}}]{NTB09}
{Nipoti}, C., {Treu}, T., \& {Bolton}, A.~S. 2009{\natexlab{b}}, \apj, 703,
  1531

\bibitem[{{O{\~n}orbe} {et~al.}(2007){O{\~n}orbe}, {Dom{\'{\i}}nguez-Tenreiro},
  {S{\'a}iz}, \& {Serna}}]{O++07}
{O{\~n}orbe}, J., {Dom{\'{\i}}nguez-Tenreiro}, R., {S{\'a}iz}, A., \& {Serna},
  A. 2007, \mnras, 376, 39

\bibitem[{{Ohyama} {et~al.}(2002){Ohyama}, {Hamana}, {Kashikawa}, {Chiba},
  {Futamase}, {Iye}, {Kawabata}, {Aoki}, {Sasaki}, {Kosugi}, \&
  {Takata}}]{Ohy++02}
{Ohyama}, Y., {Hamana}, T., {Kashikawa}, N., {Chiba}, M., {Futamase}, T.,
  {Iye}, M., {Kawabata}, K.~S., {Aoki}, K., {Sasaki}, T., {Kosugi}, G., \&
  {Takata}, T. 2002, \aj, 123, 2903

\bibitem[{{Renzini}(2006)}]{Ren06}
{Renzini}, A. 2006, \araa, 44, 141

\bibitem[{{Robertson} {et~al.}(2006){Robertson}, {Cox}, {Hernquist}, {Franx},
  {Hopkins}, {Martini}, \& {Springel}}]{Rob++06c}
{Robertson}, B., {Cox}, T.~J., {Hernquist}, L., {Franx}, M., {Hopkins}, P.~F.,
  {Martini}, P., \& {Springel}, V. 2006, \apj, 641, 21

\bibitem[{{Salpeter}(1955)}]{Salpeter1955}
{Salpeter}, E.~E. 1955, \apj, 121, 161

\bibitem[{{Saracco} {et~al.}(2009){Saracco}, {Longhetti}, \& {Andreon}}]{SLA09}
{Saracco}, P., {Longhetti}, M., \& {Andreon}, S. 2009, \mnras, 392, 718

\bibitem[{{Schlegel} {et~al.}(1998){Schlegel}, {Finkbeiner}, \&
  {Davis}}]{schlegel98}
{Schlegel}, D.~J., {Finkbeiner}, D.~P., \& {Davis}, M. 1998, \apj, 500, 525

\bibitem[{{Sheth} {et~al.}(2003){Sheth}, {Bernardi}, {Schechter}, {Burles},
  {Eisenstein}, {Finkbeiner}, {Frieman}, {Lupton}, {Schlegel}, {Subbarao},
  {Shimasaku}, {Bahcall}, {Brinkmann}, \& {Ivezi{\'c}}}]{sheth2003}
{Sheth}, R.~K., {Bernardi}, M., {Schechter}, P.~L., {Burles}, S., {Eisenstein},
  D.~J., {Finkbeiner}, D.~P., {Frieman}, J., {Lupton}, R.~H., {Schlegel},
  D.~J., {Subbarao}, M., {Shimasaku}, K., {Bahcall}, N.~A., {Brinkmann}, J., \&
  {Ivezi{\'c}}, {\v Z}. 2003, \apj, 594, 225

\bibitem[{{Suyu} {et~al.}(2010){Suyu}, {Marshall}, {Auger}, {Hilbert},
  {Blandford}, {Koopmans}, {Fassnacht}, \& {Treu}}]{Suy++10}
{Suyu}, S.~H., {Marshall}, P.~J., {Auger}, M.~W., {Hilbert}, S., {Blandford},
  R.~D., {Koopmans}, L.~V.~E., {Fassnacht}, C.~D., \& {Treu}, T. 2010, \apj,
  711, 201

\bibitem[{{Suyu} {et~al.}(2009){Suyu}, {Marshall}, {Blandford}, {Fassnacht},
  {Koopmans}, {McKean}, \& {Treu}}]{Suy++09}
{Suyu}, S.~H., {Marshall}, P.~J., {Blandford}, R.~D., {Fassnacht}, C.~D.,
  {Koopmans}, L.~V.~E., {McKean}, J.~P., \& {Treu}, T. 2009, \apj, 691, 277

\bibitem[{{Thomas} {et~al.}(2005){Thomas}, {Maraston}, {Bender}, \& {Mendes de
  Oliveira}}]{Tho++05}
{Thomas}, D., {Maraston}, C., {Bender}, R., \& {Mendes de Oliveira}, C. 2005,
  \apj, 621, 673

\bibitem[{{Trager} {et~al.}(2000){Trager}, {Faber}, {Worthey}, \&
  {Gonz{\'a}lez}}]{Tra++00a}
{Trager}, S.~C., {Faber}, S.~M., {Worthey}, G., \& {Gonz{\'a}lez}, J.~J. 2000,
  \aj, 120, 165

\bibitem[{{Treu} {et~al.}(2010){Treu}, {Auger}, {Koopmans}, {Gavazzi},
  {Marshall}, \& {Bolton}}]{Tre++10}
{Treu}, T., {Auger}, M.~W., {Koopmans}, L.~V.~E., {Gavazzi}, R., {Marshall},
  P.~J., \& {Bolton}, A.~S. 2010, \apj, 709, 1195

\bibitem[{{Treu} {et~al.}(2005){Treu}, {Ellis}, {Liao}, {van Dokkum}, {Tozzi},
  {Coil}, {Newman}, {Cooper}, \& {Davis}}]{Tre++05}
{Treu}, T., {Ellis}, R.~S., {Liao}, T.~X., {van Dokkum}, P.~G., {Tozzi}, P.,
  {Coil}, A., {Newman}, J., {Cooper}, M.~C., \& {Davis}, M. 2005, \apj, 633,
  174

\bibitem[{{Treu} {et~al.}(2009){Treu}, {Gavazzi}, {Gorecki}, {Marshall},
  {Koopmans}, {Bolton}, {Moustakas}, \& {Burles}}]{Tre++09}
{Treu}, T., {Gavazzi}, R., {Gorecki}, A., {Marshall}, P.~J., {Koopmans},
  L.~V.~E., {Bolton}, A.~S., {Moustakas}, L.~A., \& {Burles}, S. 2009, \apj,
  690, 670

\bibitem[{{Treu} {et~al.}(2006){Treu}, {Koopmans}, {Bolton}, {Burles}, \&
  {Moustakas}}]{Tre++06}
{Treu}, T., {Koopmans}, L.~V., {Bolton}, A.~S., {Burles}, S., \& {Moustakas},
  L.~A. 2006, \apj, 640, 662

\bibitem[{{Treu} \& {Koopmans}(2002)}]{T+K02a}
{Treu}, T., \& {Koopmans}, L.~V.~E. 2002, \apj, 575, 87

\bibitem[{{Treu} \& {Koopmans}(2004)}]{T+K04}
---. 2004, \apj, 611, 739

\bibitem[{{Treu} {et~al.}(1998){Treu}, {Stiavelli}, {Walker}, {Williams},
  {Baum}, {Bernstein}, {Blacker}, {Carollo}, {Casertano}, {Dickinson}, {De
  Mello}, {Ferguson}, {Fruchter}, {Lucas}, {MacKenty}, {Madau}, \&
  {Postman}}]{Tre++98}
{Treu}, T., {Stiavelli}, M., {Walker}, A.~R., {Williams}, R.~E., {Baum}, S.~A.,
  {Bernstein}, G., {Blacker}, B.~S., {Carollo}, C.~M., {Casertano}, S.,
  {Dickinson}, M.~E., {De Mello}, D.~F., {Ferguson}, H.~C., {Fruchter}, A.~S.,
  {Lucas}, R.~A., {MacKenty}, J., {Madau}, P., \& {Postman}, M. 1998, \aap,
  340, L10

\bibitem[{{van Albada} \& {Sancisi}(1986)}]{v+S86}
{van Albada}, T.~S., \& {Sancisi}, R. 1986, Royal Society of London
  Philosophical Transactions Series A, 320, 447

\bibitem[{{van der Marel}(1994)}]{vandermarel1994}
{van der Marel}, R.~P. 1994, \mnras, 270, 271

\bibitem[{{van der Wel} {et~al.}(2009){van der Wel}, {Bell}, {van den Bosch},
  {Gallazzi}, \& {Rix}}]{vdW++09}
{van der Wel}, A., {Bell}, E.~F., {van den Bosch}, F.~C., {Gallazzi}, A., \&
  {Rix}, H.-W. 2009, \apj, 698, 1232

\bibitem[{{van der Wel} {et~al.}(2005){van der Wel}, {Franx}, {van Dokkum},
  {Rix}, {Illingworth}, \& {Rosati}}]{vDW++05}
{van der Wel}, A., {Franx}, M., {van Dokkum}, P.~G., {Rix}, H.-W.,
  {Illingworth}, G.~D., \& {Rosati}, P. 2005, \apj, 631, 145

\bibitem[{{van Dokkum}(2008)}]{vdo08}
{van Dokkum}, P.~G. 2008, \apj, 674, 29

\bibitem[{{van Dokkum} {et~al.}(2008){van Dokkum}, {Franx}, {Kriek}, {Holden},
  {Illingworth}, {Magee}, {Bouwens}, {Marchesini}, {Quadri}, {Rudnick},
  {Taylor}, \& {Toft}}]{vDo++08}
{van Dokkum}, P.~G., {Franx}, M., {Kriek}, M., {Holden}, B., {Illingworth},
  G.~D., {Magee}, D., {Bouwens}, R., {Marchesini}, D., {Quadri}, R., {Rudnick},
  G., {Taylor}, E.~N., \& {Toft}, S. 2008, \apjl, 677, L5

\bibitem[{{Wucknitz} {et~al.}(2004){Wucknitz}, {Biggs}, \& {Browne}}]{WBB04}
{Wucknitz}, O., {Biggs}, A.~D., \& {Browne}, I.~W.~A. 2004, \mnras, 349, 14

\end{thebibliography}

% - - - - - - - - - - - - - - - - - - - - - - - - - - - - - - - - - - - - - - - 

% PJM: Table should go further up, but If I do that it is
% not set correctly, I don't know why... Replacing clearpage with newpage 
% has no effect. Need to flush floats? I checked its not  that the table
% has to be on the last page. Google for latex landscape...
% AJR:  Release notes (02/09/03) for emulateapj.cls:
%				(a) \rotate doesn't work (too difficult to implement).

% PJM: OK, table now set correctly, after refs. Left note further up.

\clearpage

%%%%%%%%%%%%%%%%%%%%%%%%%%%%%%%
\begin{landscape}
\renewcommand{\arraystretch}{1.50} 
\begin{deluxetable}{lcccccccccccccccc}
%\rotate
\tabletypesize{\small}
\tablecaption{\label{table:measured}
Measured SL2S galaxy-scale lens properties}
\tablehead{
       Name                    & $\zd$  & $\zs$ &  $z_{\rm s,pdf}$ & $\sigma$      & S/N         & $\REin$ & $q_{\rm mass}$ & $\Reff$ & $q_{*}$ & $\magu$ & $\magg$ & $\magr$ & $\magi$ & $\magz$ &Flag &Run \\ 
                               &        &       &                  & (km s$^{-1}$) &(\AA$^{-1}$) & (arcsec)&                & (arcsec)&         &         &         &         &         &       &  &     \\} 
\startdata
J021411$-$040502  & 0.6080 & - & $ 1.57^{+0.50}_{-0.35} $ & $ - $ & 6.1 & 0.918 &  0.33 & $ 0.94 $ & 0.89 &  22.46 & 21.08 & 19.57 & 18.78 & 19.41 & HST &2b \\ 
J021737$-$051329  & 0.6458 & 1.847 & $ 1.74^{+0.53}_{-0.37} $ & $ 257 \pm 26 $ & 14.5 & 1.268 &  0.91 & $ 0.77 $ & 0.89 &  21.57 & 20.45 & 19.51 & 19.47 & 19.16 & HST &2b,4c \\ 
J021902$-$082934  & 0.3898 & - & $ 1.30^{+0.65}_{-0.43} $ & $ 305 \pm 25 $ & 19.7 & 0.918 &  0.53 & $ 0.90 $ & 0.72 &  22.50 & 20.77 & 20.14 & 19.33 & 18.70 & CFHT &3a \\ 
J022056$-$063934  & 0.3297 & - & $ 1.47^{+0.68}_{-0.49} $ & $ 242 \pm 28 $ & 25.7 & 1.250 &  0.73 & $ 1.47 $ & 0.57 &  21.09 & 20.12 & 18.30 & 18.35 & 17.62 & CFHT &3b \\ 
J022511$-$045433  & 0.2380 & 1.1988 & $ 1.25^{+0.55}_{-0.39} $ & $ 241 \pm 12 $ & 53.0 & 1.770 &  0.58 & $ 1.90 $ & 0.75 &  19.37 & 17.91 & 17.39 & 16.84 & 16.74 & HST &5 \\ 
J022610$-$042011  & 0.4943 & 1.232 & $ 1.54^{+0.60}_{-0.40} $ & $ 266 \pm 21 $ & 16.4 & 1.153 &  0.92 & $ 0.56 $ & 0.94 &  22.04 & 20.47 & 19.66 & 18.78 & 18.58 & HST &4a \\ 
J022648$-$040610  & 0.7663 & - & $ 2.18^{+0.67}_{-0.47} $ & $ - $ & 9.4 & 1.306 &  0.82 & $ 1.20 $ & 0.37 &  23.34 & 22.15 & 21.40 & 20.00 & 19.60 & HST &2a \\ 
J022648$-$090421  & 0.4563 & - & $ 1.98^{+0.78}_{-0.57} $ & $ 301 \pm 18 $ & 30.8 & 1.582 &  0.83 & $ 1.30 $ & 0.80 &  22.47 & 20.06 & 18.14 & 18.32 & 17.73 & CFHT &4a \\ 
J023251$-$040823  & 0.3516 & - & $ 1.52^{+0.74}_{-0.53} $ & $ 264 \pm 17 $ & 22.6 & 1.102 &  0.80 & $ 0.81 $ & 0.67 &  21.87 & 20.13 & 18.41 & 18.58 & 18.52 & HST &3a \\ 
J140123$+$555705  & 0.5263 & - & $ 1.62^{+0.58}_{-0.42} $ & $ - $ & 10.0 & 1.186 &  0.49 & $ 0.76 $ & 0.73 &  22.77 & 21.35 & 20.15 & 19.14 & 18.58 & HST &1 \\ 
J140614$+$520252  & 0.4797 & - & $ - $ & $ - $ & 8.5 & - &  - & $ 2.15 $ & 0.49 &  22.05 & 20.06 & 18.47 & 18.29 & 17.93 & CFHT &1 \\ 
J141137$+$565119  & 0.3218 & 1.420 & $ 1.37^{+0.77}_{-0.52} $ & $ 228 \pm 20 $ & 34.0 & 0.924 &  0.90 & $ 0.76 $ & 0.74 &  21.08 & 19.75 & 18.59 & 18.49 & 18.32 & HST &6 \\ 
J220629$+$005728  & 0.7044 & - & $ - $ & $ - $ & 10.4 & - &  - & $ 2.25 $ & 0.99 &  23.11 & 21.47 & 20.64 & 20.59 & 19.04 & CFHT &3a \\ 
J221326$-$000946  & 0.3378 & - & $ 1.30^{+0.63}_{-0.44} $ & $ 183 \pm 34 $ & 18.9 & 1.076 &  0.19 & $ 0.41 $ & 0.30 &  23.93 & 22.31 & 20.10 & 20.00 & 19.49 & HST &5 \\ 
J221407$-$180712  & 0.6505 & - & $ 1.12^{+0.46}_{-0.22} $ & $ 167 \pm 43 $ & 8.3 & 0.411 &  0.69 & $ 0.57 $ & 0.65 &  23.91 & 21.96 & 21.04 & 20.43 & 19.79 & HST &3b \\ 
J221606$-$175131  & 0.8602 & - & $ - $ & $ 282 \pm 44 $ & 13.5 & - &  - & $ 0.93 $ & 0.87 &  23.35 & 22.39 & 21.36 & 20.68 & 19.72 & CFHT &3a \\ 
J221929$-$001743  & 0.2888 & 1.0232 & $ 0.91^{+0.54}_{-0.32} $ & $ 263 \pm 26 $ & 44.4 & 0.736 &  0.75 & $ 1.00 $ & 0.75 &  20.39 & 18.38 & 16.78 & 17.78 & 17.61 & CFHT &4a,4b \\ 
\enddata
\tablecomments{All spectroscopically-observed lens systems are shown.
$\Reff$ values are measured at the intermediate axis. The signal
to noise ratio per pixel was calculated over the rest wavelength range
4000--5000\AA. 
Typical uncertainties on $\REin$ and $\Reff$ are 5\% and 10\%, respectively. 
The column, Flag, indicates whether HST or CFHT data was used to measure 
$\REin$ and $\Reff$. 
}
\end{deluxetable}
\clearpage
\end{landscape}
%%%%%%%%%%%%%%%%%%%%%%%%%%%%%%%

%%%%%%%%%%%%%%%%%%%%%%%%%%%%%%%
\renewcommand{\arraystretch}{1.50} 
\begin{deluxetable}{lccccccccc}
\tabletypesize{\small}
\tablecaption{\label{table:inferred}
Inferred SL2S galaxy-scale lens properties}
\tablehead{
      Name    &$R_{\rm Ein}$                & $\sigmae2$  & $\sigmasie$ & $M_{*}$ &$\MEin$  & $\gamma{\prime}$     & $\fdmsre2$  &$\fdmcre2$  &        \\ 
                      & (kpc)      &  (km s$^{-1}$)      &  (km s$^{-1}$)     &     (10$^{11}M_{\odot}$) &(10$^{11}M_{\odot}$)     \\ }
\startdata
J021737$-$051329  & 8.76 & 273 $\pm$ 27 & $ 289 $ & 1.77 $^{+0.53}_{-0.37} $ & 5.35  & 1.99 $^{+0.11}_{-0.12} $ & 0.65 $^{+0.11}_{-0.08} $     & 0.80 $^{+0.11}_{-0.04} $     \\
J021902$-$082934  & 4.85 & 324 $\pm$ 26 & $ 228^{+37}_{-13} $ & 1.25 $^{+0.29}_{-0.19} $ & 1.85 $^{+0.65}_{-0.21} $ & 2.55 $^{+0.13}_{-0.17} $ & 0.71 $^{+0.07}_{-0.06} $     & 0.83 $^{+0.13}_{-0.03} $     \\
J022056$-$063934  & 5.94 & 252 $\pm$ 29 & $ 256^{+37}_{-12} $ & 3.48 $^{+0.89}_{-0.68} $ & 2.81 $^{+0.72}_{-0.25} $ & 2.04 $^{+0.19}_{-0.19} $ & 0.34 $^{+0.18}_{-0.14} $     & 0.63 $^{+0.19}_{-0.08} $     \\
J022511$-$045433  & 6.67 & 251 $\pm$ 12 & $ 287 $ & 2.82 $^{+0.59}_{-0.50} $ & 4.02  & 1.88 $^{+0.07}_{-0.07} $ & 0.55 $^{+0.10}_{-0.08} $     & 0.74 $^{+0.07}_{-0.04} $     \\
J022610$-$042011  & 6.99 & 288 $\pm$ 22 & $ 279 $ & 2.74 $^{+0.55}_{-0.49} $ & 3.99  & 2.09 $^{+0.08}_{-0.09} $ & 0.21 $^{+0.19}_{-0.15} $     & 0.56 $^{+0.08}_{-0.08} $     \\
J022648$-$090421  & 9.18 & 315 $\pm$ 18 & $ 315^{+72}_{-23} $ & 6.38 $^{+1.17}_{-1.27} $ & 6.52 $^{+2.41}_{-0.86} $ & 2.16 $^{+0.08}_{-0.09} $ & 0.24 $^{+0.16}_{-0.15} $     & 0.57 $^{+0.08}_{-0.09} $     \\
J023251$-$040823  & 5.46 & 282 $\pm$ 18 & $ 239^{+31}_{-10} $ & 1.64 $^{+0.31}_{-0.27} $ & 2.29 $^{+0.74}_{-0.21} $ & 2.27 $^{+0.10}_{-0.12} $ & 0.53 $^{+0.10}_{-0.08} $     & 0.73 $^{+0.10}_{-0.05} $     \\
J141137$+$565119  & 4.32 & 246 $\pm$ 21 & $ 214 $ & 1.33 $^{+0.29}_{-0.24} $ & 1.45  & 2.30 $^{+0.12}_{-0.14} $ & 0.45 $^{+0.13}_{-0.10} $     & 0.69 $^{+0.12}_{-0.06} $     \\
J221326$-$000946  & 5.19 & 203 $\pm$ 37 & $ 241^{+38}_{-13} $ & 0.92 $^{+0.20}_{-0.18} $ & 2.17 $^{+0.59}_{-0.21} $ & 1.84 $^{+0.19}_{-0.20} $ & 0.12 $^{+0.33}_{-0.23} $     & 0.52 $^{+0.19}_{-0.14} $     \\
J221407$-$180712  & 2.85 & 181 $\pm$ 46 & $ 171^{+29}_{-9} $ & 1.68 $^{+0.32}_{-0.23} $ & 0.57 $^{+0.16}_{-0.06} $ & 1.64 $^{+0.53}_{-0.29} $ & 0.03 $^{+0.35}_{-0.26} $     & 0.39 $^{+0.53}_{-0.18} $     \\
J221929$-$001743  & 3.19 & 278 $\pm$ 27 & $ 197 $ & 2.44 $^{+0.50}_{-0.40} $ & 0.91  & 2.66 $^{+0.13}_{-0.16} $ & 0.02 $^{+0.21}_{-0.16} $     & 0.40 $^{+0.13}_{-0.11} $     \\
\enddata
\tablecomments{Only the 11~modeled 
lenses with measured velocity dispersions are shown. 
$\sigmae2$ is the measured stellar velocity dispersion
corrected to a standard aperture. $\sigmasie$ values calculated from measured 
source redshifts are given without uncertainty and $\sigmasie$ values calculated from $\zspdf$ 
are given with the 16$\th$ and 84$\th$ percentiles (and were also 
weighted by the \citeauthor{sheth2003} fitting function and selection function).
$M_{*}$ is the total stellar mass. $\MEin$ is
the total mass enclosed at the Einstein radius. Dark matter fractions are given
at $\Reff/2$ for both \citeauthor{Salpeter1955} and \citeauthor{Chabrier2003}
IMFs.}
\end{deluxetable}
%%%%%%%%%%%%%%%%%%%%%%%%%%%%%%%

%-------------------------------------------------------------------------------

\end{document}